\documentclass[numbers]{assets/matterlab}

\usepackage{microtype}
\usepackage{graphicx}
\usepackage{booktabs}
\usepackage{float}
\usepackage{xurl}
\usepackage{hyperref}
\usepackage{titlesec}
\usepackage{titletoc}
\usepackage{cinzel}

\usepackage{amsmath}
\usepackage{amssymb}
\usepackage{mathtools}
\usepackage{amsthm}
\usepackage{bm}

\usepackage[noabbrev,nameinlink]{cleveref}

\usepackage{amsmath,amsfonts,bm}

\def\eqref#1{equation~\ref{#1}}

\def\1{\bm{1}}

\DeclareMathAlphabet{\mathsfit}{\encodingdefault}{\sfdefault}{m}{sl}
\SetMathAlphabet{\mathsfit}{bold}{\encodingdefault}{\sfdefault}{bx}{n}

\usepackage{orcidlink}
\usepackage{comment}
\usepackage[version=4]{mhchem}
\usepackage{longtable}
\usepackage{array}
\renewcommand{\arraystretch}{1.7}
\usepackage{multirow}           
\usepackage{amsfonts}           
\usepackage{nicefrac}
\usepackage{duckuments}         
\usepackage{thmtools,thm-restate}
\usepackage{enumitem}
\usepackage{xcolor,colortbl}
\usepackage{tikz}
\usepackage{tikz-cd}
\usepackage{caption}
\usepackage{subcaption}
\usepackage{listings}
\usepackage{gensymb}
\usepackage{textcomp} 

\usepackage[utf8]{inputenc}
\usepackage{lmodern}
\usepackage{microtype}
\usepackage{amsmath,amssymb}
\usepackage{booktabs}
\usepackage{array}
\usepackage{enumitem}
\usepackage{xcolor}
\usepackage{hyperref}
\usepackage{longtable}
\usepackage{tabularx}
\usepackage[inkscapelatex=false]{svg}

\newcommand{\addressCHEM}{Department of Chemistry, University of Toronto,  80 St. George St., Toronto, ON M5S 3H6, Canada}
\newcommand{\addressAC}{Acceleration Consortium, 700 University Ave., Toronto, ON M7A 2S4, Canada}
\newcommand{\addressCS}{Department of Computer Science, University of Toronto, 40 St George St., Toronto, ON M5S 2E4, Canada}
\newcommand{\addressVECTOR}{Vector Institute for Artificial Intelligence, W1140-108 College St., Schwartz Reisman Innovation Campus, Toronto, ON M5G 0C6, Canada}
\newcommand{\addressMSE}{Department of Materials Science \& Engineering, University of Toronto, 184 College St., Toronto, ON M5S 3E4, Canada}
\newcommand{\addressCHEMENG}{Department of Chemical Engineering \& Applied Chemistry, University of Toronto, 200 College St., Toronto, ON M5S 3E5, Canada}

\newcommand{\addressCIFAR}{Canadian Institute for Advanced Research (CIFAR), 661 University Ave., Toronto,
ON M5G 1M1, Canada}
\newcommand{\addressNVIDIA}{NVIDIA, 431 King St W \#6th, Toronto, ON M5V 1K4, Canada}
\newcommand{\addressMS}{Institute of Medical Science, 1 King's College Circle, Medical Sciences Building, Room 2374, Toronto, ON M5S 1A8, Canada}

\ifdefined\CASSCFMainOnly
\usepackage{xr}
\fi
\usepackage[commandnameprefix=always]{changes}
\setdeletedmarkup{\textcolor{red}{\sout{#1}}}

\title{{\cinzel El Agente Quntur}: Agentic Active-Space Selection for CASSCF}

\author[1,6,\orcidlink{0000-0003-1267-3627}]{Juan B. P\'erez-S\'anchez}
\author[10,\orcidlink{0000-0002-6614-3632}]{Viviana Roman-Ventura}
\author[11,\orcidlink{0000-0001-8754-1341}]{Olalla Nieto Faza}
\author[2,\orcidlink{0009-0004-6626-6714}]{Aiwei Yin}
\author[*,1,2,3,4,5,6,7,8,9,\orcidlink{0000-0002-8277-4434}]{Al\'an Aspuru-Guzik}
\author[*,1,2,7, \orcidlink{0000-0002-8446-7956}]{Varinia Bernales}

\affiliation[1]{\addressCHEM}
\affiliation[2]{\addressCS}
\affiliation[3]{\addressMSE}
\affiliation[4]{\addressCHEMENG}
\affiliation[5]{\addressAC}
\affiliation[6]{\addressVECTOR}
\affiliation[7]{\addressCIFAR}
\affiliation[8]{\addressNVIDIA}
\affiliation[9]{\addressMS}
\affiliation[10]{Departamento de Físico-Química Molecular, Universidad Andrés Bello, República 275, Santiago, 837014, Chile}
\affiliation[11]{Departamento de Química Orgánica, Universidad de Vigo, Campus Lagoas-Marcosende, Vigo, Spain}
\affiliation[*]{corresponding authors}

\contactdetails{{\small\textbf{Author contact details:} J.B.P.-S.: \email{juan.perez@utoronto.ca}; V.R.-V.: \email{v.romanventura@uandresbello.edu}; O.N.-F.: \email{faza@uvigo.es}; A.Y.: \email{aiwei.yin@mail.utoronto.ca}; A.A.-G.: \email{aspuru@utoronto.ca}; V.B.: \email{bernales@cs.utoronto.ca}.}}

\abstract{
Multireference calculations require time-consuming expert decisions about active spaces, state identity, and convergence. This study asks whether agents can make these decisions reliably. Using the {\cinzel El Agente Quntur} multi-agent harness with GPT-5.6 Sol and GPT-5.6 Terra as its language models, agents study 32 fixed-geometry singlet excitations, detect missing active orbitals, compare excited states, and revise state assignments as calculations progress. Prompts specify equal-weight two-singlet state-averaged complete active space self-consistent field (CASSCF) calculations followed by strongly contracted second-order N-electron valence state perturbation theory (NEVPT2) with def2-TZVPD; agents choose and refine the active spaces. Excitation energies are evaluated against published complete active space second-order perturbation theory (CASPT2) reference values at matching geometries. Agentic execution reduces mean absolute error by 48\% relative to six deterministic Active Space Finder protocols. Computational traces vary substantially across runs in selection strategies and cost, even for the same molecule. These results underscore the advantage of agentic active space selection over fixed workflows, yet leave room to improve efficiency through runtime-aware decision-making, reduced inter-agent communication overhead and redundancy, more systematic checks of state identity, and clearer stopping criteria.

}

\date{\today}
\correspondence{\email{varinia@bernales.org} and \email{alan@aspuru.com}}
\metadata[Keywords]{Agentic AI; quantum chemistry; CASSCF; excited states}

\begin{document}

\enlargethispage{1cm}
\maketitle


\section{Introduction}

Electronic-structure calculations link molecular structure to, in principle, all computable observables. Most often, energies and response spectra. In many cases, especially in many-body theory, a single electronic configuration is not a sufficient starting point for accurately describing many crucial chemical systems. Near-degenerate configurations can compete in bond breaking, open-shell systems, and for many types of electronic excitations, where a method built around one dominant configuration may describe the states unevenly \cite{helgaker2000molecular,Cohen2008,Szalay2012,Lischka2018}. The need for a multiconfigurational description is longstanding: the Heitler--London wavefunction for H$_2$ already combined electronic arrangements to represent a chemical bond \cite{Heitler1927}. In complete active-space self-consistent field (CASSCF) theory, selected active electrons are distributed among selected active orbitals in all allowed configurations while the orbitals are optimized \cite{Roos1980}. State averaging supplies a common orbital description for several electronic states \cite{Werner1981}, and a post-CASSCF treatment such as strongly contracted $n$-electron valence-state perturbation theory (SC-NEVPT2) recovers dynamical correlation outside the active space \cite{Angeli2001,Angeli2001SC}. These methods enable multireference excited-state calculations, but their predictive value depends on the chosen states and orbitals. Some practitioners criticize these approaches as not proper {\sl modern chemistries} \cite{Pople1973theoretical,HeadGordon1996Quantum}, because they involve human choice, chemical intuition, and experience in selecting the active orbitals to include in the correlated treatment. In this work, we explore whether complex agentic AI systems can make these choices automatically, so we may enter an era of {\sl agentic model chemistries} where agents reliably design, adapt, and extend active spaces themselves, helping reduce criticism from the model chemistry ``tribe''.

Choosing those orbitals has been a practical problem for as long as compact multireference models have been used \cite{Veryazov2011}. An active space is not fully specified by its number of electrons and orbitals: its members must represent the bonding, antibonding, nonbonding, charge-transfer, or diffuse degrees of freedom relevant to the property being computed \cite{Keller2015,Stein2016}. Omitting an important orbital can exclude the configuration that defines an excitation; including too many weakly relevant orbitals can make the calculation much more expensive and destabilize orbital optimization. The difficulty grows for an excitation energy because the ground and excited states must be described in a balanced space \cite{Sharma2019Valence,Sharma2019Density,Ghosh2017Generalized}. Benchmarks of CASSCF and automated multireference excitation calculations expose this sensitivity \cite{HelmichParis2019,King2022}. A numerically converged state-averaged calculation can still return the wrong excited-state root or a plausible energy from an unsuitable set of orbitals \cite{Shirazi2025ASF}. Active-space selection is therefore a model choice whose consequences must be checked in the resulting wavefunctions, not merely an input-preparation step.

Automated approaches answer different parts of the orbital-selection problem. Projection and localization schemes start from specified atomic or fragment character: atomic valence active spaces (AVAS) identify orbitals with target valence content, while subsystem projected atomic orbital decomposition (SPADE) and imposed automatic selection and localization of complete active spaces (iCAS) construct chemically targeted subspaces \cite{Sayfutyarova2017,Claudino2019,Kolodzeiski2023,Lei2021}. Occupation-based schemes draw candidates from approximate densities, including unrestricted Hartree--Fock natural orbitals and active-space selection based on first-order perturbation theory (ASS1ST) \cite{Pulay1988,Jensen1988,Keller2015,Khedkar2019,Khedkar2020}. Entanglement-based autoCAS and quantum information-assisted complete active space optimization (QICAS) use correlation diagnostics to select or refine orbitals \cite{Stein2016,Stein2019,Ding2023}. Other approaches target excitations using alternative initial orbitals or dipole-moment tests, including a recent CASCI-based dipole screen \cite{Bao2018,Bao2019,Kaufold2023,Kaufold2026CASCI}. Orbital rankings and learned selection rules provide further options \cite{King2021,Jeong2020,Golub2021}. Each approach proposes a candidate space; whether that space retains the intended orbital character after optimization and represents the intended excited state must still be manually checked in the completed calculation.

The Active Space Finder (ASF) uses correlated preliminary calculations, a two-electron cumulant, and one-orbital entropies to construct candidate spaces \cite{HQSActiveSpaceFinder}. Its recent excitation-energy study compares six choices of initial orbitals and selection thresholds under a common state-averaged CASSCF/SC-NEVPT2 protocol \cite{Shirazi2025ASF}. Excited-state reference collections and comparisons of post-CASSCF methods supply essential context for interpreting such benchmarks \cite{Loos2018,Veril2021,Sarkar2022}. These methods draw on different wave-function-derived quantities and make active-space selection more systematic, but the selected space remains a hypothesis about which electronic configurations matter, and that hypothesis needs testing and possible revision. If an excitation energy is poor or a calculation fails, should the active space be enlarged, the initial orbitals or selection criteria changed, or additional roots included? Similarly, a fixed fallback rule may address a known failure pattern but cannot cover all possible problems: omitted active orbitals, root misassignment, convergence failures, or limitations of the electronic-structure method itself. Moreover, each additional calculation incurs a computational cost, so the central question is whether using information from completed calculations to guide subsequent decisions can improve accuracy and reliability sufficiently to justify that cost, compared with simply accepting the initial result of an automatic active-space selector.

AI scientists are currently built primarily as so-called harnesses \cite{Li2026Harness}. A harness is a complex software system that combines tool access, careful context and procedural engineering, safety, and cloud access with large-language models. Examples of harnesses made for coding include Codex \cite{openai_codex}, Claude Code \cite{anthropics_claude_code}, or Pi \cite{earendil_works_pi}. AI scientific harnesses include commercial options such as Claude Science \cite{Anthropic2026ClaudeScience} and Kosmos \cite{Mitchener2025kosmos}, or in our case, the academic-hosted family of \href{https://elagente.ca}{{\cinzel El Agente}} systems \cite{Zou2025,Cao2025Automating,gustin2026elagentecuantico,Perez-Zhou2026ElAgenteQuntur,Zhang2026Forjador,choi2026elagenteEstructural,Ko2026ElAgentePotente,mueller2026elagenteFarmaco,Kang2026ElAgenteSeguro,bai-aldossary2026grafico,Govind2026El}. One may ask whether harnesses built for coding are enough for science, and our work shows that there is a gap in using raw models in coding harnesses vs. specialized AI scientists \cite{Perez-Zhou2026ElAgenteQuntur,Zhang2026Forjador,Ko2026ElAgentePotente,Kang2026ElAgenteSeguro}. Skeptics should be pointed to the fact that Claude offers both Claude Code and a specialist AI scientist harness: Claude Science.

\begin{figure}[!htbp]
\centering
\includegraphics[width=0.96\linewidth,height=0.40\textheight,keepaspectratio]{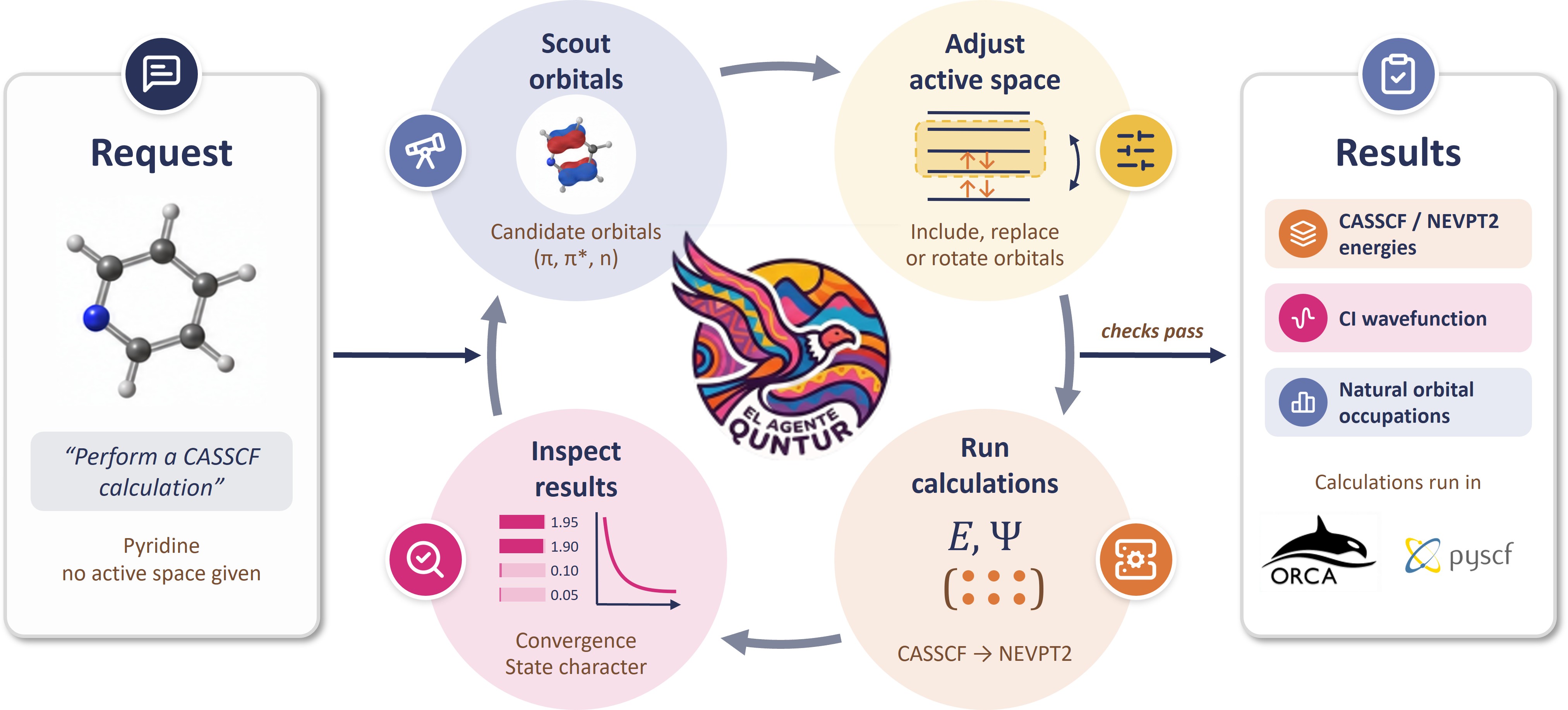}
\caption{Conceptual workflow for agent-guided multireference excited-state calculations. Given a molecular geometry and target excitation, language-model agents can inspect preliminary orbitals, choose and revise an active space, plan and run CASSCF/SC-NEVPT2 calculations, and use convergence and state-character evidence to decide whether another calculation is needed. The reported result includes an excitation energy, the selected active space, and the computed state assignment. The loop illustrates available decisions; the examples and benchmarks studied in this article did not prescribe a common sequence of steps for every case.}
\label{fig:intro-agent-workflow}
\end{figure}

Rather than applying a single orbital ranking or threshold to every case, AI scientists can consider orbital character, occupations and entropies, convergence behaviour, and the identity of the computed states together. It can then decide whether to change the active orbitals, include more singlet roots, retain the existing calculation, or stop because the likely gain does not justify another job. Figure~\ref{fig:intro-agent-workflow} depicts this cycle of orbital selection, calculation, and evidence-based revision. The {\cinzel El Agente} family has applied model-driven decisions and scientific tools to quantum chemistry \cite{Zou2025,Perez-Zhou2026ElAgenteQuntur}, molecular editing \cite{choi2026elagenteEstructural}, chemical safety \cite{Kang2026ElAgenteSeguro}, solid-state, quantum, and atomistic simulations \cite{gustin2026elagentecuantico,Govind2026El,Ko2026ElAgentePotente}, and the organization of scientific execution and reusable tools \cite{Zhang2026Forjador,bai-aldossary2026grafico}. In its original benchmark, {\cinzel El Agente Quntur} could perform CASSCF calculations, but it left a more demanding question open: can model-driven agents propose appropriate active orbitals, examine the resulting states and orbitals, and revise their calculations when the evidence challenges an initial choice? If so, how do they compare with deterministic active-space selectors?

Lee and Rondinelli's multireference agent inspects calculation outputs and can revise active spaces, state targeting, and convergence strategies; an expert-written decision ladder guides but does not execute those choices \cite{Lee2026Multireference}. Li and Sun's CAFES also uses completed-task diagnostics to propose refinements, including active-space review, within a registered workflow with qualification checks and approval gates \cite{Li2026CAFES}. Its lutein CASSCF example maintains a selected active space across geometries, while its explicit demonstration of adaptive task selection concerns a Hubbard-model scan. For the present study, what matters is whether such conditional choices yield better predictions than a fixed automatic-selection workflow, and at what computational cost.

Evaluating this approach requires examining both the excitation energies and the decisions used to obtain them. Calculation records can show whether agents recognize missing orbitals, distinguish competing excited states, and revise calculations when the results challenge their initial choices. They also reveal which additional calculations resolve a scientific question and which add work without changing the outcome.

Here, we ask whether agents making decisions during a CASSCF calculation obtain better predictions than a fixed automatic active-space selection workflow, and what additional effort those decisions require. Using {\cinzel El Agente Quntur}~\cite{Perez-Zhou2026ElAgenteQuntur}, we examine 32 fixed geometries from the Active Space Finder excitation benchmark~\cite{Shirazi2025ASF}. Each prompt requests an active-space choice and a vertical excitation energy from state-averaged CASSCF followed by SC-NEVPT2. We compare the reported gaps with CASPT2 values at matching geometries~\cite{Schreiber2008,Hoyer2016} and with six published ASF protocols~\cite{Shirazi2025ASF}. We then examine the trajectories to establish which decisions the calculations support, where state or protocol qualifications remain, and how much work the agents used to reach their answers. Finally, we repeat agent runs three times for two representative benchmark questions to study LLM variability and its advantages and disadvantages.

\section{Methods}

\subsection{The AI scientist harness and base models}

{\cinzel El Agente Quntur} is a hierarchical multi-agent system that simultaneously performs deep-research and computational quantum chemistry calculations given a high-level chemistry request \citep{Perez-Zhou2026ElAgenteQuntur}. A computational chemist plans the study and obtains user approval, while subagents coordinate geometry preparation, method-specific input generation, execution, and analysis to specialist agents. {\cinzel Quntur}'s execution backend in this work includes ORCA 6.0 \cite{Neese2012The,Neese_ORCA6}, and PySCF \cite{Qiming2018PySCF}. It also has access to useful Python packages for multireference calculations like ASF \cite{HQSActiveSpaceFinder} and MOKIT \cite{zoujingxiang2026MOKIT}.

For this study, we use GPT-5.6 Sol for scientific planning and specialist decisions, and GPT-5.6 Terra for supporting roles, including software execution and file analysis. Across the 32 selected sessions, the recorded workflow contains 5,471 Sol calls and 3,201 Terra calls. Supporting Information Table~\ref{tab:quntur-agents} lists the configured model for each role. We note that several other open-weight models and frontier models could be employed with the harness and that further ablation over different models is not necessary for the focus of this paper. Models evolve weekly, and comparing models is a rabbit hole that scientific publications should avoid. Between submission and review, several model families could have been released, and models, both open-weight and proprietary, are approaching similar capabilities \cite{Stanford2025technical,Stanford2026technical,Artificialanalysis2026openweights,Arena2026leaderboard,Arcprize2026leaderboard,Chiang2024chatbotarena,Chollet2025arcagi2,hle2026}.

For CASSCF tasks, the input-file service consults a dedicated expert on chemically relevant active orbitals, active-electron counts, state averaging, and orbital-index selection. The expert can assess orbital evidence and recommend focused diagnostics when the choice is uncertain. The service prepares an ORCA~6.0 input file \cite{Neese2012The,Neese_ORCA6} or a PySCF script \cite{Qiming2018PySCF}, and dedicated runners execute and check the calculation before scientific review.

\begin{figure}[!ht]
\centering
\includegraphics[width=0.90\linewidth,height=0.38\textheight,keepaspectratio]{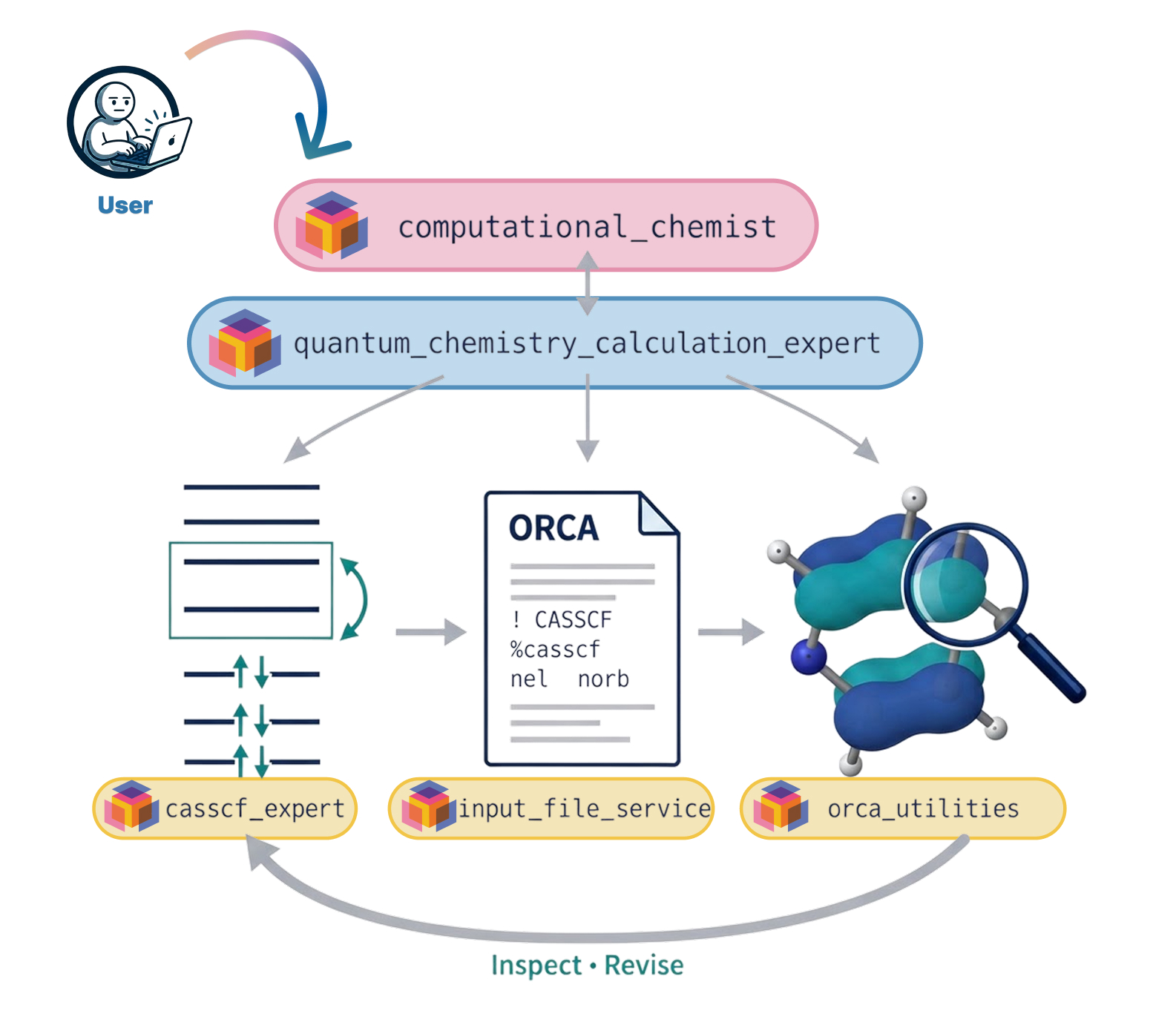}
\caption{Illustration of {\cinzel El Agente Quntur} roles involved in active-space selection, input preparation, and orbital inspection. {\cinzel Quntur} currently supports 59 agents, but the image only represents the most relevant for this work. The return arrow indicates possible revision from calculated evidence, not a required sequence; execution runners and other specialists are omitted.}
\label{fig:quntur-casscf-architecture}
\end{figure}

These roles distribute scientific decisions (Figure~\ref{fig:quntur-casscf-architecture}). The \texttt{casscf\_expert} proposes and revises the active-space and state-averaging choices; \texttt{input\_file\_service} routes its advice into executable inputs and returns new evidence or unresolved questions for further review. The \texttt{quantum\_chemistry\_calculation\_expert} coordinates the calculation stages and diagnostic requests, while \texttt{computational\_chemist} frames the objective, reviews the resulting evidence, and decides what to report or test next. The \texttt{orca\_utilities} agent can generate orbital images with \texttt{orca\_plot}; these are evidence for the other agents' decisions, not an automatic active-space choice. Execution runners like \texttt{orca\_runner} execute jobs and check their status and convergence.

We kept the CASSCF expert's fixed context deliberately minimal: it specified broad chemical criteria and suggested possible diagnostics, but prescribed neither a case-specific active space nor a calculation sequence. This left orbital selection, tests, and revisions largely to the model-driven agents' reasoning within {\cinzel Quntur}'s orchestration. The \texttt{elagente} environment made PySCF, \texttt{pyscf-dmrgscf}/\texttt{block2}, and ASE available to the workflow; ORCA calculations were accessed through designated runners. Availability does not imply use in any given trajectory.

The available agent roles and registered tools are listed in Appendices~\ref{app:quntur-agents} and~\ref{app:quntur-tools}; the case traces establish which ones were actually used.

\subsection{Benchmarks}

Our benchmarks follow the 32 entries listed in the Supporting Information of Shirazi et al. \citep{Shirazi2025ASF}. Their geometries and the source-paper calculations used here as reference values come from Schreiber and co-workers and Hoyer and co-workers \citep{Schreiber2008,Hoyer2016}. The 32 rows represent 31 molecule names: the two all-\textit{E}-hexatriene rows use different fixed geometries and reference values. All systems are neutral singlets.

We checked the fixed coordinates against Schreiber's Supporting Information for entries 01--15 and against the deposited EE23 geometries for entries 16--32 \citep{Schreiber2008,VermaTruhlar2019Geometries}. The ASF article states that it used the literature geometries but does not supply its calculation inputs for a coordinate-by-coordinate check. We independently selected the lowest singlet excitation listed for each case in the source paper's calculation column at the matching geometry: Schreiber's current-work MS-CASPT2/TZVP column for entries 01--15 and Hoyer's CASPT2 column for entries 16--32 \citep{Schreiber2008,Hoyer2016}. These source papers tabulate selected transitions, so their lowest listed state does not necessarily establish the global $S_1$; root character is evaluated separately. Their author-designated ``Best est.'' and ``reference'' columns incorporate heterogeneous literature or experimental values and are not uniformly calculations at the fixed geometry. The ASF columns in Table~\ref{tab:asf-quntur-results-32} are transcribed from the ASF Supporting Information, while Ref. is independently transcribed from the original papers.

Each benchmark prompt supplies the fixed XYZ geometry, charge, and multiplicity, and specifies the target quantity:
\begin{quote}\small
For [molecule] at the fixed XYZ geometry [path] (charge 0, multiplicity 1), calculate the vertical $S_0$-to-$S_1$ excitation energy using equal-weight two-singlet SA-CASSCF followed by SC-NEVPT2 with def2-TZVPD and density fitting. Select the active space and report the gap in eV.    
\end{quote}
The \texttt{casscf\_expert} context asked for a chemically justified active space and state average based on the relevant electronic processes, orbital classes, and active-electron count; it suggested using energies, occupations, and atom/orbital populations before targeted plots or screens; and it asked the specialist to identify what later calculation would resolve a remaining ambiguity. It supplied no molecule-specific orbital list, prescribed CAS size, or fixed sequence of calculations (see Supporting Information Section \ref{si:Agent_context}).

To assess within-question variability, we retained the original benchmark session and three independent repeats for each of $p$-nitroaniline (026) and pyrimidine (030). Only the original sessions enter the 32-case benchmark statistics. For each run, elapsed time is measured from the input of the first prompt to the reporting of the calculated gap; total job counts include submitted ORCA calculations and Python Slurm jobs for orbital preparation or result analysis, including failed submissions.

The requested numerical result is the difference between root-specific correlated total energies,
\begin{equation}
\Delta E_i^{\mathrm{SC\mbox{-}NEVPT2}}=E_i^{\mathrm{SC\mbox{-}NEVPT2}}(S_1)-E_i^{\mathrm{SC\mbox{-}NEVPT2}}(S_0),
\label{eq:gap}
\end{equation}
in electronvolts. We separately evaluated whether the final orbitals, state averaging, convergence, and root character support the assigned transition.

\subsection{Comparison and evidence sources}

We compare {\cinzel Quntur} against six ASF selections from the published production protocol \citep{Shirazi2025ASF}, spanning different entropy thresholds and singlet, triplet, or quasi-restricted starting orbitals (QRO) \cite{Neese2006Importance}. Their selected CAS sizes are comparison outcomes, not prescribed correct spaces for the {\cinzel Quntur} runs. We compute absolute differences from the ASF gap values and the reported {\cinzel Quntur} gaps, using the same independently selected source-paper CASPT2 reference (see Table~\ref{tab:asf-quntur-results-32}). Following the ASF convention, a failed ASF CASSCF calculation contributes an error equal to the reference excitation energy to the penalized mean absolute error (MAE), while the converged-only MAE excludes failures. A {\cinzel Quntur} result with uncertain state assignment is highlighted as such but retained as reported without a failure penalty.

For each case, the scientific audit reconstructs the sequence of decisions from the runner trace where available, submitted inputs, ORCA outputs, orbital and restart records, and job metadata. The 32-row evidence matrix in Supporting Information Section~\ref{app:decisions} records one consequential decision per trajectory. Independent trajectory audits count input and output tokens, model calls, and tool calls for the final-answer session of each case. Finally, the files generated during the 32 benchmark runs, together with the prompts, agent traces, and LLM-assisted audit reports, are available in our data repository \citep{Perez2026QunturCASSCFFiles}.

\subsection{Statistical Analysis}

Each fixed geometry was one observation in the primary comparison ($n=32$); the repeat runs for cases 026 and 030 were summarized separately. Absolute differences from the geometry-matched reference gaps were summarized by mean absolute error, with failed ASF calculations treated as described above. The plotted medians and interquartile ranges were calculated in Python using NumPy, and Matplotlib generated the distribution figure. No significance tests or outlier filtering were applied (see Figure \ref{fig:benchmark-error-distribution}).

\section{Results}

\subsection{Performance of {\cinzel Quntur} and ASF on the benchmark set}

\begin{figure}[!t]
\centering
\includegraphics[width=0.96\linewidth,keepaspectratio]{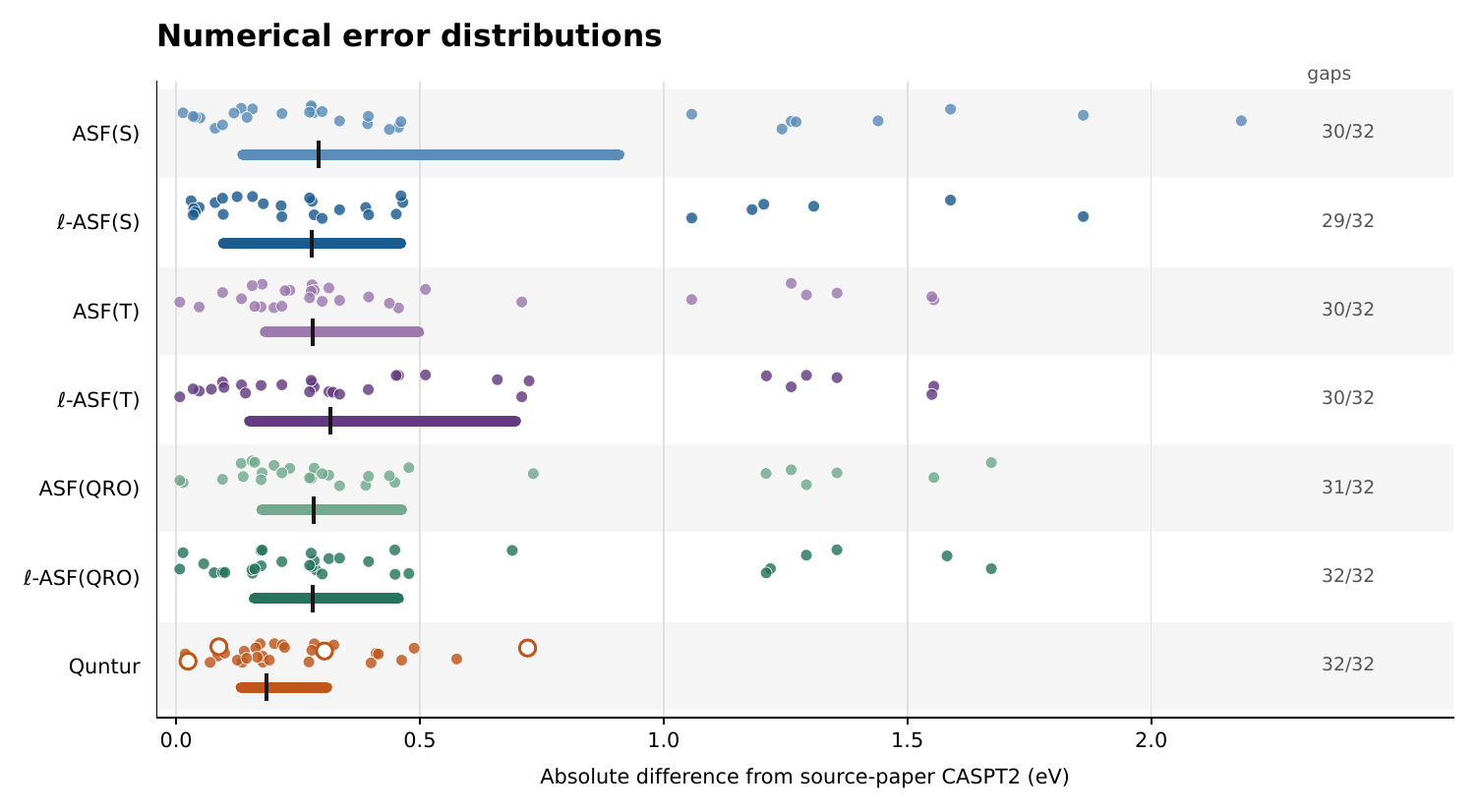}
\caption{Distribution of absolute numerical differences between reported SC-NEVPT2 excitation gaps by Quntur and all 6 ASF methods, and the geometry-matched source-paper CASPT2 values for the 32 fixed geometries. ASF(S) and ASF(T) select orbitals using singlet and triplet MP2 starting points, respectively. ASF(QRO) starts from the triplet selection and rotates the selected space toward quasi-restricted orbitals. The corresponding \(\ell\)-ASF variants use a lower orbital-entropy threshold (0.11 rather than approximately 0.14), allowing larger active spaces. Supporting Information Table~\ref{tab:asf-quntur-results-32} gives every reference, gap, and active-space size. Each dot represents one completed calculation; its vertical displacement is only for visibility. Colored bars span the interquartile range of the absolute differences for completed calculations, and black ticks mark their medians. The fractions at right count completed numerical gaps out of 32; failed ASF CASSCF jobs have no dot and do not enter these summaries. Hollow {\cinzel Quntur} circles mark cases 005, 019, 023, and 025, whose audited roots differ from the source-paper transitions. The other {\cinzel Quntur} roots are not thereby independently established as the intended $S_1$, and per-case ASF root character is unavailable. The plot measures numerical agreement, not state-validated accuracy; Table~\ref{tab:asf-quntur-mae-32} reports MAE with the stated ASF failure convention.}
\label{fig:benchmark-error-distribution}
\end{figure}

Supporting Information Table~\ref{tab:asf-quntur-results-32} lists all 32 independently selected CASPT2 references and the six published ASF selections alongside {\cinzel Quntur}-reported gaps and active-space sizes. Figure~\ref{fig:benchmark-error-distribution} compares the distributions of numerical error across these seven approaches. Every completed {\cinzel Quntur} session yielded a numerical value for the $S_0\to S_1$ gap.

Relative to these source-paper values, the 32-case raw MAE is 0.239~eV for {\cinzel Quntur}-reported gaps, 0.221~eV (48\%) lower than the 0.459~eV MAE of $\ell$-ASF(QRO) (a specific variant of the ASF method with lowered threshold and quasi-restricted starting orbitals), the best of the six fixed protocols applied uniformly (Table~\ref{tab:asf-quntur-mae-32}). {\cinzel Quntur}'s raw MAE is below that of every ASF variant. The best ASF variant and {\cinzel Quntur} both returned a gap on all 32 entries, so this comparison does not depend on an ASF failure penalty. Excluding $p$-benzoquinone, propanamide, and 4-(dimethylamino)benzonitrile, for which the ASF study notes that more than two states may need to be averaged, gives MAEs of 0.247~eV for {\cinzel Quntur} and 0.364~eV for $\ell$-ASF(QRO). Across individual cases, {\cinzel Quntur} has the smaller numerical error in 16 and $\ell$-ASF(QRO) in 16; their median absolute errors are 0.185 and 0.281~eV, respectively. {\cinzel Quntur} has no numerical error above 1~eV, whereas $\ell$-ASF(QRO) has six. The benchmark-wide improvement in numerical agreement is therefore accompanied by substantial case-level variation. State identity and computational effort are assessed separately.

\begin{table}[ht]
\centering
\caption{Numerical mean absolute error (MAE) relative to the geometry-matched source-paper CASPT2 references in Table~\ref{tab:asf-quntur-results-32}. The {\cinzel Quntur} row compares all 32 reported gaps with those values; the state and protocol checks are reported separately.}
\label{tab:asf-quntur-mae-32}
\small
\setlength{\tabcolsep}{6pt}
\renewcommand{\arraystretch}{1.14}
\begin{tabular}{@{}lrrr@{}}
\toprule
Method & Cases without gap & Penalized MAE (eV) & Reported-gap MAE (eV) \\
\midrule
ASF(S) & 2/32 & 0.873 & 0.563 \\
\rowcolor{black!7}
$\ell$-ASF(S) & 3/32 & 0.786 & 0.455 \\
ASF(T) & 2/32 & 0.816 & 0.482 \\
\rowcolor{black!7}
$\ell$-ASF(T) & 2/32 & 0.840 & 0.508 \\
ASF(QRO) & 1/32 & 0.635 & 0.484 \\
\rowcolor{black!7}
$\ell$-ASF(QRO) & 0/32 & 0.459 & 0.459 \\
\midrule
All ASF (pooled) & 10/192 & 0.735 & 0.492 \\
\rowcolor{black!7}
{\cinzel Quntur}, reported gaps\textsuperscript{a} & 0/32 & 0.239 & 0.239 \\
\bottomrule
\end{tabular}
\par\smallskip
\begin{minipage}{0.94\linewidth}\footnotesize
\emph{Notes.} An ASF case has no gap when its selected CASSCF calculation fails. Following the failure convention of Shirazi et al., the penalized MAE assigns that case an error equal to its independently selected reference gap; the reported-gap MAE omits it. The pooled ASF row weights all six protocols and all 32 systems equally. \textsuperscript{a}{\cinzel Quntur} reported a numerical gap for every case: 0/32 counts cases without a gap, not failed intermediate jobs. Its two MAEs therefore coincide. Orbital and configuration analyses assign cases 005, 019, 023, and 025 to transitions of different character from their source-paper references. Cases 006 and 014 used a different final state-average treatment, while 029 compared separately optimized state pairs. These state and method qualifications receive no missing-gap penalty; the {\cinzel Quntur} MAE measures numerical agreement rather than validated accuracy for the intended state and protocol. Further evidence appears in Supporting Information Section~\ref{app:decisions}.
\end{minipage}
\end{table}

\subsection{Trajectory analysis, key decision points, and agentic active space selection}

We also examine the agent trajectories to identify the decisions underlying these results and clarify their interpretation. Although the trajectories are complex, we highlight key decisions that shaped the reported results (see Supporting Information \ref{app:trajectory-analyses}). In cytosine, the agent detected the loss of intended lone pairs during orbital optimization and adopted a fallback calculation that retained them. For pyridine, calculations of competing $S_1$ states needed to be resolved for the final assignment. Other cases reveal limits in such state identification. The reported states for cyclopropene, E-butadiene, furan, and naphthalene differ from the transitions associated with their reference values, while pyrrole retains an unresolved assignment. Some agents also changed the prescribed state average or used separately optimized state pairs. Besides the reported gaps and reference-state mismatches, the Supporting Information Section~\ref{app:decisions} records the state and method qualifications for each trajectory.

We illustrate six orbital montages that, together with the corresponding root and energy diagnostics, illustrate complementary decisions: retaining or revising an active space, establishing which state was computed, and checking how basis changes or correlation affect the choice. The molecular orbitals depicted correspond to the active orbitals of the CAS space employed in the SC-NEVPT2 calculations. The corresponding orbital occupation numbers can be obtained directly from the output files of these calculations. Ordered by benchmark ID, the captions summarize key agent decisions, including recovery from orbital loss in cytosine (006), an active-space expansion test in acetone (017), a close numerical match to a different state in furan (023), multi-agent orbital-selection review in $p$-nitroaniline (026), a correlation-driven state reversal in pyridine (029), and a root-order change first observed at basis projection in pyrimidine (030). The numerical results and trajectory analyses for all 32 cases are reported in the Supporting Information Section~\ref{app:trajectory-analyses}.

\begin{figure}[!htbp]
\centering\captionsetup{font=small}\includegraphics[width=0.75\linewidth,height=0.3\textheight,keepaspectratio]{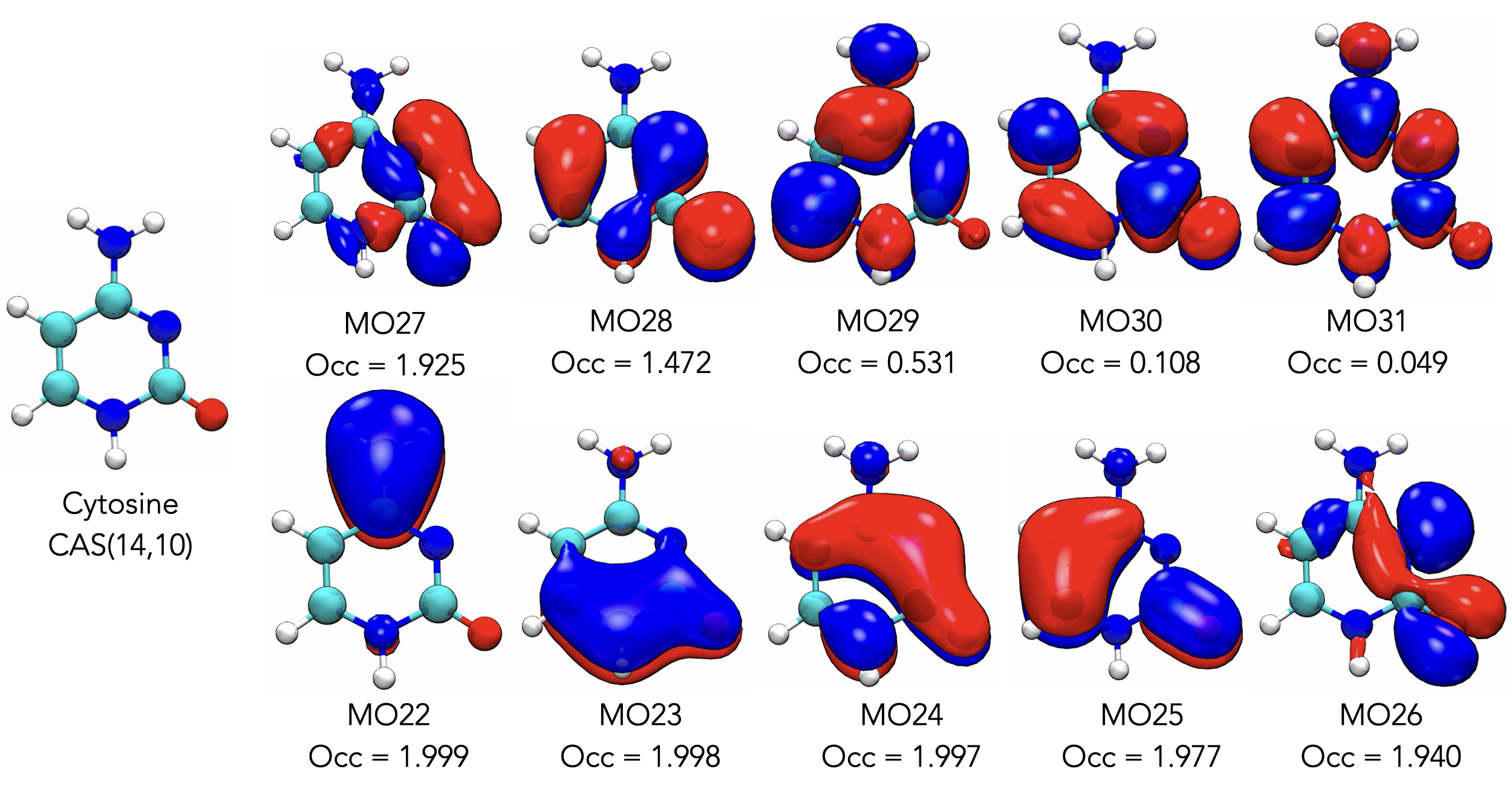}
\caption{\textbf{Cytosine (006): orbital escape after normal termination.} The pictured four-root CAS(14,10) space includes the intended in-plane O/N lone-pair directions (MOs 26 and 27). In the first two-root optimization, those directions instead became inactive and $\sigma$/framework orbitals entered the active space; that rejected orbital set is not pictured. The two-root job converged numerically and completed SC-NEVPT2, but the agent rejected its wrong-space result. Three recovery attempts did not establish a stationary intended-space two-root solution. Returning to the four-root orbitals yielded a qualified fixed-orbital CASCI/SC-NEVPT2 gap of 4.661~eV, versus the 4.68~eV benchmark and a 0.047~eV error for the closest ASF result. This adaptation supplied a defensible reported gap at added cost; its 0.019~eV error does not establish what the prescribed two-root protocol would predict.}
\label{fig:trajectory-cytosine}
\end{figure}

\begin{figure}[!htbp]
\centering\captionsetup{font=small}\includegraphics[width=1\linewidth,height=0.4\textheight,keepaspectratio]{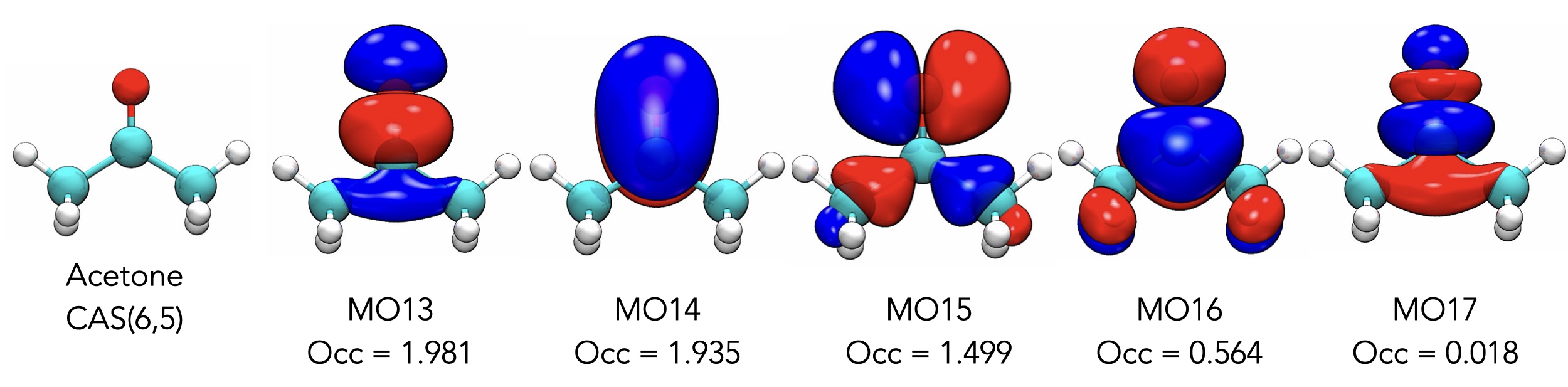}
\caption{\textbf{Acetone (017): testing a larger space.} The panels show all five final def2-TZVPD active orbitals, including the carbonyl $\sigma/\sigma^*$ pair retained after a separate def2-SVP test. That predeclared CASSCF comparison shifted the gap by $+0.337$~eV and led the agent from CAS(4,3) to CAS(6,5); final orbital and configuration checks support an $n\pi^*$ $S_1$. After two failed production optimizations, a damped rerun from the RHF seed converged in energy; a subsequent Newton--Raphson restart met the gradient criterion. The final SC-NEVPT2 gap is 4.530~eV, 0.090~eV above the 4.44~eV benchmark; the closest ASF result is 4.479~eV (0.039~eV high). The expansion was not compared at SC-NEVPT2 level in the same basis, leaving its necessity unresolved.}
\label{fig:trajectory-acetone}
\end{figure}

\begin{figure}[!htbp]
\centering\captionsetup{font=small}\includegraphics[width=0.75\linewidth,height=0.3\textheight,keepaspectratio]{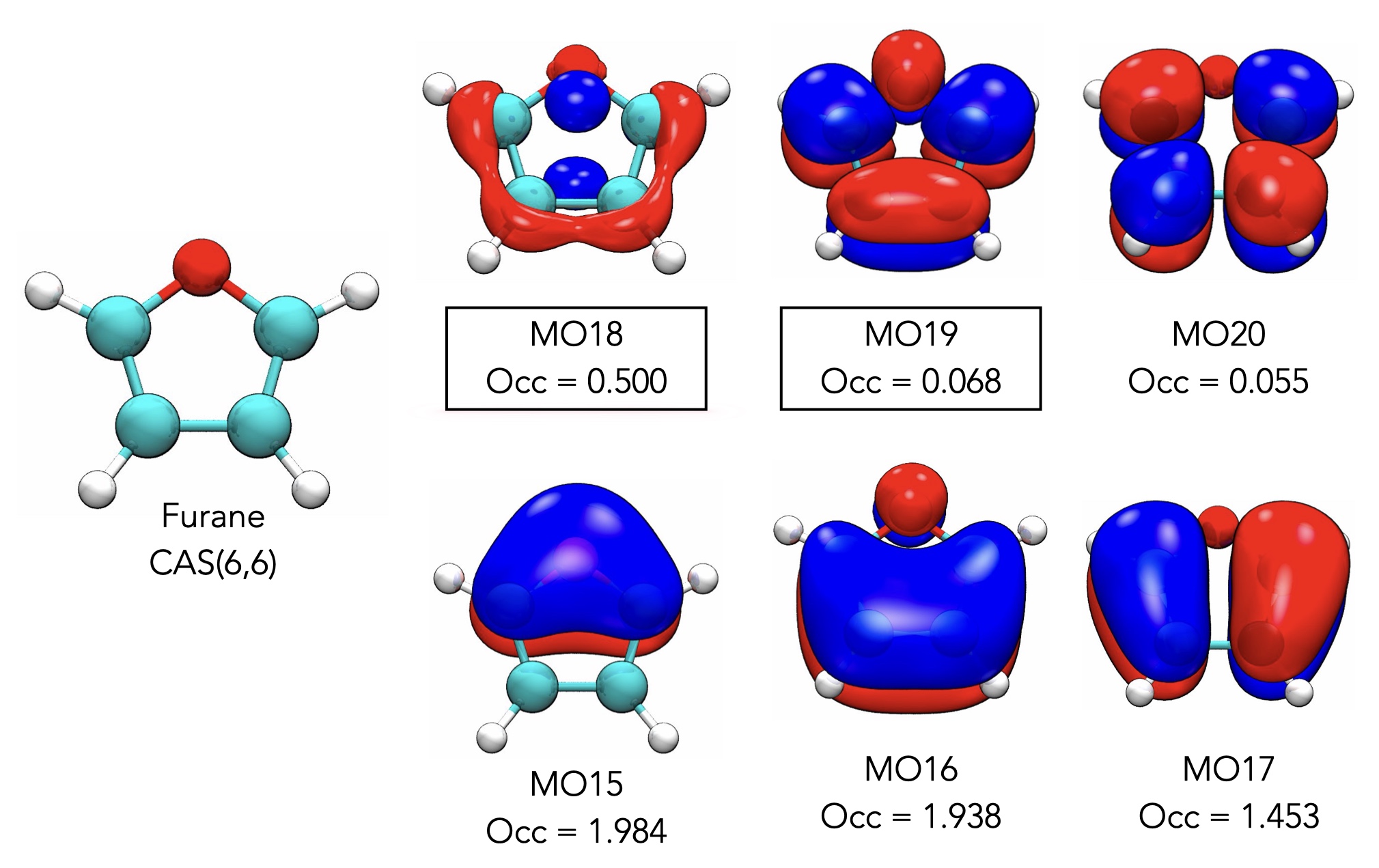}
\caption{\textbf{Furan (023): a close gap for a different reference state.} The final CAS(6,6) orbital montage includes the diffuse $3s$ acceptor (MO18) and two valence $\pi^*$ acceptors (MOs 19 and 20). A separate four-root CASSCF screen identified competing $\pi\!\to\!3s$ and mixed $\pi\pi^*$ states. After a converged first validation, the agent found an in-plane orbital in place of a deeper occupied $\pi$ orbital and repaired the selection. Its final two-root $\pi\!\to\!3s$ SC-NEVPT2 gap, 6.344~eV, lies 0.024~eV from the 6.32~eV benchmark, Hoyer's CASPT2 value for a \emph{valence} $\pi\pi^*$ state. ASF gives 6.768--6.831~eV, without published root assignments. The numerical match does not validate the target state; correlated ordering of both states remains untested.}
\label{fig:trajectory-furan}
\end{figure}

\begin{figure}[!htbp]
\centering\captionsetup{font=small}\includegraphics[width=0.8\linewidth,height=0.35\textheight,keepaspectratio]{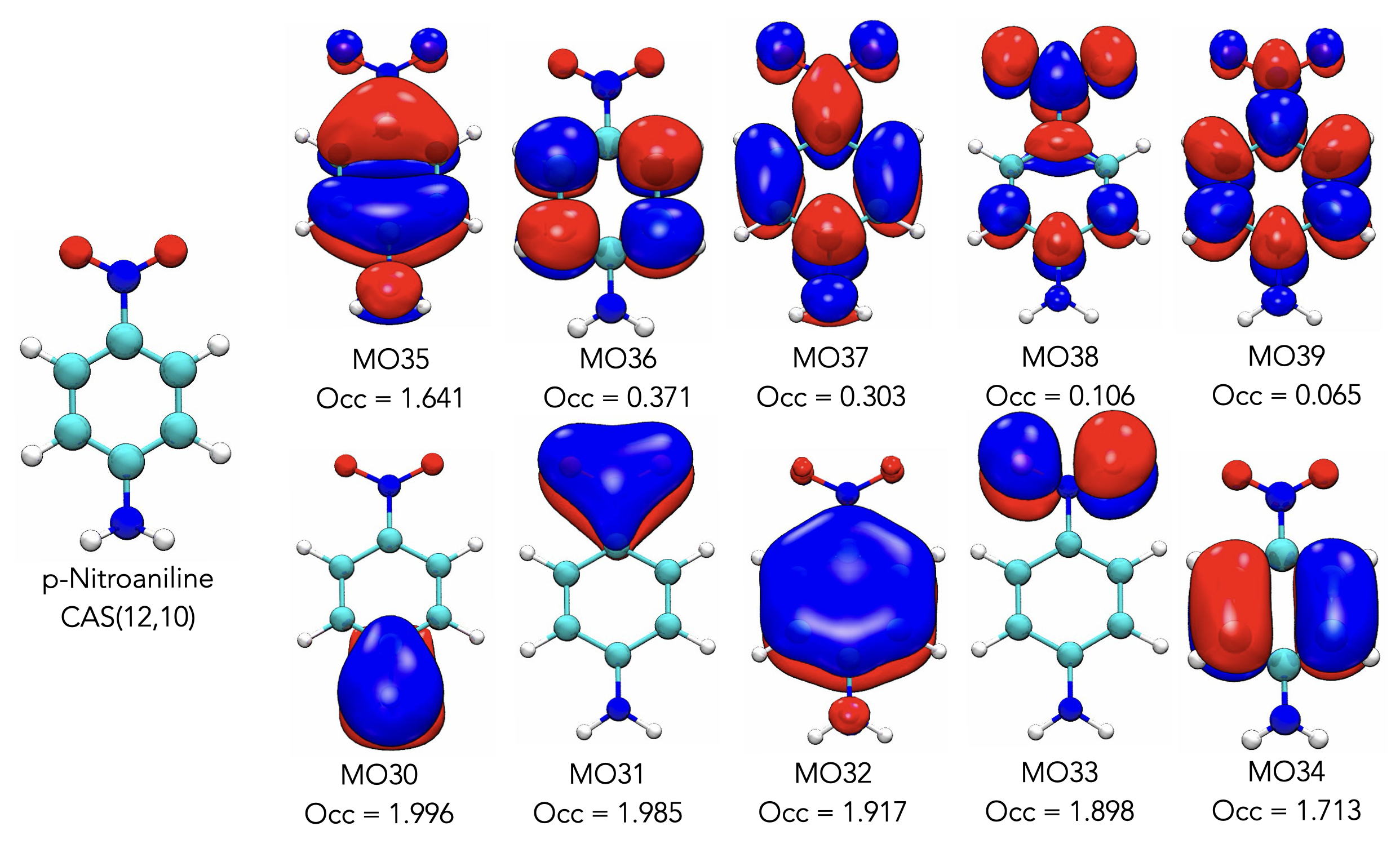}
\caption{\textbf{$p$-Nitroaniline (026): final orbital choice.} The montage shows the final CAS(12,10) orbitals, with occupied donor/ring combinations (MOs 30--35) and ring/nitro acceptors (MOs 36--39). The \texttt{casscf\_expert} provisionally approved a projection drawing on all 394 def2-TZVPD virtuals, but withdrew it after the \texttt{quantum\_chemistry\_calculation\_expert} questioned whether the projected orbitals had established valence character. The disputed orbitals and multi-agent review appear in Section~\ref{subsec:pna-review} and Fig.~\ref{fig:pna-rejected-projection}; a bounded orbital transfer supplied the starting space for the final orbitals shown here. The final gap is 4.943~eV versus a 4.62~eV benchmark (0.323~eV high); ASF errors are 0.437--0.451~eV. Two failed preparation helpers added work.}
\label{fig:trajectory-pna}
\end{figure}

\begin{figure}[!htbp]
\centering\captionsetup{font=small}\includegraphics[width=0.75\linewidth,height=0.3\textheight,keepaspectratio]{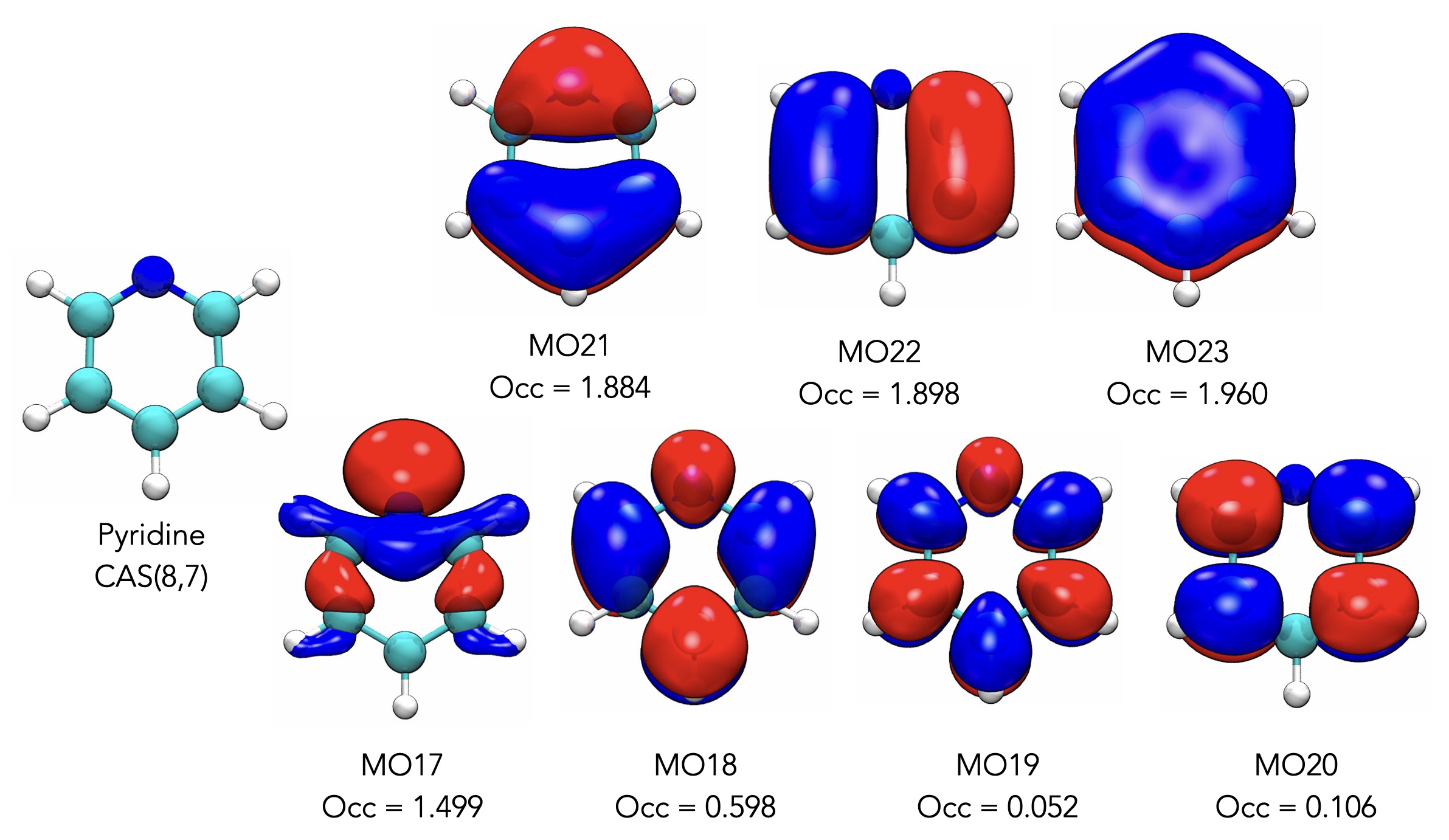}
\caption{\textbf{Pyridine (029): correlation reverses a state choice.} The pictured CAS(8,7) orbitals for the selected $n\pi^*$ pair include the N lone pair (MO17) and ring $\pi/\pi^*$ orbitals (MOs 18--23). After an unsymmetrized test lost the $n\pi^*$ state, the agent used separately optimized symmetry-blocked state pairs; the competing $\pi\pi^*$ pair is not pictured. CASSCF put $\pi\pi^*$ lower by 0.178~eV, but matched SC-NEVPT2 gaps reversed the order: 5.186~eV for $n\pi^*$ and 5.402~eV for $\pi\pi^*$. The selected $n\pi^*$ gap is 0.126~eV above the 5.06~eV benchmark; ASF gaps are about 5.395~eV (0.335~eV high), with unreported state character. The diagnostic changed the choice, although the separate averages do not reproduce one common-orbital two-state protocol.}
\label{fig:trajectory-pyridine}
\end{figure}

\begin{figure}[!htbp]
\centering\captionsetup{font=small}\includegraphics[width=0.75\linewidth,height=0.3\textheight,keepaspectratio]{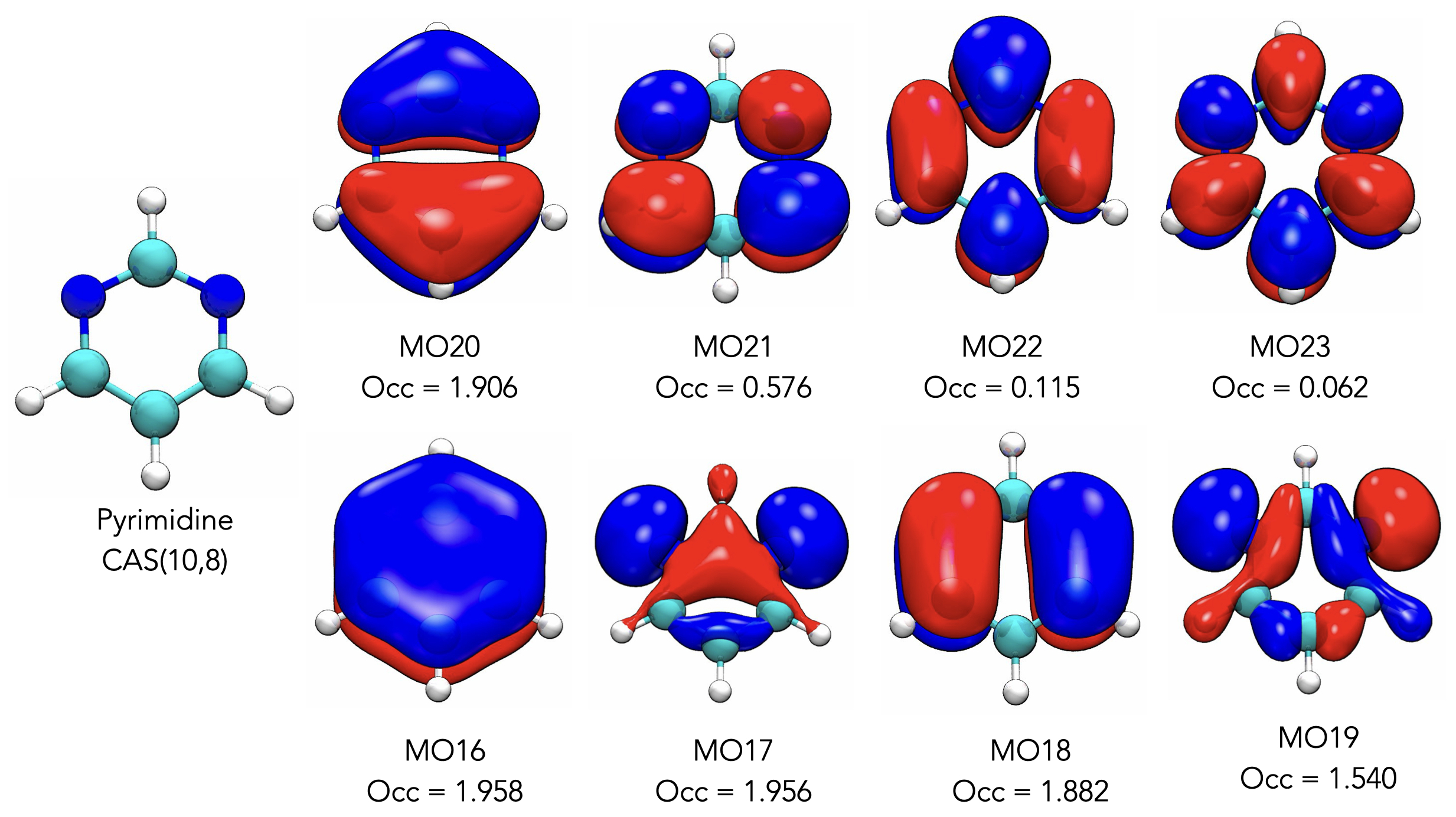}
\caption{\textbf{Pyrimidine (030): a root-order change at basis projection.} The montage shows the eight CAS(10,8) active orbitals, including two N lone pairs and six ring $\pi/\pi^*$ orbitals. An initial four-root CASCI screen and relaxed def2-SVP CASSCF put mixed $\pi\pi^*$ below $n\pi^*$; the order first reversed in a fixed-orbital def2-TZVPD projection, before target-basis optimization. The agent checked configurations, rejected a GBW replay with inconsistent reference energy, and reconverged CASSCF with SC-NEVPT2 in one job. Its $n\pi^*$ gap is 4.571~eV, 0.191~eV above the 4.38~eV benchmark; the closest ASF gap is 5.069~eV (0.689~eV high). Fifteen ORCA jobs were used; no larger CAS or separately targeted diffuse-state calculation was tested.}
\label{fig:trajectory-pyrimidine}
\end{figure}

\FloatBarrier

\subsection{Run-to-run variability at fixed geometry}

To examine how run-to-run variability in LLM-guided decisions affects active-space selection, we compared four independent executions of the same prompt for each of two representative benchmark questions, $p$-nitroaniline and pyrimidine (Table~\ref{tab:run-variability}). We discuss these results in Section~\ref{sec:variability_discusion}. The Original and rep1--rep3 labels identify each run's trajectory and calculation files in the data repository \citep{Perez2026QunturCASSCFFiles}.

\begin{table}[H]
\centering
\caption{Independent runs of two fixed prompts. The reported gaps span 0.0048~eV for $p$-nitroaniline and 0.0011~eV for pyrimidine despite different calculation paths and job counts. Submitted jobs include ORCA calculations and Python preparation or analysis jobs.}
\label{tab:run-variability}
\footnotesize
\setlength{\tabcolsep}{3pt}
\renewcommand{\arraystretch}{1.14}
\begin{tabular}{@{}llclrrr@{}}
\toprule
Run & Orbital-selection route & Final CAS & Assigned $S_1$ character & Gap (eV) & Elapsed (min) & Submitted jobs \\
\midrule
\multicolumn{7}{@{}l}{\textit{$p$-Nitroaniline (026)}} \\
Original & Bounded transfer & (12,10) & Weak, ring-local valence & 4.943442 & 236.6 & 9 \\
\rowcolor{black!7}
rep1 & MP2 natural orbitals & (14,11) & Weak, mixed valence & 4.948070 & 140.6 & 11 \\
rep2 & AVAS in target basis & (12,10) & Weak, mixed valence & 4.944673 & 77.1 & 7 \\
\rowcolor{black!7}
rep3 & AVAS after basis change & (12,10) & Weak, mixed valence & 4.943318 & 183.4 & 20 \\
\midrule
\multicolumn{7}{@{}l}{\textit{Pyrimidine (030)}} \\
Original & RHF orbital reorder & (10,8) & Valence $n\pi^*$ & 4.571334 & 95.2 & 17 \\
\rowcolor{black!7}
rep1 & MP2 natural orbitals & (10,8) & Valence $n\pi^*$ & 4.570588 & 167.5 & 12 \\
rep2 & Rotation after CAM-B3LYP TDA & (10,8) & Valence $n\pi^*$ & 4.571700 & 108.0 & 8 \\
\rowcolor{black!7}
rep3 & Rotation after PBE0 TDA & (10,8) & Valence $n\pi^*$ & 4.570927 & 31.6 & 5 \\
\bottomrule
\end{tabular}
\end{table}

\section{Discussion}

{\cinzel El Agente Quntur} combines active space selection with inspection and revision of the resulting calculations. Its reported excitation energies have lower average error than those from each of the six predefined ASF protocols under the stated scoring convention. The calculation records show how agents respond to missing orbital character, competing excited states, and inconsistent outputs. These responses change the calculation strategy and, in several cases, the reported state or energy.

\subsection{An active space as a hypothesis to test}

In pyrimidine (see Fig.~\ref{fig:trajectory-pyrimidine}), the coordinating agent proposed CAS(10,8): both N lone pairs and six ring $\pi/\pi^*$ orbitals, intended to describe competing $n\pi^*$ and $\pi\pi^*$ excitations. The first four-root CASCI screen placed mixed $\pi\pi^*$ at $S_1$ and clear $n\pi^*$ at $S_2$, which seemed contrary to the agent's hypothesized order. The quantum-chemistry calculation specialist checked the configurations and retained CAS(10,8), because the intended orbitals and states were present; the problem was their ordering. It replaced the planned direct two-root follow-on with four-root orbital optimization and state tracking. The order reversed on projection into def2-TZVPD, putting $n\pi^*$ at $S_1$. Final orbital populations still showed the proposed lone-pair and ring $\pi/\pi^*$ characters. The space was therefore retained for the final calculation, with the two lowest roots averaged and two more tracked diagnostically through the SC-NEVPT2 energy check.

In $s$-tetrazine (see Supporting Information Section~\ref{app:case-031}), the \texttt{input\_file\_service} proposed CAS(14,10); the \texttt{casscf\_expert} checked symmetry and orbital populations after the screen and audited the validation output before the coordinator retained that space for production. No alternative CAS was tested. In acetone (see Fig.~\ref{fig:trajectory-acetone}), the \texttt{input\_file\_service} planned a comparison between an initial $n,\pi,\pi^*$ CAS(4,3) and CAS(6,5) with an added carbonyl $\sigma/\sigma^*$ pair. The \texttt{casscf\_expert} identified that pair and prepared the expanded-space input; the orbital-plotting agent helped check its character. The added pair shifted the def2-SVP CASSCF gap by 0.337~eV, so the coordinator kept it for the final def2-TZVPD calculation. The comparison was not repeated at SC-NEVPT2 in the final basis, leaving the necessity of that expansion unresolved.

\subsection{Importance of multi-agent critical review during active space selection}
\label{subsec:pna-review}

\begin{figure}[H]
\centering
\captionsetup{font=small}
\includegraphics[width=0.96\linewidth,height=0.45\textheight,keepaspectratio]{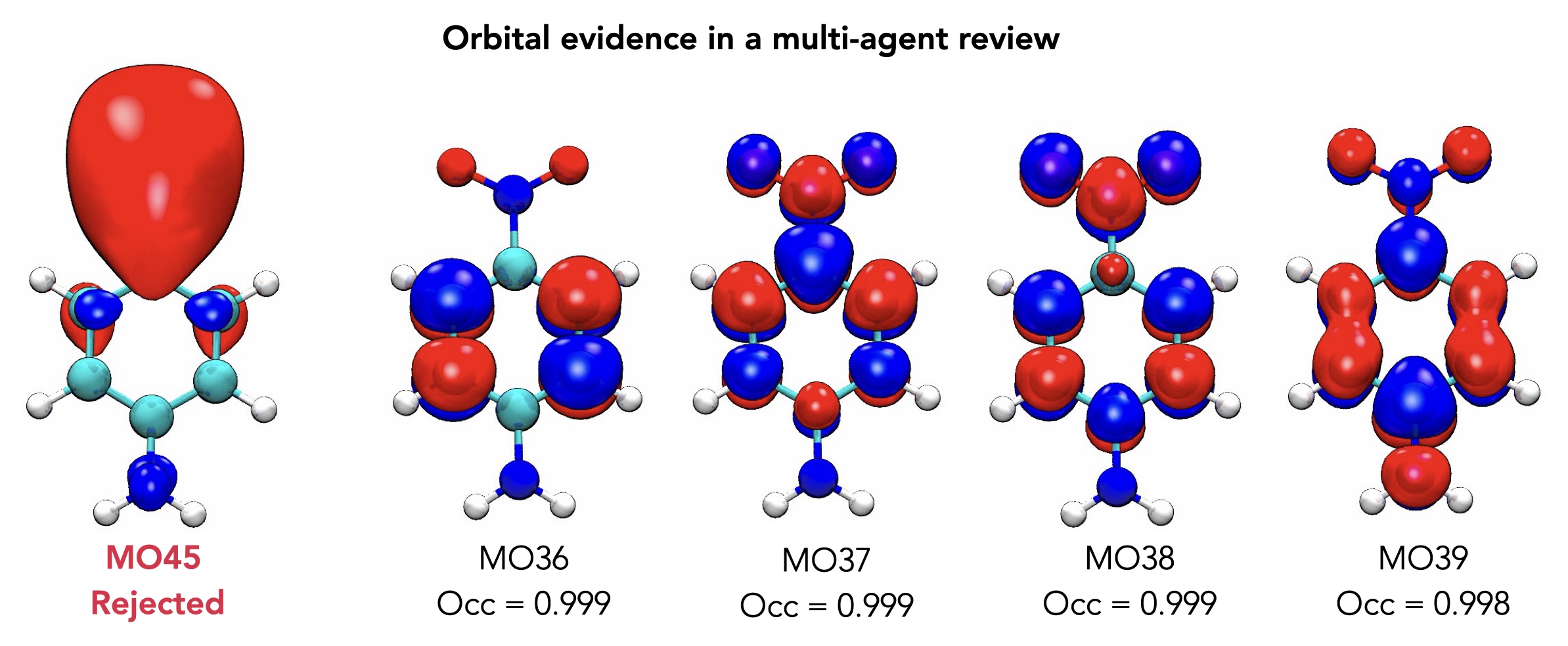}
\caption{\textbf{$p$-Nitroaniline (026): orbital evidence in the specialist--coordinator review.} Canonical MO45 (left) was rejected as a mixed/diffuse fourth virtual. A subsequent repair yielded four projected virtuals (right; zero-based MOs 36--39) that received provisional approval but were never tested by CASSCF due to an energy-based rejection. The final orbitals appear in Fig.~\ref{fig:trajectory-pna}.}
\label{fig:pna-rejected-projection}
\end{figure}

In $p$-nitroaniline, the \texttt{casscf\_expert} and \texttt{input\_file\_service} subagents critically reviewed two successive orbital choices before running CASSCF. The subagents proposed CAS(12,10) for the ring, amino donor, and nitro group. An ORCA PBE0/def2-TZVPD scout found only three clean canonical virtual $\pi^*$ orbitals; the candidate fourth, MO45, was mixed and diffuse, so the \texttt{casscf\_expert} subagent rejected a simple orbital permutation (Fig.~\ref{fig:pna-rejected-projection}, left). It then used PySCF to construct a separate candidate, by projecting ten out-of-plane heavy-atom $2p$ directions over all 394 def2-TZVPD virtual orbitals; MOKIT converted the orbital files between the PySCF and ORCA steps. The four resulting virtuals looked compact and had target-projector scores of 0.998--1.000 (Fig.~\ref{fig:pna-rejected-projection}, right). The \texttt{casscf\_expert} provisionally approved this new active space for a two-root CASSCF test. The \texttt{quantum\_chemistry\_calculation\_expert} then requested a re-audit, in which the \texttt{input\_file\_service} objected that a high projector score shows overlap with the chosen atomic targets, not that the fourth orbital is a low-energy valence $\pi^*$ direction \citep{Sayfutyarova2017}. The selected ten MOs captured 7.994 of the ten target-projector trace units; other MOs retained 2.006. The virtual energy expectations were $+0.313$ to $+0.382$~hartree. Such high energy expectations and unrestricted virtual-orbital ancestry left the proposed valence character insufficiently established, triggering the specialist to withdraw the approval and to disable the prepared CASSCF input before it was tested.

For the third choice, the \texttt{casscf\_expert} ran another ORCA PBE0 scout in def2-TZVP (no diffuse functions), identified four valence-like $\pi^*$ directions within a 20-orbital reference window, and used PySCF cross-basis overlaps to transfer those directions into a 32-orbital def2-TZVPD window. MOKIT handled orbital-file conversion, while PySCF generated images of the orbitals for visual character inspection. The target-basis orbital-energy expectations were $-0.079$, $-0.033$, $+0.006$, and $+0.128$~hartree, lower than the values with the previous selection (these energies are selection diagnostics, not excitation energies). Plots and overlap checks supported the transferred valence character, although the highest virtual, MO39, retained an appreciable diffuse tail. After the specialist approved this active space for a CASSCF test, ORCA ran the two-root calculation, a five-root state screen, and the final two-root SC-NEVPT2 calculation, yielding 4.943~eV (Fig.~\ref{fig:trajectory-pna}). The revised route required a new reference and two failed preparation-helper submissions. This exchange shows the \texttt{quantum\_chemistry\_calculation\_expert} requesting critical review, the \texttt{input\_file\_service} questioning whether the numerical evidence supported the chemical claim, and the \texttt{casscf\_expert} revising its plan without a prompt-level correction. Because the rejected spaces were not tested by CASSCF, their adequacy and the benefit of the stricter gate remain unknown. Further records and agent exchanges appear in the Supporting Information Section~\ref{app:case-026}, and our data repository \citep{Perez2026QunturCASSCFFiles}.

\subsection{Run-to-run variability in active-space selection}\label{sec:variability_discusion}

Repeating the same prompt revealed a stable chemical starting hypothesis but different plans for testing it. While ASF fixes its orbital-selection route in advance, LLM agents can instead choose orbital sources, state screens, and repairs as intermediate results arrived. The \texttt{casscf\_expert} context permitted this flexibility by requesting chemical justification and relevant diagnostics without prescribing a workflow. The two cases below show that different agent-chosen routes can yield nearly identical gaps, but with different job counts and unequal checks of state ordering.

Table~\ref{tab:run-variability} compares the four runs for $p$-nitroaniline and pyrimidine. For $p$-nitroaniline, all four runs began by proposing CAS(12,10) for the donor, ring, and nitro valence system, before inspecting any orbitals (Table~\ref{tab:run-variability}). Each consulted the $p$-nitroaniline literature, but the subsequent screens differed. Two runs inspected PBE0 orbitals; the other two used CAM-B3LYP time-dependent density functional theory in the Tamm--Dancoff approximation (TDA) to examine low-lying states. Orbital selection then proceeded by bounded transfer between bases, RI-MP2 natural-orbital occupations, or AVAS; one AVAS space had to be rebuilt in def2-TZVPD after a failed basis projection. The MP2-natural-orbital route (rep1) expanded the final space to CAS(14,11), while the others retained CAS(12,10). Despite these different procedures and 5--20 ORCA jobs, the four reported gaps differ by less than 0.005~eV.

For pyrimidine, the four runs likewise proposed CAS(10,8) with both N lone pairs and the ring $\pi/\pi^*$ manifold before calculation. The original used RHF orbitals and a four-root CASCI screen; the repeats used PBE0/TDDFT with MP2 natural orbitals and symmetry targeting, CAM-B3LYP/TDA with orbital rotations and seven-root screening, or PBE0/TDA with a six-root test. Rep1 tested smaller spaces and rep2 attempted to add diffuse challengers, yet all retained CAS(10,8) and reported an $n\pi^*$ first excited state. The symmetry-targeted rep1 did not correct competing singlet irreducible representations, so the four calculations did not test the global state ordering to the same extent. Their SC-NEVPT2 gaps differ by only 0.0011~eV across 5--15 ORCA jobs.

The final orbital images reinforce the similarity of the outcomes: all four pyrimidine spaces contain the same lone-pair and ring $\pi/\pi^*$ orbital types with nearly identical occupations, despite differences in orbital phase (Fig.~\ref{fig:trajectory-pyrimidine}; Supporting Information Figs.~\ref{fig:si-active-030-rep1}--\ref{fig:si-active-030-rep3}). For $p$-nitroaniline, the three CAS(12,10) spaces look similar, while rep1 includes one additional orbital with occupation 1.999; this makes it a plausible active/inactive boundary choice (Fig.~\ref{fig:trajectory-pna}; Supporting Information Figs.~\ref{fig:si-active-026-rep1}--\ref{fig:si-active-026-rep3}). Four runs per molecule show reproducibility in these examples, but cannot rule out divergent selections or failed calculations if further repetitions are run.

\subsection{The importance of state identification}

The most consequential choice was often which excited state the active space and optimization actually represented. In pyridine (see Fig.~\ref{fig:trajectory-pyridine}), the agent included the N lone pair in CAS(8,7) and considered $n\pi^*$ from the outset. Separate symmetry-blocked CASSCF averages nevertheless put $\pi\pi^*$ 0.178~eV lower. Its 0.15~eV rule selected $\pi\pi^*$, and production for that pair was prepared. Before either SC-NEVPT2 job ran, the agent added a matched $n\pi^*$ test and submitted both together. Correlation reversed the choice: the gaps were 5.186~eV for $n\pi^*$ and 5.402~eV for $\pi\pi^*$. The matched test addressed the CASSCF margin, although the separately optimized pairs do not reproduce one common-orbital average of the lowest two singlets. In uracil (see Supporting Information Section~\ref{app:case-015}) and pyrazine (see Supporting Information Section~\ref{app:case-027}), configuration and occupation checks similarly exposed converged $\pi\pi^*$ roots where the agents sought $n\pi^*$; symmetry-targeted calculations then recovered the intended character, with other possible low roots still untested.

Furan (see Fig.~\ref{fig:trajectory-furan}) shows why the target state must accompany every error statistic. The agent anticipated a diffuse 3s candidate, corrected an orbital-mapping error, and verified a $\pi\!\to\!3s$ root. The benchmark's 6.32~eV value is Hoyer et al.'s CASPT2 energy for a valence $^1B_2\,\pi\pi^*$ state \citep{Hoyer2016}; their separate reference column lists 6.06~eV. Its numerical proximity to the computed Rydberg gap does not establish accuracy for the valence state. The figure contrasts the diffuse and valence acceptors. In naphthalene (see Supporting Information Section~\ref{app:case-025}), the agent selected a well-motivated full $\pi$ CAS and initially obtained a $^1L_b$-consistent state. It then reinterpreted literature symmetry labels in another axis convention and reported a $^1L_a$ state whose number was closer to the benchmark. Symmetry conventions, configuration weights, and state ordering are part of the scientific decision, not just checks on CAS size.

\subsection{Checking results beyond numerical convergence}

Convergence alone does not guarantee that a calculation supports the intended result. In cytosine (see Fig.~\ref{fig:trajectory-cytosine}), the two-root CASSCF calculation converged and completed SC-NEVPT2, but inspection showed that the intended in-plane lone-pair orbitals had left the active space. The agent rejected that gap. Its recovery attempts did not establish a stationary two-root solution retaining those orbitals, so it reported a gap from the four-root orbitals that did retain them, explicitly identifying this as a fixed-orbital CASCI/SC-NEVPT2 fallback rather than the requested two-root optimized result.

Benzene and pyrimidine exposed problems with numerical evidence after an orbital choice had been made. In benzene (see Supporting Information Section~\ref{app:case-018}), the first CASSCF job terminated normally but missed its orbital-gradient threshold. After a converged restart, a separate SC-NEVPT2 job recovered a different CASSCF reference energy; the agent discarded that result and recomputed CASSCF and SC-NEVPT2 together. In pyrimidine (see Fig.~\ref{fig:trajectory-pyrimidine}), a fixed-orbital replay likewise recovered an inconsistent reference energy from a saved orbital file. Running CASSCF and SC-NEVPT2 in one job supplied a consistent energy pair, although it did not explain the replay mismatch. These revisions addressed distinct checks: whether the active orbitals remained chemically appropriate, whether optimization was stationary, and whether the perturbative correction used the intended reference.

\subsection{Decision efficiency and avoidable work}

Across the 32 selected final-answer sessions, model and tool activity varied widely, with some trajectories requiring extensive searches and repeated calls (Table~\ref{tab:overall-usage-32}). ORCA jobs occupied 9.8~h in total (median 4.4~min per case); time outside ORCA jobs totaled 37.9~h (median 51.4~min), including model calls, other tools, and waits. The numerical gain therefore came with substantial, uneven diagnostic work, some of which did not resolve the question that prompted it. The ASF paper provides no comparable per-case resource accounting, so we cannot determine whether the gain justifies that effort or claim an efficiency advantage for either approach.

The highest-usage trajectories distinguish informative diagnostics from work that did not resolve the task (Supporting Information Table~\ref{tab:usage-ledger-32}). Cytosine (see Fig.~\ref{fig:trajectory-cytosine}) illustrates a useful but costly recovery. The agent caught lone-pair loss in a normally terminated two-root calculation. Three recovery jobs tested ways to retain the intended space, but none established a stationary two-root solution. The full trajectory used 11 ORCA jobs, including four retries across screening and recovery. Returning to the four-root orbitals preserved the lone pairs and yielded a gap much closer to the reference than any of the ASF calculations (see Supporting Information Table~\ref{tab:asf-quntur-results-32}). The $p$-nitroaniline review described above also had a resource consequence: after the untested projected space was withdrawn, the bounded repair required two failed preparation-helper submissions before the CASSCF jobs could proceed. Its five ORCA jobs were accompanied by 1,468 in-trace tool calls.

In pyrimidine (see Fig.~\ref{fig:trajectory-pyrimidine}), the agent spent 15 ORCA jobs on four-root screens, orbital convergence, and restart checks. The initial screen changed its planned path, the target-basis projection changed the observed root order, and a GBW replay exposed an inconsistent reference energy. Reconverging CASSCF and running SC-NEVPT2 in one job supplied a consistent energy pair without explaining the replay mismatch. Further replay and density-analysis attempts contributed to the cost, including one failed analysis job. In pyrazine (see Supporting Information Section~\ref{app:case-027}), the agent likewise needed a symmetry-targeted calculation after configurations exposed a wrong-character root, but it also ran two RHF restarts over printed convergence fields that had negligible effect on energy and retried a production input after an unsupported keyword caused a parser error. The two cases made 140 and 141 in-trace web-search tool calls, respectively, overwhelmingly on ORCA operation and syntax. Repeated searches, model calls, and large inputs contributed to the high token usage (see Table~\ref{tab:overall-usage-32}).

The 32-case experiment establishes an aggregate numerical advantage for the agent-driven workflow over every tested static ASF variant, including the one with no failed CASSCF jobs (Table~\ref{tab:asf-quntur-mae-32}). The trajectories also establish that the agents made chemically consequential, testable revisions beyond executing a selected input. The workflow incurred substantial model, tool, and search activity as well as extra scientific jobs, but some of that extensive work did not answer its diagnostic question.

\begin{table}[ht]
\centering
\caption{Aggregate resources across the 32 selected {\cinzel Quntur} final-answer sessions. Medians and ranges show variation among cases.}
\label{tab:overall-usage-32}
\small
\setlength{\tabcolsep}{5pt}
\renewcommand{\arraystretch}{1.14}
\begin{tabular}{@{}lrrr@{}}
\toprule
Measure & Total & Median per session & Range per session \\
\midrule
In-trace input tokens & 483,251,290 & 10,530,456.5 & 1,683,926--62,071,649 \\
\rowcolor{black!7}
In-trace output tokens & 7,705,631 & 202,591 & 79,294--761,392 \\
In-trace total tokens & 490,956,921 & 10,713,592.5 & 1,763,220--62,677,852 \\
\rowcolor{black!7}
Matched detached-search tokens & 273,147,313 & 6,895,222 & 3,602,820--17,001,286 \\
Combined tokens & 764,104,234 & 18,221,464.5 & 7,536,313--73,613,587 \\
\midrule
\rowcolor{black!7}
In-trace model calls & 8,672 & 227.5 & 73--768 \\
In-trace tool calls & 12,756 & 318.5 & 108--1,468 \\
\rowcolor{black!7}
Matched detached-search calls & 4,056 & 104.5 & 54--258 \\
ORCA jobs & 219 & 6 & 1--15 \\
\midrule
\rowcolor{black!7}
ORCA calculation time (min) & 585.5 & 4.4 & 0.6--353.8 \\
Time outside ORCA jobs (min) & 2,275.6 & 51.4 & 17.9--502.2 \\
\bottomrule
\end{tabular}
\par\smallskip
\begin{minipage}{0.94\linewidth}\footnotesize
\emph{Notes.} In-trace input and output tokens sum to the in-trace total; detached-search tokens are added once in the combined total. Times sum active work across cases. Time outside ORCA jobs includes planning, searches, file inspection, other tools, and waits; it does not measure model execution time alone. Case-level token and call counts appear in Supporting Information Table~\ref{tab:usage-ledger-32}.
\end{minipage}
\end{table}

\FloatBarrier

\section{Outlook}

{\cinzel El Agente} is a comprehensive effort to build an AI scientist for computational and experimental chemistry in the context of self-driving laboratories and manual science. Our team is working tirelessly to release the platform to the community this calendar year. As part of this effort, we must carefully evaluate and optimize its core capabilities for multireference calculations. This work represents an important milestone towards understanding the state-of-the-art of not only {\cinzel El Agente} \cite{Zou2025,Cao2025Automating,gustin2026elagentecuantico,Perez-Zhou2026ElAgenteQuntur,Zhang2026Forjador,choi2026elagenteEstructural,Ko2026ElAgentePotente,mueller2026elagenteFarmaco,Kang2026ElAgenteSeguro,bai-aldossary2026grafico,Govind2026El,Kim2026materealize}, but also AI scientists developed by the community for multireference chemistry work \cite{Lee2026Multireference,Li2026CAFES}.

Multireference calculations require choosing an electronic model, obtaining its states, and checking that they support the scientific claim. In 32 fixed-geometry excitation problems, {\cinzel El Agente Quntur} used chemical and literature knowledge to propose active spaces, inspect orbitals and roots, run diagnostics, and revise calculations in response to evidence. Their reported gaps have lower aggregate numerical error than the tested static ASF selections. These trajectories show elements of scientific reasoning in CASSCF work, while extensive and uneven model activity and additional scientific jobs make efficiency a central challenge.

Reducing that effort calls for a few explicit decision/reasoning protocols suggested by domain experts: before each job, an agent could state which uncertainty it will resolve and what evidence would justify stopping or revising. Existing active-space guidance and protocols can inform those checks \citep{Veryazov2011,Keller2015,Rowell2023ActiveSpace,Seal2025WASP}. This is active work, and great progress in typed messages and workflow-based thinking has been demonstrated by efforts such as {\cinzel El Agente Gráfico} \cite{bai-aldossary2026grafico}. Bridging {\cinzel Quntur}'s generalist capability with {\cinzel Gráfico}'s workflow-oriented nature, which saves considerable AI token costs, is an active direction that will make these methods affordable for the general public.

Finally, training small specialized open-weight models for active-space selection on audited trajectories can improve efficiency, increase numerical accuracy, and further reduce token usage.

The importance of academically backed AI scientists who are described in the open literature, with benchmarks and trajectories available for community analysis, is crucial to keep the academic enterprise competitive with commercial efforts.


\section*{Acknowledgments}
We gratefully acknowledge the longstanding contributions of the Matter Lab's current
and past group members (\url{matter.toronto.edu}), and in particular from El Agente's team. We also thank Ignacio Gustin and Magali Gimeno for useful discussions. 
A.A.-G. thanks Anders G. Fr{\o}seth for his generous support.

\section*{Funding}
This work was supported by the Defense Advanced Research Projects Agency (DARPA) under Agreement No. HR0011262E022. This work was also supported by the AI2050 program of Schmidt Sciences.

J.B.P.S. acknowledges funding of this project by the Natural Sciences and Engineering Research Council of Canada (NSERC) Alliance Grant \#ALLRP587593-23 (Quantamole), and from the Council for Science, Technology and Innovation (CSTI), Cross-ministerial Strategic Innovation Promotion Program (SIP), ``Promoting the application of advanced quantum technology platforms to social issues'' (Funding agency: QST). V. R.-V. acknowledges support from the National Agency for Research and Development (ANID)/Scholarship Program/BECAS DOCTORADO NACIONAL/2024-21241825.

O.N.F. acknowledges support from the Spanish Ministry of Science, Innovation and Universities (MICIU) through a Salvador de Madariaga research mobility fellowship (grant PRX24/00448).

A.A.-G. and V.B. acknowledge the generous support of Natural Resources Canada and the Canada 150 Research Chairs program and the University of Toronto's Acceleration Consortium, which receives funding from the CFREF-2022-00042 Canada First Research Excellence Fund.

\section*{Resource Availability}

\subsection*{Data and code availability}

The benchmark inputs, calculation files, agent traces, and analysis materials supporting this study are available at \url{https://doi.org/10.5683/SP4/B8MB9Y} \citep{Perez2026QunturCASSCFFiles}. The source code of {\cinzel El Agente Quntur} is not publicly available.

\subsection*{Lead contact}

Requests for further information and resources should be directed to the lead contacts, Varinia Bernales and Al\'an Aspuru-Guzik.

\section*{Declaration of Interests}

The authors declare no competing interests. At the time of submission, no author holds equity, receives royalties or licensing income, or has a commercial agreement related to {\cinzel El Agente Quntur}. Its source code is presently non-public for reasons unrelated to commercialization.

\section*{Declaration of Generative AI and AI-Assisted Technologies in the Writing Process}

During manuscript preparation, the authors used OpenAI ChatGPT and Codex to help revise and draft portions of the text and figure captions, improve readability and organization, and assist with  reference formatting checks. The authors reviewed and edited this material and take full responsibility for the final manuscript.


{
\small
\bibliography{references}

@article{Heitler1927,
  author  = {Heitler, Walter and London, Fritz},
  title   = {Wechselwirkung neutraler Atome und homöopolare Bindung nach der Quantenmechanik},
  journal = {Zeitschrift für Physik},
  year    = {1927},
  volume  = {44},
  pages   = {455--472},
  doi     = {10.1007/BF01397394},
  url     = {https://doi.org/10.1007/BF01397394}
}

@article{Roos1980,
  author  = {Roos, Björn O. and Taylor, Peter R. and Siegbahn, Per E. M.},
  title   = {A Complete Active Space {SCF} Method ({CASSCF}) Using a Density Matrix Formulated Super-{CI} Approach},
  journal = {Chemical Physics},
  year    = {1980},
  volume  = {48},
  pages   = {157--173},
  doi     = {10.1016/0301-0104(80)80045-0},
  url     = {https://doi.org/10.1016/0301-0104(80)80045-0}
}

@article{Cohen2008,
  author  = {Cohen, Aron J. and Mori-Sánchez, Paula and Yang, Weitao},
  title   = {Insights into Current Limitations of Density Functional Theory},
  journal = {Science},
  year    = {2008},
  volume  = {321},
  pages   = {792--794},
  doi     = {10.1126/science.1158722},
  url     = {https://doi.org/10.1126/science.1158722}
}

@article{Szalay2012,
  author  = {Szalay, Péter G. and Müller, Thomas and Gidofalvi, Gergely and Lischka, Hans and Shepard, Ron},
  title   = {Multiconfiguration Self-Consistent Field and Multireference Configuration Interaction Methods and Applications},
  journal = {Chemical Reviews},
  year    = {2012},
  volume  = {112},
  pages   = {108--181},
  doi     = {10.1021/cr200137a},
  url     = {https://doi.org/10.1021/cr200137a}
}

@article{Angeli2001,
  author  = {Angeli, Celestino and Cimiraglia, Renzo and Evangelisti, Stefano and Leininger, Thierry and Malrieu, Jean-Paul},
  title   = {Introduction of {$n$}-Electron Valence States for Multireference Perturbation Theory},
  journal = {The Journal of Chemical Physics},
  year    = {2001},
  volume  = {114},
  pages   = {10252--10264},
  doi     = {10.1063/1.1361246},
  url     = {https://doi.org/10.1063/1.1361246}
}

@article{Keller2015,
  author  = {Keller, Sebastian and Boguslawski, Katharina and Janowski, Tomasz and Reiher, Markus and Pulay, Peter},
  title   = {Selection of Active Spaces for Multiconfigurational Wavefunctions},
  journal = {The Journal of Chemical Physics},
  year    = {2015},
  volume  = {142},
  pages   = {244104},
  doi     = {10.1063/1.4922352},
  url     = {https://doi.org/10.1063/1.4922352}
}

@article{Stein2016,
  author  = {Stein, Christopher J. and Reiher, Markus},
  title   = {Automated Selection of Active Orbital Spaces},
  journal = {Journal of Chemical Theory and Computation},
  year    = {2016},
  volume  = {12},
  pages   = {1760--1771},
  doi     = {10.1021/acs.jctc.6b00156},
  url     = {https://doi.org/10.1021/acs.jctc.6b00156}
}

@article{Sayfutyarova2017,
  author  = {Sayfutyarova, Elvira R. and Sun, Qiming and Chan, Garnet Kin-Lic and Knizia, Gerald},
  title   = {Automated Construction of Molecular Active Spaces from Atomic Valence Orbitals},
  journal = {Journal of Chemical Theory and Computation},
  year    = {2017},
  volume  = {13},
  pages   = {4063--4078},
  doi     = {10.1021/acs.jctc.7b00128},
  url     = {https://doi.org/10.1021/acs.jctc.7b00128}
}

@article{Stein2019,
author = {Stein, Christopher J. and Reiher, Markus},
title = {auto{CAS}: A Program for Fully Automated Multiconfigurational Calculations},
journal = {Journal of Computational Chemistry},
volume = {40},
number = {25},
pages = {2216-2226},
doi = {https://doi.org/10.1002/jcc.25869},
url = {https://onlinelibrary.wiley.com/doi/abs/10.1002/jcc.25869},
eprint = {https://onlinelibrary.wiley.com/doi/pdf/10.1002/jcc.25869},
year = {2019}
}

@article{Khedkar2019,
  author = {Khedkar, Abhishek and Roemelt, Michael},
  title = {Active Space Selection Based on Natural Orbital Occupation Numbers from {$n$}-Electron Valence Perturbation Theory},
  journal = {Journal of Chemical Theory and Computation},
  year = {2019},
  volume = {15},
  number = {6},
  pages = {3522--3536},
  doi = {10.1021/acs.jctc.8b01293},
  url = {https://doi.org/10.1021/acs.jctc.8b01293},
}

@article{Ding2023,
  author  = {Ding, Lexin and Knecht, Stefan and Schilling, Christian},
  title   = {Quantum Information-Assisted Complete Active Space Optimization ({QICAS})},
  journal = {The Journal of Physical Chemistry Letters},
  year    = {2023},
  volume  = {14},
  pages   = {11022--11029},
  doi     = {10.1021/acs.jpclett.3c02536},
  url     = {https://doi.org/10.1021/acs.jpclett.3c02536}
}

@article{Zou2025,
  author  = {Zou, Yunheng and Cheng, Austin H. and Aldossary, Abdulrahman and Bai, Jiaru and Leong, Shi Xuan and Campos-Gonzalez-Angulo, Jorge Arturo and Choi, Changhyeok and Ser, Cher Tian and Tom, Gary and Wang, Andrew and Zhang, Zijian and Yakavets, Ilya and Hao, Han and Crebolder, Chris and Bernales, Varinia and Aspuru-Guzik, Alán},
  title   = {El Agente: An Autonomous Agent for Quantum Chemistry},
  journal = {Matter},
  year    = {2025},
  volume  = {8},
  pages   = {102263},
  doi     = {10.1016/j.matt.2025.102263},
  url     = {https://doi.org/10.1016/j.matt.2025.102263}
}

@article{Neese2012The,
author = {Neese, Frank},
title = {The {ORCA} program system},
journal = {WIREs Computational Molecular Science},
volume = {2},
number = {1},
pages = {73-78},
doi = {https://doi.org/10.1002/wcms.81},
url = {https://wires.onlinelibrary.wiley.com/doi/abs/10.1002/wcms.81},
year = {2012}
}

@article{Neese_ORCA6,
  author       = {Neese, Frank},
  title        = {Software Update: The {ORCA} Program System—Version 6.0},
  journal      = {WIREs Computational Molecular Science},
  year         = {2025},
  volume       = {15},
  number       = {2},
  pages        = {e70019},
  doi          = {10.1002/wcms.70019},
  url          = {https://wires.onlinelibrary.wiley.com/doi/10.1002/wcms.70019}
}

@misc{Mitchener2025kosmos,
      title={Kosmos: An {AI} Scientist for Autonomous Discovery. {P}reprint at https://arxiv.org/abs/2511.02824}, 
      author={Ludovico Mitchener and Angela Yiu and Benjamin Chang and Mathieu Bourdenx and Tyler Nadolski and Arvis Sulovari and Eric C. Landsness and Daniel L. Barabasi and Siddharth Narayanan and Nicky Evans and Shriya Reddy and Martha Foiani and Aizad Kamal and Leah P. Shriver and Fang Cao and Asmamaw T. Wassie and Jon M. Laurent and Edwin Melville-Green and Mayk Caldas and Albert Bou and Kaleigh F. Roberts and Sladjana Zagorac and Timothy C. Orr and Miranda E. Orr and Kevin J. Zwezdaryk and Ali E. Ghareeb and Laurie McCoy and Bruna Gomes and Euan A. Ashley and Karen E. Duff and Tonio Buonassisi and Tom Rainforth and Randall J. Bateman and Michael Skarlinski and Samuel G. Rodriques and Michaela M. Hinks and Andrew D. White},
      year={2025},
      eprint={2511.02824},
      archivePrefix={arXiv},
      primaryClass={cs.AI},
      url={https://arxiv.org/abs/2511.02824}, 
}

@misc{Govind2026El,
      title={El {A}gente {S}\'olido: A New Age(nt) for Solid State Simulations. {P}reprint available at https://arxiv.org/abs/2602.17886},
      author={Sai Govind Hari Kumar and Yunheng Zou and Andrew Wang and Jesús Valdés-Hernández and Tsz Wai Ko and Nathan Yue and Olivia Leng and Hanyong Xu and Chris Crebolder and Alán Aspuru-Guzik and Varinia Bernales},
      year={2026},
      eprint={2602.17886},
      archivePrefix={arXiv},
      primaryClass={cond-mat.mtrl-sci},
      url={https://arxiv.org/abs/2602.17886}, 
}

@unpublished{mueller2026elagenteFarmaco,
  title = {El {A}gente {F}\'armaco: {{Agentic AI}} for Autonomous in-Silico Drug Discovery},
  author = {M{\"u}ller, Marcel and Vakili, Mohammad Ghazi and Rajaonson, Ella Miray and Ashoori, Ella and Zou, Yunheng and Bernales, Varinia and {Aspuru-Guzik}, Al{\'a}n},
  year = 2026,
  note = {Manuscript in preparation}
}

@misc{bai-aldossary2026grafico,
      title={El {A}gente {G}r\'afico: A Semantic Execution Runtime for Scientific Agents. {P}reprint available at https://arxiv.org/abs/2602.17902},
      author={Jiaru Bai and Abdulrahman Aldossary and Thomas Swanick and Marcel Müller and Yeonghun Kang and Changhyeok Choi and Naruki Yoshikawa and Zijian Zhang and Jin Won Lee and Tsz Wai Ko and Aiwei Yin and Mohammad Ghazi Vakili and Chris Crebolder and Varinia Bernales and Alán Aspuru-Guzik},
      year={2026},
      eprint={2602.17902},
      archivePrefix={arXiv},
      primaryClass={cs.AI},
      url={https://arxiv.org/abs/2602.17902}, 
}

@misc{choi2026elagenteEstructural,
      title={El {A}gente {E}structural: An Artificially Intelligent Molecular Editor. {P}reprint available at https://arxiv.org/abs/2602.04849},
      author={Changhyeok Choi and Yunheng Zou and Marcel Müller and Han Hao and Yeonghun Kang and Juan B. Pérez-Sánchez and Ignacio Gustin and Hanyong Xu and Andrew Wang and Mohammad Ghazi Vakili and Chris Crebolder and Alán Aspuru-Guzik and Varinia Bernales},
      year={2026},
      eprint={2602.04849},
      archivePrefix={arXiv},
      primaryClass={physics.chem-ph},
      url={https://arxiv.org/abs/2602.04849}, 
}

@article{gustin2026elagentecuantico,
      title={El {A}gente {C}u\'antico: Automating Quantum Simulations},
      author={Ignacio Gustin and Luis Mantilla Calderón and Juan B. Pérez-Sánchez and Chris Crebolder and Jérôme F. Gonthier and Mohammad Ghazi Vakili and Yuma Nakamura and Karthik Panicker and Manav Ramprasad and Aiwei Yin and Zijian Zhang and Yunheng Zou and Varinia Bernales and Alán Aspuru-Guzik},
      journal={Reports on Progress in Physics},
      volume={89},
      number={7},
      pages={077602},
      year={2026},
      doi={10.1088/1361-6633/ae8933},
      url={https://doi.org/10.1088/1361-6633/ae8933},
}

@article{Qiming2018PySCF,
author = {Sun, Qiming and Berkelbach, Timothy C. and Blunt, Nick S. and Booth, George H. and Guo, Sheng and Li, Zhendong and Liu, Junzi and McClain, James D. and Sayfutyarova, Elvira R. and Sharma, Sandeep and Wouters, Sebastian and Chan, Garnet Kin-Lic},
title = {Py{SCF}: the Python-based simulations of chemistry framework},
journal = {WIREs Computational Molecular Science},
volume = {8},
number = {1},
pages = {e1340},
doi = {https://doi.org/10.1002/wcms.1340},
url = {https://wires.onlinelibrary.wiley.com/doi/abs/10.1002/wcms.1340},
eprint = {https://wires.onlinelibrary.wiley.com/doi/pdf/10.1002/wcms.1340},
year = {2018}
}

@misc{Kim2026materealize,
      title={Materealize: a multi-agent deliberation system for end-to-end material design and synthesis. {P}reprint available at https://arxiv.org/abs/2601.15743}, 
      author={Seongmin Kim and Jaehwan Choi and Kunik Jang and Junkil Park and Varinia Bernales and Alán Aspuru-Guzik and Yousung Jung},
      year={2026},
      eprint={2601.15743},
      archivePrefix={arXiv},
      primaryClass={cond-mat.mtrl-sci},
      url={https://arxiv.org/abs/2601.15743}, 
}

@misc{Perez2026QunturCASSCFFiles,
  author       = {Perez Sanchez, Juan},
  title        = {Replication Data for: Agentic Decision-Making versus Static Active-Space Selection for {CASSCF} Calculations},
  year         = {2026},
  month        = {October},
  publisher    = {Borealis},
  doi          = {10.5683/SP4/B8MB9Y},
  url          = {https://borealisdata.ca/dataset.xhtml?persistentId=doi:10.5683/SP4/B8MB9Y}
}

@misc{Shirazi2025ASF,
  author        = {Shirazi, Reza G. and Zech, Alexander and Pinski, Peter and Rybkin, Vladimir V.},
  title         = {Performance of Automatic Active Space Selection for Electronic Excitation Energies. {P}reprint available at https://arxiv.org/abs/2511.05732},
  year          = {2025},
  eprint        = {2511.05732},
  archivePrefix = {arXiv},
  primaryClass  = {physics.chem-ph},
  note          = {Version 2},
  url           = {https://arxiv.org/abs/2511.05732}
}

@article{Perez-Zhou2026ElAgenteQuntur,
  author        = {P{\'e}rez-S{\'a}nchez, Juan B. and Zou, Yunheng and Campos-Gonzalez-Angulo, Jorge A. and M{\"u}ller, Marcel and Gustin, Ignacio and Wang, Andrew and Hao, Han and Ko, Tsz Wai and Choi, Changhyeok and Isbrandt, Eric S. and Vakili, Mohammad Ghazi and Xu, Hanyong and Crebolder, Chris and Bernales, Varinia and Aspuru-Guzik, Al{\'a}n},
  title         = {{El Agente Quntur}: A Research Collaborator Agent for Quantum Chemistry},
  journal       = {Chem},
  pages         = {103259},
  year          = {2026},
  doi           = {10.1016/j.chempr.2026.103259},
  url           = {https://doi.org/10.1016/j.chempr.2026.103259}
}

@misc{Ko2026ElAgentePotente,
  author        = {Ko, Tsz Wai and Bai, Jiaru and Swanick, Thomas and Kang, Yeonghun and Choi, Changhyeok and Jiang, Angelina Qihong and Yin, Aiwei and Bernales, Varinia and Aspuru-Guzik, Al{\'a}n},
  title         = {{El Agente Potente}: High-Throughput Agentic Atomistic Simulations. {P}reprint available at https://arxiv.org/abs/2609.14840},
  year          = {2026},
  eprint        = {2609.14840},
  archivePrefix = {arXiv},
  primaryClass  = {cs.AI},
  url           = {https://arxiv.org/abs/2609.14840}
}

@misc{Kang2026ElAgenteSeguro,
  author        = {Kang, Yeonghun and Leong, Shi Xuan and Park, Hyun Suk and Bai, Jiaru and Papidocha, Sven and Zhao, Yuchi and Zhang, Rui and Gimeno, Magali and Lo, Stanley and Isbrandt, Eric S. and Hao, Han and Cao, Yang and Katzenburg, Felix and Du, Fangshi and Aspuru-Guzik, Al{\'a}n and Bernales, Varinia},
  title         = {{El Agente Seguro}: An Agent for Chemical Safety},
  year          = {2026},
  doi           = {10.26434/chemrxiv.15004378/v2},
  note          = {ChemRxiv preprint, version 2},
  url           = {https://doi.org/10.26434/chemrxiv.15004378/v2}
}

@misc{Lee2026Multireference,
  author        = {Lee, Victor Chang and Rondinelli, James M.},
  title         = {Can Autonomous {LLM} Agents Execute Multireference Quantum Chemistry Calculations?. {P}reprint available at https://arxiv.org/abs/2609.13357},
  year          = {2026},
  eprint        = {2609.13357},
  archivePrefix = {arXiv},
  primaryClass  = {physics.chem-ph},
  url           = {https://arxiv.org/abs/2609.13357}
}

@misc{Li2026CAFES,
  author        = {Li, Tenghui and Sun, Chong},
  title         = {Automated Many-Body Simulations of Strongly Correlated Systems Using a Correlation-Aware Agentic Framework. {P}reprint available at https://arxiv.org/abs/2610.00943},
  year          = {2026},
  eprint        = {2610.00943},
  archivePrefix = {arXiv},
  primaryClass  = {physics.comp-ph},
  url           = {https://arxiv.org/abs/2610.00943}
}

@article{Schreiber2008,
  author  = {Schreiber, Marko and Silva-Junior, Mario R. and Sauer, Stephan P. A. and Thiel, Walter},
  title   = {Benchmarks for Electronically Excited States: {CASPT2}, {CC2}, {CCSD}, and {CC3}},
  journal = {The Journal of Chemical Physics},
  year    = {2008},
  volume  = {128},
  pages   = {134110},
  doi     = {10.1063/1.2889385}
}

@article{Hoyer2016,
  author  = {Hoyer, Chad E. and Ghosh, Soumen and Truhlar, Donald G. and Gagliardi, Laura},
  title   = {Multiconfiguration Pair-Density Functional Theory Is as Accurate as {CASPT2} for Electronic Excitation},
  journal = {The Journal of Physical Chemistry Letters},
  year    = {2016},
  volume  = {7},
  pages   = {586--591},
  doi     = {10.1021/acs.jpclett.5b02773}
}

@article{Lischka2018,
  author = {Lischka, Hans and Nachtigallov{\'a}, Dana and Aquino, Ad{\'e}lia J. A. and Szalay, P{\'e}ter G. and Plasser, Felix and Machado, Francisco B. C. and Barbatti, Mario},
  title = {{Multireference Approaches for Excited States of Molecules}},
  journal = {Chemical Reviews},
  year = {2018},
  volume = {118},
  number = {15},
  pages = {7293-7361},
  doi = {10.1021/acs.chemrev.8b00244},
  url = {https://doi.org/10.1021/acs.chemrev.8b00244},
}

@article{Veryazov2011,
  author = {Veryazov, Valera and Malmqvist, Per {\AA}ke and Roos, Bj{\"o}rn O.},
  title = {{How to select active space for multiconfigurational quantum chemistry?}},
  journal = {International Journal of Quantum Chemistry},
  year = {2011},
  volume = {111},
  number = {13},
  pages = {3329-3338},
  doi = {10.1002/qua.23068},
  url = {https://doi.org/10.1002/qua.23068},
}

@article{Pulay1988,
  author = {Pulay, Peter and Hamilton, Tracy P.},
  title = {{UHF natural orbitals for defining and starting MC-SCF calculations}},
  journal = {The Journal of Chemical Physics},
  year = {1988},
  volume = {88},
  number = {8},
  pages = {4926-4933},
  doi = {10.1063/1.454704},
  url = {https://doi.org/10.1063/1.454704},
}

@article{Jensen1988,
  author = {Jensen, Hans J{\o}rgen Aa. and J{\o}rgensen, Poul and {\AA}gren, Hans and Olsen, Jeppe},
  title = {{Second-order M{\o}ller--Plesset perturbation theory as a configuration and orbital generator in multiconfiguration self-consistent field calculations}},
  journal = {The Journal of Chemical Physics},
  year = {1988},
  volume = {88},
  number = {6},
  pages = {3834-3839},
  doi = {10.1063/1.453884},
  url = {https://doi.org/10.1063/1.453884},
}

@article{Bao2018,
  author = {Bao, Jie J. and Dong, Sijia S. and Gagliardi, Laura and Truhlar, Donald G.},
  title = {{Automatic Selection of an Active Space for Calculating Electronic Excitation Spectra by MS-CASPT2 or MC-PDFT}},
  journal = {Journal of Chemical Theory and Computation},
  year = {2018},
  volume = {14},
  number = {4},
  pages = {2017-2025},
  doi = {10.1021/acs.jctc.8b00032},
  url = {https://doi.org/10.1021/acs.jctc.8b00032},
}

@article{Bao2019,
  author = {Bao, Jie J. and Truhlar, Donald G.},
  title = {{Automatic Active Space Selection for Calculating Electronic Excitation Energies Based on High-Spin Unrestricted Hartree--Fock Orbitals}},
  journal = {Journal of Chemical Theory and Computation},
  year = {2019},
  volume = {15},
  number = {10},
  pages = {5308-5318},
  doi = {10.1021/acs.jctc.9b00535},
  url = {https://doi.org/10.1021/acs.jctc.9b00535},
}

@article{Khedkar2020,
  author = {Khedkar, Abhishek and Roemelt, Michael},
  title = {{Extending the ASS1ST Active Space Selection Scheme to Large Molecules and Excited States}},
  journal = {Journal of Chemical Theory and Computation},
  year = {2020},
  volume = {16},
  number = {8},
  pages = {4993-5005},
  doi = {10.1021/acs.jctc.0c00332},
  url = {https://doi.org/10.1021/acs.jctc.0c00332},
}

@article{Claudino2019,
  author = {Claudino, Daniel and Mayhall, Nicholas J.},
  title = {{Automatic Partition of Orbital Spaces Based on Singular Value Decomposition in the Context of Embedding Theories}},
  journal = {Journal of Chemical Theory and Computation},
  year = {2019},
  volume = {15},
  number = {2},
  pages = {1053-1064},
  doi = {10.1021/acs.jctc.8b01112},
  url = {https://doi.org/10.1021/acs.jctc.8b01112},
}

@article{Kolodzeiski2023,
  author = {Kolodzeiski, Elena and Stein, Christopher J.},
  title = {{Automated, Consistent, and Even-Handed Selection of Active Orbital Spaces for Quantum Embedding}},
  journal = {Journal of Chemical Theory and Computation},
  year = {2023},
  volume = {19},
  number = {19},
  pages = {6643-6655},
  doi = {10.1021/acs.jctc.3c00653},
  url = {https://doi.org/10.1021/acs.jctc.3c00653},
}

@article{Lei2021,
  author = {Lei, Yibo and Suo, Bingbing and Liu, Wenjian},
  title = {{iCAS: Imposed Automatic Selection and Localization of Complete Active Spaces}},
  journal = {Journal of Chemical Theory and Computation},
  year = {2021},
  volume = {17},
  number = {8},
  pages = {4846-4859},
  doi = {10.1021/acs.jctc.1c00456},
  url = {https://doi.org/10.1021/acs.jctc.1c00456},
}

@article{King2021,
  author = {King, Daniel S. and Gagliardi, Laura},
  title = {{A Ranked-Orbital Approach to Select Active Spaces for High-Throughput Multireference Computation}},
  journal = {Journal of Chemical Theory and Computation},
  year = {2021},
  volume = {17},
  number = {5},
  pages = {2817-2831},
  doi = {10.1021/acs.jctc.1c00037},
  url = {https://doi.org/10.1021/acs.jctc.1c00037},
}

@article{King2022,
  author = {King, Daniel S. and Hermes, Matthew R. and Truhlar, Donald G. and Gagliardi, Laura},
  title = {{Large-Scale Benchmarking of Multireference Vertical-Excitation Calculations via Automated Active-Space Selection}},
  journal = {Journal of Chemical Theory and Computation},
  year = {2022},
  volume = {18},
  number = {10},
  pages = {6065-6076},
  doi = {10.1021/acs.jctc.2c00630},
  url = {https://doi.org/10.1021/acs.jctc.2c00630},
}

@article{Jeong2020,
  author = {Jeong, WooSeok and Stoneburner, Samuel J. and King, Daniel and Li, Ruye and Walker, Andrew and Lindh, Roland and Gagliardi, Laura},
  title = {{Automation of Active Space Selection for Multireference Methods via Machine Learning on Chemical Bond Dissociation}},
  journal = {Journal of Chemical Theory and Computation},
  year = {2020},
  volume = {16},
  number = {4},
  pages = {2389-2399},
  doi = {10.1021/acs.jctc.9b01297},
  url = {https://doi.org/10.1021/acs.jctc.9b01297},
}

@article{Golub2021,
  author = {Golub, Pavlo and Antalik, Andrej and Veis, Libor and Brabec, Jiri},
  title = {{Machine Learning-Assisted Selection of Active Spaces for Strongly Correlated Transition Metal Systems}},
  journal = {Journal of Chemical Theory and Computation},
  year = {2021},
  volume = {17},
  number = {10},
  pages = {6053-6072},
  doi = {10.1021/acs.jctc.1c00235},
  url = {https://doi.org/10.1021/acs.jctc.1c00235},
}

@article{Kaufold2023,
  author = {Kaufold, Benjamin W. and Chintala, Nithin and Pandeya, Pratima and Dong, Sijia S.},
  title = {{Automated Active Space Selection with Dipole Moments}},
  journal = {Journal of Chemical Theory and Computation},
  year = {2023},
  volume = {19},
  number = {9},
  pages = {2469-2483},
  doi = {10.1021/acs.jctc.2c01128},
  url = {https://doi.org/10.1021/acs.jctc.2c01128},
}

@article{Kaufold2026CASCI,
  author = {Kaufold, Benjamin W. and Dong, Sijia S.},
  title = {Automated Active Space Selection with {CASCI} Dipole Moments},
  journal = {Journal of Chemical Theory and Computation},
  year = {2026},
  volume = {22},
  number = {15},
  pages = {7624--7641},
  doi = {10.1021/acs.jctc.6c00473},
  url = {https://doi.org/10.1021/acs.jctc.6c00473}
}

@article{HelmichParis2019,
  author = {Helmich-Paris, Benjamin},
  title = {{Benchmarks for Electronically Excited States with CASSCF Methods}},
  journal = {Journal of Chemical Theory and Computation},
  year = {2019},
  volume = {15},
  number = {7},
  pages = {4170-4179},
  doi = {10.1021/acs.jctc.9b00325},
  url = {https://doi.org/10.1021/acs.jctc.9b00325},
}

@article{Veril2021,
  author = {V{\'e}ril, Mickaël and Scemama, Anthony and Caffarel, Michel and Lipparini, Filippo and Boggio-Pasqua, Martial and Jacquemin, Denis and Loos, Pierre-François},
  title = {{QUESTDB: A database of highly accurate excitation energies for the electronic structure community}},
  journal = {WIREs Computational Molecular Science},
  year = {2021},
  volume = {11},
  number = {5},
  pages = {e1517},
  doi = {10.1002/wcms.1517},
  url = {https://doi.org/10.1002/wcms.1517},
}

@article{Loos2018,
  author = {Loos, Pierre-François and Scemama, Anthony and Blondel, Aymeric and Garniron, Yann and Caffarel, Michel and Jacquemin, Denis},
  title = {{A Mountaineering Strategy to Excited States: Highly Accurate Reference Energies and Benchmarks}},
  journal = {Journal of Chemical Theory and Computation},
  year = {2018},
  volume = {14},
  number = {8},
  pages = {4360-4379},
  doi = {10.1021/acs.jctc.8b00406},
  url = {https://doi.org/10.1021/acs.jctc.8b00406},
}

@article{Werner1981,
  author = {Werner, Hans-Joachim and Meyer, Wilfried},
  title = {{A quadratically convergent MCSCF method for the simultaneous optimization of several states}},
  journal = {The Journal of Chemical Physics},
  year = {1981},
  volume = {74},
  number = {10},
  pages = {5794-5801},
  doi = {10.1063/1.440892},
  url = {https://doi.org/10.1063/1.440892},
}

@article{Angeli2001SC,
  author = {Angeli, Celestino and Cimiraglia, Renzo and Malrieu, Jean-Paul},
  title = {{N-electron valence state perturbation theory: a fast implementation of the strongly contracted variant}},
  journal = {Chemical Physics Letters},
  year = {2001},
  volume = {350},
  number = {3-4},
  pages = {297-305},
  doi = {10.1016/s0009-2614(01)01303-3},
  url = {https://doi.org/10.1016/s0009-2614(01)01303-3},
}

@article{Sarkar2022,
  author = {Sarkar, Rudraditya and Loos, Pierre-François and Boggio-Pasqua, Martial and Jacquemin, Denis},
  title = {{Assessing the Performances of CASPT2 and NEVPT2 for Vertical Excitation Energies}},
  journal = {Journal of Chemical Theory and Computation},
  year = {2022},
  volume = {18},
  number = {4},
  pages = {2418-2436},
  doi = {10.1021/acs.jctc.1c01197},
  url = {https://doi.org/10.1021/acs.jctc.1c01197},
}

@misc{VermaTruhlar2019Geometries,
  author = {Verma, Pragya and Truhlar, Donald G.},
  title = {{Geometries for Minnesota Database 2019}},
  year = {2019},
  publisher = {Data Repository for the University of Minnesota},
  doi = {10.13020/217y-8g32},
  url = {https://doi.org/10.13020/217y-8g32},
}

@misc{Rowell2023ActiveSpace,
  author = {Rowell, Keiran},
  title = {Selecting an active space},
  year = {2023},
  howpublished = {Online guide},
  url = {https://keiran-rowell.github.io/guide/2023-04-12-picking-an-active-space/},
  note = {Accessed 10 October 2026}
}

@article{Seal2025WASP,
  author = {Seal, Aniruddha and Perego, Simone and Hennefarth, Matthew R. and Raucci, Umberto and Bonati, Luigi and Ferguson, Andrew L. and Parrinello, Michele and Gagliardi, Laura},
  title = {Weighted active space protocol for multireference machine-learned potentials},
  journal = {Proceedings of the National Academy of Sciences},
  year = {2025},
  volume = {122},
  number = {38},
  pages = {e2513693122},
  doi = {10.1073/pnas.2513693122},
  url = {https://doi.org/10.1073/pnas.2513693122}
}

@book{helgaker2000molecular,
  author    = {Helgaker, Trygve and J{\o}rgensen, Poul and Olsen, Jeppe},
  title     = {Molecular Electronic-Structure Theory},
  publisher = {John Wiley \& Sons},
  address   = {Chichester, UK},
  year      = {2000},
  isbn      = {978-0-471-96755-2}
}

@article{Sharma2019Valence,
    author = {Sharma, Prachi and Bernales, Varinia and Truhlar, Donald G. and Gagliardi, Laura},
    title = {Valence $\pi\pi^*$ Excitations in Benzene Studied
by Multiconfiguration Pair-Density Functional Theory},
    journal = {The Journal of Physical Chemistry Letters},
    volume = {10},
    number = {1},
    pages = {75-81},
    year = {2019},
    month = {01},
    issn = {1948-7185},
    doi = {10.1021/acs.jpclett.8b03277},
    url = {https://doi.org/10.1021/acs.jpclett.8b03277},
    eprint = {https://pubs.acs.org/jpclcd/article-pdf/10/1/75/6446861/jz8b03277.pdf},
}

@article{Sharma2019Density,
    author = {Sharma, Prachi and Bernales, Varinia and Knecht, Stefan and Truhlar, Donald G. and Gagliardi, Laura},
    title = {Density matrix renormalization group pair-density functional theory (DMRG-PDFT): singlet–triplet gaps in polyacenes and polyacetylenes},
    journal = {Chemical Science},
    volume = {10},
    number = {6},
    pages = {1716-1723},
    year = {2019},
    month = {02},
    issn = {2041-6520},
    doi = {10.1039/c8sc03569e},
    url = {https://doi.org/10.1039/c8sc03569e},
    eprint = {https://pubs.rsc.org/sc/article-pdf/10/6/1716/6316205/c8sc03569e.pdf},
}

@article{Ghosh2017Generalized,
    author = {Ghosh, Soumen and Cramer, Christopher J. and Truhlar, Donald G. and Gagliardi, Laura},
    title = {Generalized-active-space pair-density functional theory: an efficient method to study large, strongly correlated, conjugated systems},
    journal = {Chemical Science},
    volume = {8},
    number = {4},
    pages = {2741-2750},
    year = {2017},
    month = {04},
    issn = {2041-6520},
    doi = {10.1039/c6sc05036k},
    url = {https://doi.org/10.1039/c6sc05036k},
    eprint = {https://pubs.rsc.org/sc/article-pdf/8/4/2741/5218459/c6sc05036k.pdf},
}

@misc{HQSActiveSpaceFinder,
  author       = {{HQS Quantum Simulations GmbH}},
  title        = {{Active Space Finder (ASF)}},
  howpublished = {GitHub repository},
  url          = {https://github.com/HQSquantumsimulations/ActiveSpaceFinder},
  year      = {2026}
}

@misc{zoujingxiang2026MOKIT,
	title = {Molecular {Orbital} {Kit} ({MOKIT})},
	url = {https://gitlab.com/jxzou/mokit},
	urldate = {2026-10-18},
	author = {Zou, Jingxiang},
	month = aug,
}

@article{Neese2006Importance,
    author = {Neese, Frank},
    title = {Importance of Direct Spin-Spin Coupling and Spin-Flip
Excitations for the Zero-Field Splittings of Transition Metal Complexes: A Case Study},
    journal = {Journal of the American Chemical Society},
    volume = {128},
    number = {31},
    pages = {10213-10222},
    year = {2006},
    month = {08},
    issn = {0002-7863},
    doi = {10.1021/ja061798a},
    url = {https://doi.org/10.1021/ja061798a},
    eprint = {https://pubs.acs.org/jacsat/article-pdf/128/31/10213/38981197/ja061798a.pdf},
}

@inproceedings{Pople1973theoretical,
  author    = {Pople, John A.},
  title     = {Theoretical Models for Chemistry},
  booktitle = {Energy, Structure, and Reactivity: Proceedings of the
               1972 Boulder Summer Research Conference on Theoretical Chemistry},
  editor    = {Smith, Darwin W. and McRae, Walter B.},
  publisher = {John Wiley \& Sons},
  address   = {New York},
  year      = {1973},
  pages     = {51--61}
}

@article{HeadGordon1996Quantum,
    author = {Head-Gordon, Martin},
    title = {Quantum Chemistry and Molecular Processes},
    journal = {The Journal of Physical Chemistry},
    volume = {100},
    number = {31},
    pages = {13213-13225},
    year = {1996},
    month = {08},
    issn = {0022-3654},
    doi = {10.1021/jp953665+},
    url = {https://doi.org/10.1021/jp953665+},
    eprint = {https://pubs.acs.org/jpchax/article-pdf/100/31/13213/16072561/jp953665%2b.pdf},
}

@article{Li2026Harness,
	doi = {10.20944/preprints202606.2203.v1},
	url = {https://doi.org/10.20944/preprints202606.2203.v1},
	year = 2026,
	month = {June},
	publisher = {Preprints},
	author = {Jinzhe Li and Yuan Wu and Yi Chang},
	title = {Harness Engineering for LLM Agents: A Survey of Harness Component Taxonomy, Evaluation, and Model–Harness Coevolution},
	journal = {Preprints}
}

@misc{openai_codex,
  author       = {{openai}},
  title        = {{openai/codex}},
  howpublished = {GitHub repository},
  url          = {https://github.com/openai/codex},
  note         = {Accessed 2026-10-27}
}

@misc{anthropics_claude_code,
  author       = {{anthropics}},
  title        = {{anthropics/claude-code}},
  howpublished = {GitHub repository},
  url          = {https://github.com/anthropics/claude-code},
  note         = {Accessed 2026-10-27}
}

@misc{earendil_works_pi,
  author       = {{earendil-works}},
  title        = {{earendil-works/pi}},
  howpublished = {GitHub repository},
  url          = {https://github.com/earendil-works/pi},
  note         = {Accessed 2026-10-27}
}

@misc{Anthropic2026ClaudeScience,
  author       = {{Anthropic}},
  title        = {{Claude Science}, an AI Workbench for Scientists,
                  Is Now Available},
  year         = {2026},
  month        = jun,
  howpublished = {Product announcement},
  url          = {https://www.anthropic.com/news/claude-science-ai-workbench},
  note         = {Published June 30, 2026; accessed October 27, 2026}
}

@misc{Zhang2026Forjador,
      title={El Agente Forjador: Task-Driven Agent Generation for Quantum Simulation. {P}reprint available at https://arxiv.org/abs/2604.14609},
      author={Zijian Zhang and Aiwei Yin and Amaan Baweja and Jiaru Bai and Ignacio Gustin and Varinia Bernales and Alán Aspuru-Guzik},
      year={2026},
      eprint={2604.14609},
      archivePrefix={arXiv},
      primaryClass={cs.AI},
      url={https://arxiv.org/abs/2604.14609}, 
}

@misc{Stanford2025technical,
  author        = {{Stanford HAI}},
  title         = {{Technical Performance: The 2025 AI Index Report}},
  year          = {2025},
  howpublished  = {AI Index Report chapter webpage},
  url           = {https://hai.stanford.edu/ai-index/2025-ai-index-report/technical-performance},
  urldate       = {2026-10-27},
  note          = {Accessed October 27, 2026}
}

@misc{Stanford2026technical,
  author        = {{Stanford HAI}},
  title         = {{Technical Performance: The 2026 AI Index Report}},
  year          = {2026},
  howpublished  = {AI Index Report chapter webpage},
  url           = {https://hai.stanford.edu/ai-index/2026-ai-index-report/technical-performance},
  urldate       = {2026-10-27},
  note          = {Accessed October 27, 2026}
}

@misc{Artificialanalysis2026openweights,
  author        = {{Artificial Analysis}},
  title         = {{Recent open weights model launches}},
  year          = {2026},
  month         = {April},
  day           = {30},
  howpublished  = {Artificial Analysis article},
  url           = {https://artificialanalysis.ai/articles/recent-open-weights-model-launches},
  urldate       = {2026-10-27},
  note          = {Published April 30, 2026; accessed October 27, 2026}
}

@misc{Arena2026leaderboard,
  author        = {{Arena}},
  title         = {{Text Arena: LLM Rankings \& AI Model Benchmarks}},
  howpublished  = {Online leaderboard},
  url           = {https://arena.ai/leaderboard/chat/text},
  urldate       = {2026-10-27},
  note          = {Accessed October 27, 2026}
}

@misc{Arcprize2026leaderboard,
  author        = {{ARC Prize}},
  title         = {{ARC Prize - Leaderboard}},
  howpublished  = {Verified online leaderboard},
  url           = {https://arcprize.org/leaderboard},
  urldate       = {2026-10-27},
  note          = {Accessed October 27, 2026}
}

@inproceedings{Chiang2024chatbotarena,
  author    = {Chiang, Wei-Lin and Zheng, Lianmin and Sheng, Ying and Angelopoulos, Anastasios Nikolas and Li, Tianle and Li, Dacheng and Zhu, Banghua and Zhang, Hao and Jordan, Michael and Gonzalez, Joseph E. and Stoica, Ion},
  title     = {Chatbot Arena: An Open Platform for Evaluating {LLM}s by Human Preference},
  booktitle = {Proceedings of the 41st International Conference on Machine Learning},
  series    = {Proceedings of Machine Learning Research},
  volume    = {235},
  pages     = {8359--8388},
  year      = {2024},
  publisher = {PMLR},
  url       = {https://proceedings.mlr.press/v235/chiang24b.html}
}

@misc{Chollet2025arcagi2,
  author        = {Chollet, Francois and Knoop, Mike and Kamradt, Gregory and Landers, Bryan and Pinkard, Henry},
  title         = {{ARC-AGI-2: A New Challenge for Frontier AI Reasoning Systems}. {P}reprint available at https://arxiv.org/abs/2505.11831},
  year          = {2025},
  publisher     = {arXiv},
  eprint        = {2505.11831},
  archivePrefix = {arXiv},
  primaryClass  = {cs.AI},
  doi           = {10.48550/arxiv.2505.11831},
  url           = {https://arxiv.org/abs/2505.11831}
}

@article{hle2026, 
title={A benchmark of expert-level academic questions to assess {AI} capabilities}, 
volume={649}, 
ISSN={1476-4687}, 
url={http://dx.doi.org/10.1038/s41586-025-09962-4}, 
DOI={10.1038/s41586-025-09962-4}, 
number={8099}, 
journal={Nature}, 
publisher={Springer Science and Business Media LLC}, 
author={{Center for AI Safety} and {Scale AI} and {HLE Contributors Consortium}}, 
year={2026}, 
month=jan, 
pages={1139--1146} 
}

@Article{Cao2025Automating,
author={Cao, Shuxiang
and Zhang, Zijian
and Alghadeer, Mohammed
and Fasciati, Simone D.
and Piscitelli, Michele
and Bakr, Mustafa
and Leek, Peter
and Aspuru-Guzik, Al{\'a}n},
title={Automating quantum computing laboratory experiments with an agent-based AI framework},
journal={Patterns},
year={2025},
month={Oct},
day={10},
publisher={Elsevier},
volume={6},
number={10},
issn={2666-3899},
doi={10.1016/j.patter.2025.101372},
url={https://doi.org/10.1016/j.patter.2025.101372}
}
\bibliographystyle{unsrtnat}
}


\ifdefined\CASSCFMainOnly\else
\clearpage
\appendix
\setcounter{figure}{0}
\setcounter{table}{0}
\renewcommand{\thefigure}{S\arabic{figure}}
\renewcommand{\thetable}{S\arabic{table}}
\renewcommand{\theHfigure}{SI.\arabic{figure}}
\renewcommand{\theHtable}{SI.\arabic{table}}
\startcontents[si]
\section*{Supporting Information}
\subsection*{Contents}
\printcontents[si]{}{0}{\setcounter{tocdepth}{3}}
\clearpage


\section{Benchmark provenance and scoring details}
\label{app:benchmark}

\subsection{Source records}

The curated benchmark is a row-level transcription of the 32 fixed-geometry entries in the Supporting Information of Shirazi and co-workers \citep{Shirazi2025ASF}. Each record has an entry identifier, molecule label, XYZ geometry, geometry provenance, charge, multiplicity, published ASF benchmark target, and six ASF outputs. A separate transcription records the independently selected source-paper CASPT2 value, table column, state label, and orbital character used for each Ref. value in Table~\ref{tab:asf-quntur-results-32}. The two all-\textit{E}-hexatriene records have different geometries and values and remain separate entries.

\subsection{Agent context}\label{si:Agent_context}

The CASSCF-specific portion of the expert context is reproduced below as written in {\cinzel Quntur}'s agent definition. Shared domain-expert and ORCA instructions are separate from this text.

\begin{quote}\small
Develop a chemically justified active-space and CASSCF-family recommendation for the requested states and backend. Begin with the electronic processes that matter, the coupled orbital classes they require, the active-electron count, and a suitable state-averaging strategy. Use existing orbital and state data to map those classes to MO indices when possible. For orbital characterization, start with energies, occupations, and atom/AO populations; use selected coefficients, cube values, or rendered orbitals when they resolve a specific uncertainty. Keep index conventions and numerical evidence clear. ASF and AVAS can also help resolve a specific selection question; interpret their suggestions in chemical context and keep diagnostic effort proportional to the decision it informs.

Your usual handoff is a chemistry-level proposal or an index-resolved selection, with the relevant ORCA or PySCF settings for input\_file\_service. The parent and execution agents own calculation submission and validation; ORCA calculations are routed through the designated runner. When more evidence is needed, recommend the focused preparation or multiroot screening job that would discriminate between candidates. If the available evidence remains ambiguous, give the smallest chemically justified space, a sensible expanded alternative, and the precise orbital or root behavior that the parent should check. Treat that clear handoff as completion rather than continuing exploratory calculations for certainty.

For a repeat input review, compare the changed CASSCF settings and new calculation evidence with your earlier recommendation. Reuse earlier manual findings for unchanged settings. Consult the manual when a decision-critical detail remains unresolved; a focused search can include worked examples as well as reference sections. Treat a relevant result or explicit nonconfirmation as evidence for the handoff, and distinguish documentation evidence from an actual parser check. Report any remaining uncertainty to input\_file\_service with the next validation step.

For ORCA input tasks, provide accurate CASSCF keywords or blocks; for PySCF tasks, provide the corresponding method and orbital-selection guidance. Describe any useful orbital-transfer steps with MOKIT and the ORCA-version-appropriate utilities. Report what is established from existing evidence, what is provisional, and what a later calculation should validate. Keep the response concise.
\end{quote}

\subsection{ASF comparators and numerical scoring}

The six ASF selection protocols use different entropy thresholds and singlet, triplet, or quasi-restricted starting orbitals \citep{Shirazi2025ASF}. They all target equal-weight two-singlet SA-CASSCF and SC-NEVPT2 with def2-TZVPD and density fitting. The larger-space variants use the lower entropy threshold of 0.11. The resulting CAS sizes are outcomes to compare, not unique correct active spaces for the agents.

For each converged ASF row, we recompute the absolute difference between the printed excitation energy and the independently transcribed, geometry-matched CASPT2 value in Table~\ref{tab:asf-quntur-results-32}. A miss means an error strictly greater than 1~eV. Following the ASF paper's failure convention, a failed CASSCF calculation contributes that case's independently selected reference energy to the penalized MAE. The published combined ASF procedure has only aggregate statistics, so no per-entry combined energies are imputed. Our revised references can change miss counts relative to the ASF publication; row-level classifications follow the stated numerical rule and our reference column. Cyclopropene $\ell$-ASF(T) is marked as a CASSCF failure despite the printed zero in its energy column, consistent with the paper's failure accounting.

\subsection{Per-case excitation energies and active spaces}

\begingroup
\scriptsize
\setlength{\tabcolsep}{3pt}
\renewcommand{\arraystretch}{1.14}
\begin{longtable}{@{}r l r *{7}{c}@{}}
\caption{Geometry-matched, source-paper CASPT2 reference energies and SC-NEVPT2 results for the six ASF selections from Shirazi et al. and the {\cinzel Quntur}-reported result. For each molecule, the first line gives gaps in eV and the second gives active-space sizes $(n_e,n_o)$, i.e., CAS$(n_e,n_o)$.}\label{tab:asf-quntur-results-32}\\
\toprule
ID & Molecule & Ref. & ASF(S) & $\ell$-ASF(S) & ASF(T) & $\ell$-ASF(T) & ASF(QRO) & $\ell$-ASF(QRO) &{\cinzel Quntur} \\
\midrule
\endfirsthead
\caption[]{ASF and {\cinzel Quntur} benchmark results (continued).}\\
\toprule
ID & Molecule & Ref. & ASF(S) & $\ell$-ASF(S) & ASF(T) & $\ell$-ASF(T) & ASF(QRO) & $\ell$-ASF(QRO) &{\cinzel Quntur} \\
\midrule
\endhead
\midrule\multicolumn{10}{r}{\emph{Continued on next page}}\\
\endfoot
\bottomrule
\endlastfoot
01 & acetamide & 5.80 & 5.6434 & 5.6433 & 7.0928 & 7.0928 & 7.0928 & 7.0927 & 5.8858 \\*
 &  &  & (8,7) & (8,7) & (4,4) & (4,4) & (4,4) & (4,4) & (6,4) \\
\rowcolor{black!7}
02 & adenine & 5.20 & 5.6561 & 5.5889 & 5.6562 & 5.6562 & 5.5890 & 5.4864 & 5.4723 \\*
\rowcolor{black!7}
 &  &  & (8,8) & (10,9) & (8,8) & (8,8) & (10,9) & (12,10) & (18,13) \\
03 & $p$-benzoquinone & 2.78\textsuperscript{b} & 4.3684 & 4.3683 & 4.3342 & 4.3342 & 4.3342 & 4.3612 & 2.9154 \\*
 &  &  & (16,12) & (16,12) & (8,8) & (8,8) & (8,8) & (14,11) & (12,10) \\
\rowcolor{black!7}
04 & cyclopentadiene & 5.51 & 6.7716 & 5.5406 & 6.7715 & 6.7715 & 6.7715 & 6.7289 & 5.0474 \\*
\rowcolor{black!7}
 &  &  & (4,4) & (8,8) & (4,4) & (4,4) & (4,4) & (8,8) & (4,4) \\
05 & cyclopropene & 6.76 & 6.7742 & 6.8566 & \emph{fail} & \emph{fail} & 6.7742 & 6.7741 & 7.4809\textsuperscript{a} \\*
 &  &  & (4,4) & (10,10) & (6,5) & (6,5) & (4,4) & (4,4) & (2,3) \\
\rowcolor{black!7}
06 & cytosine & 4.68 & 4.8135 & 4.7273 & 4.7273 & 4.7273 & 4.8133 & 4.7582 & 4.6614 \\*
\rowcolor{black!7}
 &  &  & (6,6) & (8,7) & (8,7) & (8,7) & (6,6) & (10,8) & (14,10) \\
07 & formamide & 5.63 & 5.5499 & 5.5499 & 7.1801 & 7.1801 & 7.3020 & 7.3020 & 5.8022 \\*
 &  &  & (8,7) & (8,7) & (6,6) & (6,6) & (4,4) & (4,4) & (6,4) \\
\rowcolor{black!7}
08 & all-$E$-hexatriene (Thiel) & 5.31 & 5.5889 & 5.5889 & 5.5889 & 5.5889 & 5.5889 & 5.5889 & 5.5934 \\*
\rowcolor{black!7}
 &  &  & (6,6) & (6,6) & (6,6) & (6,6) & (6,6) & (6,6) & (6,6) \\
09 & norbornadiene & 5.34 & \emph{fail} & 5.1612 & 5.2060 & 5.2060 & \emph{fail} & 5.1662 & 4.8519 \\*
 &  &  & (6,5) & (6,5) & (4,4) & (4,4) & (6,5) & (6,5) & (4,4) \\
\rowcolor{black!7}
10 & all-$E$-octatetraene & 4.64 & 4.7352 & 4.7352 & 4.7352 & 4.7352 & 4.7352 & 4.7352 & 4.7396 \\*
\rowcolor{black!7}
 &  &  & (8,8) & (8,8) & (8,8) & (8,8) & (8,8) & (8,8) & (8,8) \\
11 & propanamide & 5.72 & \emph{fail} & \emph{fail} & 7.0755 & 7.0755 & 7.0754 & 7.0754 & 5.9216 \\*
 &  &  & (8,5) & (8,5) & (4,4) & (4,4) & (4,4) & (4,4) & (6,4) \\
\rowcolor{black!7}
12 & pyrrole & 6.31 & 6.5931 & 6.5931 & 6.5931 & 6.5931 & 6.5931 & 6.5931 & 5.7345 \\*
\rowcolor{black!7}
 &  &  & (6,5) & (6,5) & (6,5) & (6,5) & (6,5) & (6,5) & (6,6) \\
13 & thymine & 4.94 & 5.2169 & 5.1554 & 5.2169 & 5.2169 & 5.2169 & 5.2169 & 5.1179 \\*
 &  &  & (8,7) & (10,8) & (8,7) & (8,7) & (8,7) & (8,7) & (12,9) \\
\rowcolor{black!7}
14 & $s$-triazine & 4.60 & 6.0398 & 5.9079 & 4.9133 & 4.9133 & 4.9132 & 4.9133 & 4.1899 \\*
\rowcolor{black!7}
 &  &  & (6,6) & (10,10) & (10,8) & (10,8) & (10,8) & (10,8) & (12,9) \\
15 & uracil & 4.90 & 5.2928 & 5.3651 & \emph{fail} & \emph{fail} & 5.3774 & 5.3774 & 5.0633 \\*
 &  &  & (10,8) & (12,10) & (18,12) & (18,12) & (8,7) & (8,7) & (14,10) \\
\rowcolor{black!7}
16 & acetaldehyde & 4.27 & 4.3190 & 4.3061 & 4.4707 & 4.4123 & 4.4707 & 4.4267 & 4.4883 \\*
\rowcolor{black!7}
 &  &  & (8,6) & (8,7) & (4,3) & (8,8) & (4,3) & (10,9) & (4,3) \\
17 & acetone & 4.44 & 4.4793 & 4.4793 & 4.6165 & 4.5381 & 4.6164 & 4.6165 & 4.5303 \\*
 &  &  & (8,7) & (8,7) & (4,3) & (10,9) & (4,3) & (4,3) & (6,5) \\
\rowcolor{black!7}
18 & benzene & 4.83 & 5.2242 & 5.2248 & 5.2249 & 5.2242 & 5.2246 & 5.2248 & 5.2297 \\*
\rowcolor{black!7}
 &  &  & (6,6) & (6,6) & (6,6) & (6,6) & (6,6) & (6,6) & (6,6) \\
19 & $E$-butadiene & 6.51 & 6.8095 & 6.8095 & 6.8096 & 6.8314 & 6.8096 & 6.8096 & 6.8143\textsuperscript{a} \\*
 &  &  & (4,4) & (4,4) & (4,4) & (6,6) & (4,4) & (4,4) & (4,4) \\
\rowcolor{black!7}
20 & DMABN & 4.87 & 3.8124 & 3.8124 & 3.8125 & 3.6593 & 3.6600 & 3.6597 & 4.7304 \\*
\rowcolor{black!7}
 &  &  & (12,12) & (12,12) & (12,12) & (14,13) & (14,13) & (14,13) & (12,11) \\
21 & ethene (ethylene) & 8.16 & 8.0146 & 8.0349 & 8.0041 & 8.0877 & 8.0045 & 8.0045 & 8.5748 \\*
 &  &  & (4,3) & (6,5) & (4,3) & (8,8) & (4,3) & (4,3) & (2,2) \\
\rowcolor{black!7}
22 & formaldehyde & 3.92 & 3.9549 & 3.9549 & 4.1533 & 3.9549 & 4.1533 & 3.9766 & 4.1424 \\*
\rowcolor{black!7}
 &  &  & (8,7) & (8,7) & (4,3) & (8,7) & (4,3) & (6,5) & (4,3) \\
23 & furan & 6.32 & 6.7812 & 6.7812 & 6.8314 & 6.8314 & 6.7685 & 6.7685 & 6.3444\textsuperscript{a} \\*
 &  &  & (8,7) & (8,7) & (4,4) & (4,4) & (6,5) & (6,5) & (6,6) \\
\rowcolor{black!7}
24 & all-$E$-hexatriene (Hoyer) & 5.32 & 5.5938 & 5.5937 & 5.5937 & 5.5937 & 5.5937 & 5.5937 & 5.5982 \\*
\rowcolor{black!7}
 &  &  & (6,6) & (6,6) & (6,6) & (6,6) & (6,6) & (6,6) & (6,6) \\
25 & naphthalene & 4.21 & 4.4272 & 4.4269 & 4.4269 & 4.4269 & 4.4269 & 4.4269 & 4.2977\textsuperscript{a} \\*
 &  &  & (10,10) & (10,10) & (10,10) & (10,10) & (10,10) & (10,10) & (10,10) \\
\rowcolor{black!7}
26 & $p$-nitroaniline (pNA) & 4.62 & 5.0574 & 5.0715 & 5.0574 & 5.0714 & 5.0574 & 5.0691 & 4.9434 \\*
\rowcolor{black!7}
 &  &  & (8,8) & (12,11) & (8,8) & (12,11) & (8,8) & (14,12) & (12,10) \\
27 & pyrazine & 4.04 & 5.2826 & 5.2213 & 4.2142 & 4.2142 & 4.2142 & 4.2142 & 4.2180 \\*
 &  &  & (6,6) & (8,8) & (10,8) & (10,8) & (10,8) & (10,8) & (10,8) \\
\rowcolor{black!7}
28 & pyridazine & 3.67 & 5.8547 & \emph{fail} & 3.8312 & 4.3289 & 3.8312 & 3.8312 & 3.8363 \\*
\rowcolor{black!7}
 &  &  & (6,5) & (14,11) & (10,8) & (10,9) & (10,8) & (10,8) & (10,8) \\
29 & pyridine & 5.06 & 5.3952 & 5.3952 & 5.3953 & 5.3953 & 5.3952 & 5.3952 & 5.1856 \\*
 &  &  & (6,6) & (6,6) & (6,6) & (6,6) & (6,6) & (6,6) & (8,7) \\
\rowcolor{black!7}
30 & pyrimidine & 4.38 & 5.6516 & 5.5854 & 5.0891 & 5.0890 & 5.1125 & 5.0694 & 4.5713 \\*
\rowcolor{black!7}
 &  &  & (12,9) & (14,11) & (10,9) & (10,9) & (8,7) & (6,6) & (10,8) \\
31 & $s$-tetrazine & 2.56 & 2.6786 & \emph{fail} & 2.3365 & 3.2839 & 2.4224 & 2.6599 & 2.4153 \\*
 &  &  & (10,8) & (18,15) & (10,8) & (16,15) & (12,9) & (16,14) & (14,10) \\
\rowcolor{black!7}
32 & water & 7.61 & 9.4707 & 9.4707 & 7.6025 & 7.6025 & 7.6025 & 7.6025 & 7.6797 \\*
\rowcolor{black!7}
 &  &  & (6,6) & (6,6) & (4,5) & (4,5) & (4,5) & (4,5) & (8,10) \\
\end{longtable}
\noindent\parbox{\linewidth}{\scriptsize\raggedright\emph{Notes.} Ref. is the lowest singlet excitation listed for each case in the geometry-matched calculation column of its original source: current-work MS-CASPT2/TZVP for rows 01--15 \citep{Schreiber2008}, and CASPT2 for rows 16--32 \citep{Hoyer2016}. These are calculated reference values, not the source papers' heterogeneous ``Best est.'' or ``reference'' columns. The source tables list selected transitions, so Ref. does not independently establish the global $S_1$ in every system. \textsuperscript{b}For $p$-benzoquinone, Schreiber et al. also report the nearby $1\,{}^1A_u$ $n\pi^*$ state at 2.80~eV; the tabulated 2.78~eV is $1\,{}^1B_{1g}$ $n\pi^*$. \emph{Fail} denotes a failed ASF-selected CASSCF calculation; the source table's printed zero for cyclopropene $\ell$-ASF(T) is counted as a failure. {\cinzel Quntur} gaps were checked against the trajectory-audit summaries. \textsuperscript{a}The audited {\cinzel Quntur} root differs in character from the source-paper transition; the marker makes no claim about the ASF root, for which per-row character diagnostics are unavailable. Unmarked gaps are not independently confirmed as the intended $S_1$. The two all-$E$-hexatriene rows use distinct geometries.}
\endgroup
\par

\subsection{Trajectory analyses}
\label{app:trajectory-analyses}

The complete case materials are deposited in the associated data repository \citep{Perez2026QunturCASSCFFiles}. The deposit includes the benchmark prompts, source-paper reference transcription, exported agent traces, the files from each calculation session, and the trajectory-analysis reports with their supporting files. Case identifiers 001--032 connect the deposit to the results tables and decision matrix. For cases 026 and 030, the Original and rep1--rep3 labels in Table~\ref{tab:run-variability} identify the four separate trajectories for each molecule.

Each scientific report identifies the final-answer session, distinct attempts, submitted inputs, decisive outputs, orbital and state evidence, and unresolved questions. Where a full action trace is absent, inputs and outputs establish what ran but cannot prove the agent's earlier reasoning or human direction. The decision matrix therefore describes an action as prospective only when an earlier trace, plan, or input supports its timing. Case-level telemetry covers executed model and tool calls and their input and output tokens in the selected final-answer session; matched detached web-search tokens are listed separately. Runner-monitor traces and abandoned sessions are excluded from the combined usage figure. Cached-read and reasoning tokens are subtypes of input and output tokens, respectively, not additional tokens. Token counts measure model activity rather than ORCA compute cost.

Numerical comparisons use the independently transcribed Ref. values in Table~\ref{tab:asf-quntur-results-32}. The displayed active-space montages were prepared from saved orbitals after the benchmark runs.

\subsubsection{001: Acetamide}

\begin{figure}[H]
\centering
\captionsetup{font=small}
\includegraphics[width=0.75\linewidth,height=0.3\textheight,keepaspectratio]{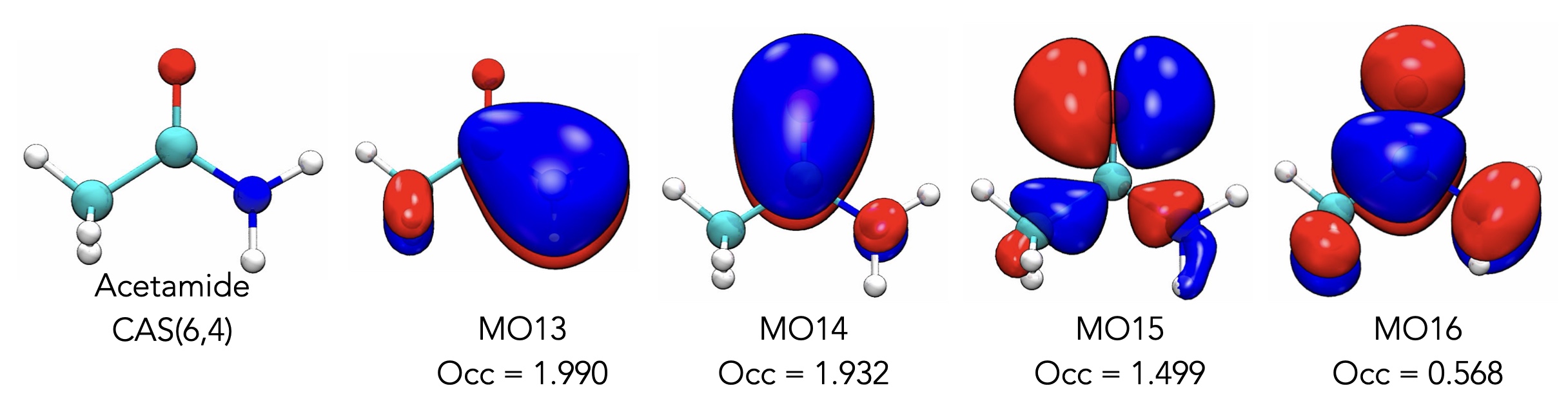}
\caption{The agent compared compact carbonyl and amide spaces before carrying CAS(6,4) into the final basis. The smaller CAS(4,3) pilot had an adverse active--inactive orbital ordering, whereas CAS(6,4) included the nearly occupied N $p$ partner as well as the O lone pair, bonding $\pi$, and valence $\pi^*$. An explicit orbital swap kept the compact acceptor in the active space despite a lower-lying diffuse virtual orbital. Configurations, root-specific occupations, and a fixed-orbital four-root survey supported an $n_{\mathrm O}\pi^*$ assignment for the computed first excited root. The final SC-NEVPT2 calculation reused a previously converged two-root CASSCF orbital set.}
\label{fig:si-active-001}
\end{figure}

\subsubsection{002: Adenine}

\begin{figure}[H]
\centering
\captionsetup{font=small}
\includegraphics[width=1\linewidth,height=0.4\textheight,keepaspectratio]{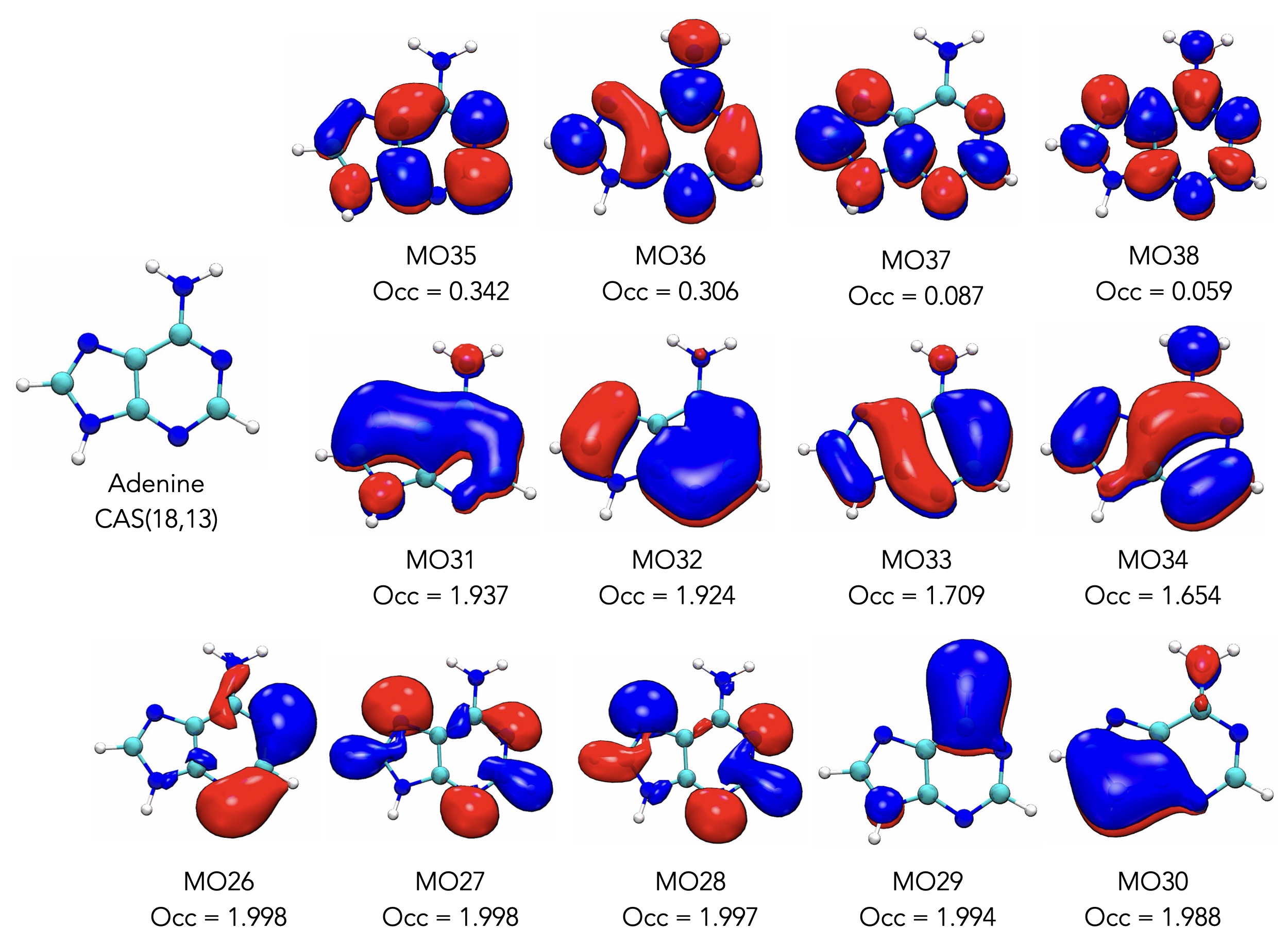}
\caption{Preparatory natural-orbital occupations and TDA transition characters informed a large CAS(18,13) containing the conjugated $\pi/\pi^*$ system and N lone-pair directions. The agent then remapped its selected orbitals into def2-TZVPD and checked the final configurations and state-specific energies. The equal-weight two-root CASSCF and SC-NEVPT2 calculation completed, and the reported gap follows from the correlated totals. The substantial space is chemically defensible, but the record does not establish why this size was preferred over a smaller tested alternative. In particular, a preparatory TDA calculation indicated a lower $n\pi^*$ candidate; no wider-root correlated screen demonstrated that the final two-root state was the globally lowest singlet. Thus the run documents orbital-aware preparation and a reproducible prediction, with a remaining state-ordering question.}
\label{fig:si-active-002}
\end{figure}

\subsubsection{003: $p$-Benzoquinone}

\begin{figure}[H]
\centering
\captionsetup{font=small}
\includegraphics[width=0.75\linewidth,height=0.3\textheight,keepaspectratio]{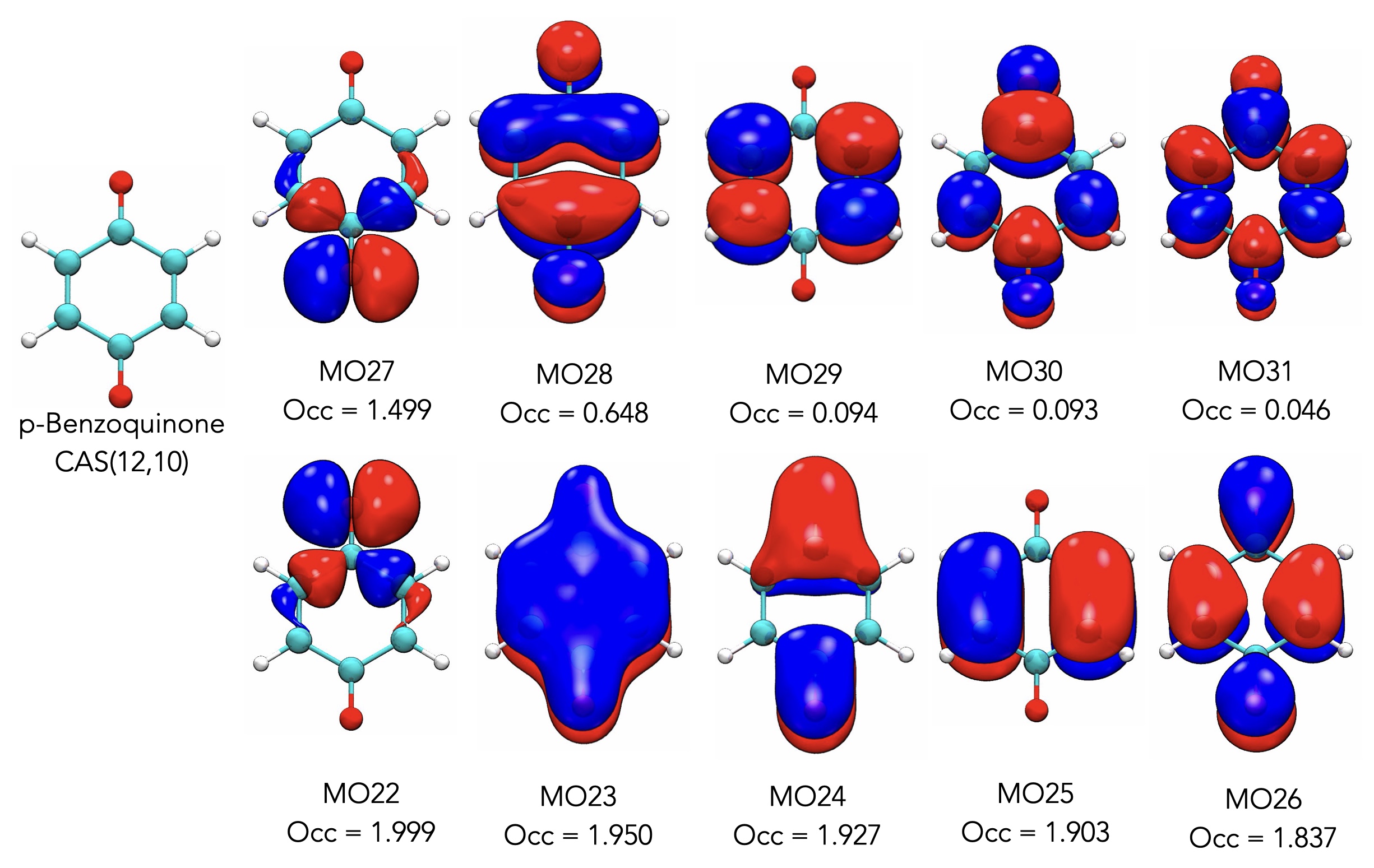}
\caption{The \textit{initial} active space was largely literature-led rather than selected through a controlled comparison among alternative orbital inventories. The selected CAS(12,10) was assembled with explicit orbital reordering and checked through small-basis multiroot screens and target-basis validation. The agent rejected an inconsistent post-CASSCF output and accepted a retry whose two state-specific SC-NEVPT2 totals were internally consistent. Those actions protect the provenance of the reported gap. A near-degenerate pair observed under broader state averaging raises a question about how robust the lowest-state assignment remains under the prescribed two-root optimization.}
\label{fig:si-active-003}
\end{figure}

\subsubsection{004: Cyclopentadiene}

\begin{figure}[H]
\centering
\captionsetup{font=small}
\includegraphics[width=0.75\linewidth,height=0.3\textheight,keepaspectratio]{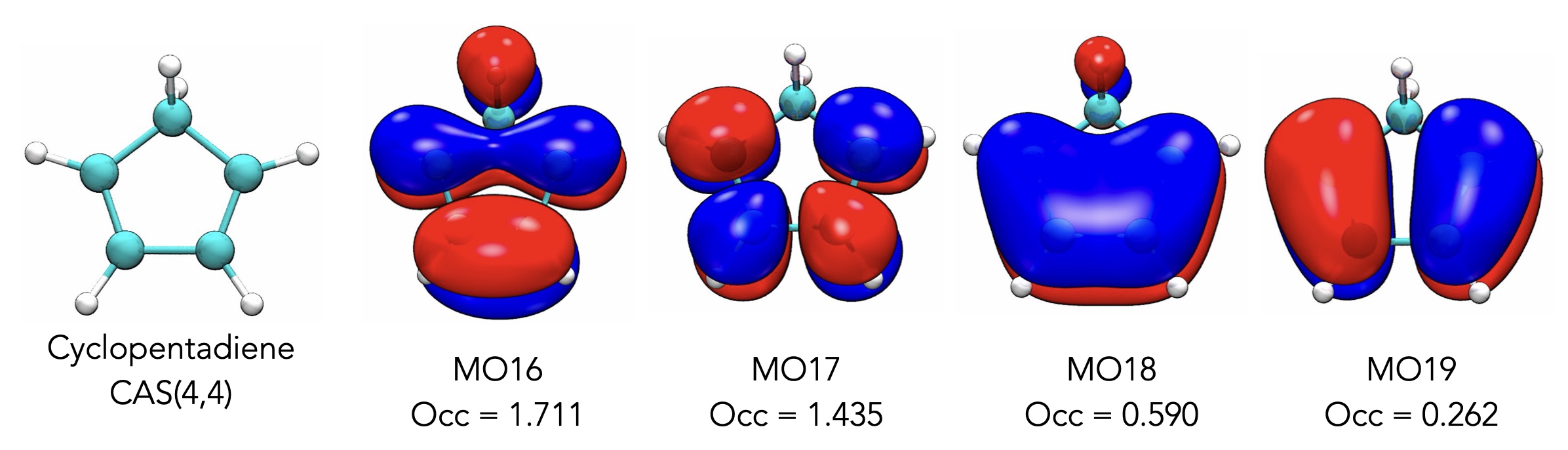}
\caption{The first completed calculation gave a higher-root assignment rather than a secure $S_1$. Subsequent symmetry and root-character checks led the agent to replace it with an $A_1/B_2$ state pair in a four-orbital $\pi$ space. The recovery calculation supplied a lower, reproducible SC-NEVPT2 gap. This revision illustrates why a normally completed excited-state job need not answer the requested state question. On the recovered orbitals, a CASCI check placed the $B_2$ candidate lowest among the inspected excited states, but the competing second $A_1$ root was not given a matched NEVPT2 correction. The final assignment is consequently supported at the level examined, rather than proven to be the lowest correlated singlet across all relevant roots.}
\label{fig:si-active-004}
\end{figure}

\subsubsection{005: Cyclopropene}

\begin{figure}[H]
\centering
\captionsetup{font=small}
\includegraphics[width=0.75\linewidth,height=0.3\textheight,keepaspectratio]{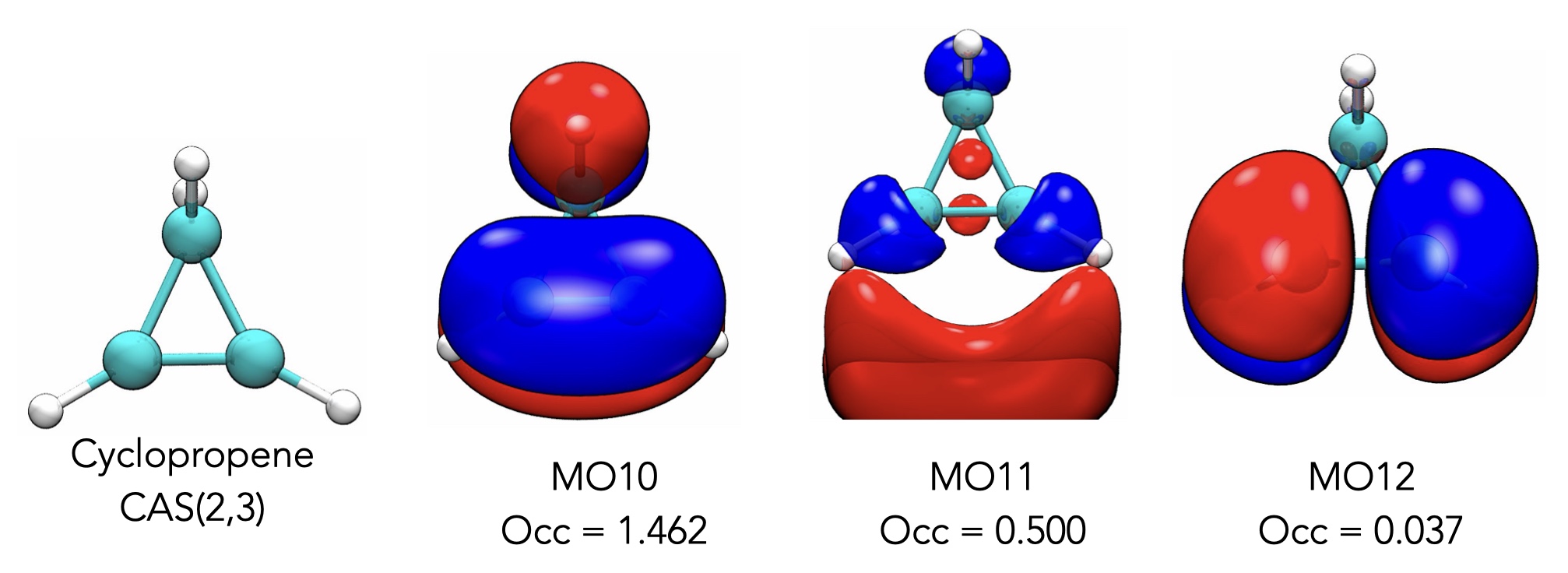}
\caption{The agent identified a diffuse acceptor and built a small Rydberg-capable CAS(2,3). After input, convergence, and restart problems, it obtained a consistent two-root SC-NEVPT2 calculation. Orbital coefficients and symmetry support a $\pi\!\to\!3s$ interpretation of the reported state. The benchmark reference, however, concerns a valence transition requiring an occupied in-plane $\sigma$ orbital that the selected space omits. A numerical difference from that reference therefore does not validate the state assignment. The repair sequence improved the technical credibility of the Rydberg result but did not address the principal chemical omission; a balanced space containing the relevant $\sigma$ and diffuse directions was not tested.}
\label{fig:si-active-005}
\end{figure}

\subsubsection{006: Cytosine}

\begin{figure}[H]
\centering
\captionsetup{font=small}
\includegraphics[width=0.75\linewidth,height=0.3\textheight,keepaspectratio]{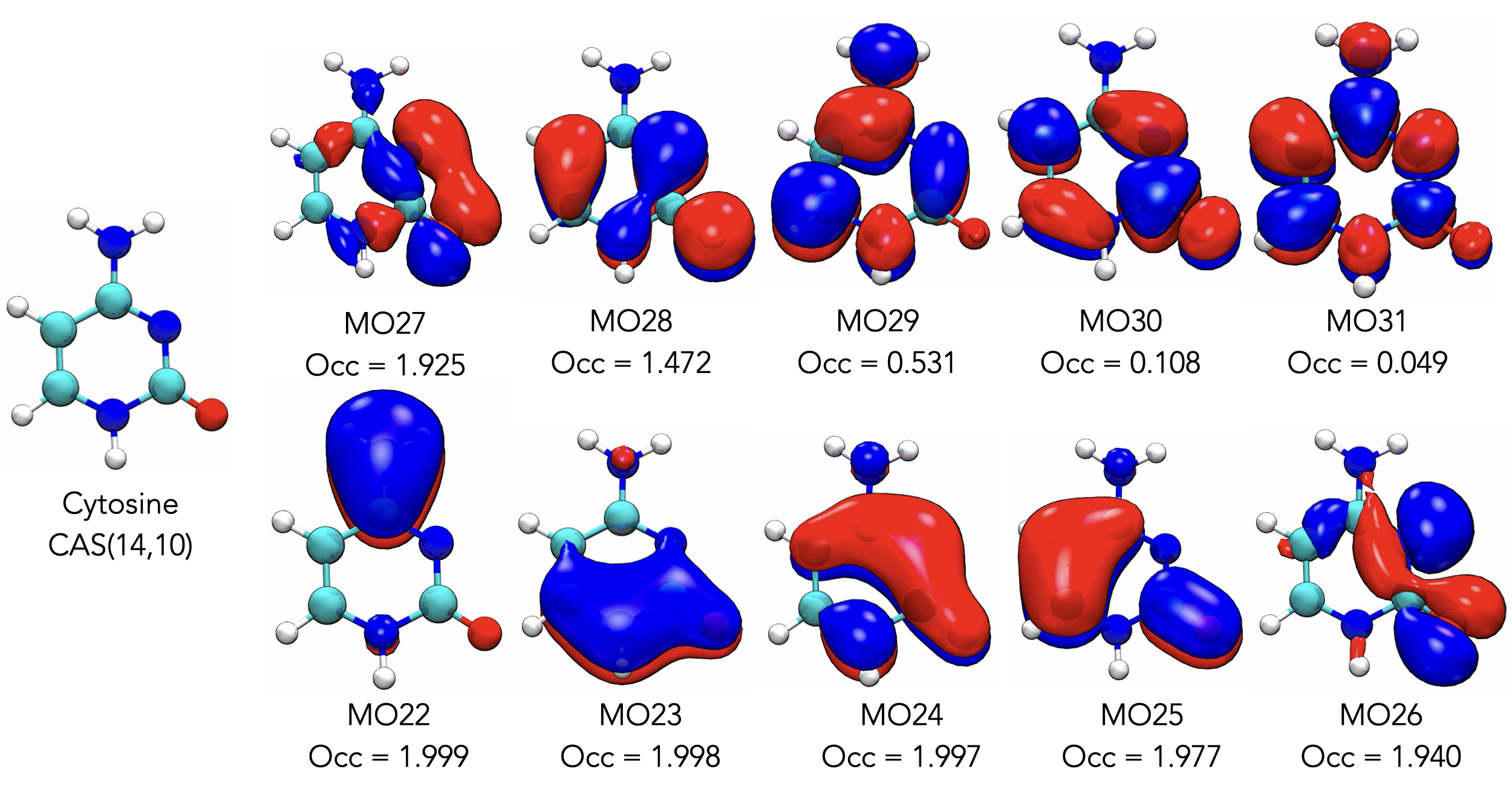}
\caption{Independent orbital starts and four-root screens led to CAS(14,10) with the intended O/N lone-pair directions. The first two-root CASSCF optimization converged and completed SC-NEVPT2, yet inspection showed that those lone-pair orbitals had left the active subspace. The agent declined to use that apparently valid gap, attempted to recover an intended-space two-root solution, and stopped the follow-on correction when the recovery failed its stationary-gradient check. Its reported gap instead came from a four-root common-orbital CASCI/SC-NEVPT2 calculation on orbitals that retained the lone pairs. This was a useful and explicitly qualified response to orbital drift: it yielded a defensible result for the fallback model, but not the requested stationary two-root calculation. The additional jobs were the cost of testing that distinction; the exact two-root result for the intended space remains unknown.}
\label{fig:si-active-006}
\end{figure}

\subsubsection{007: Formamide}

\begin{figure}[H]
\centering
\captionsetup{font=small}
\includegraphics[width=0.75\linewidth,height=0.3\textheight,keepaspectratio]{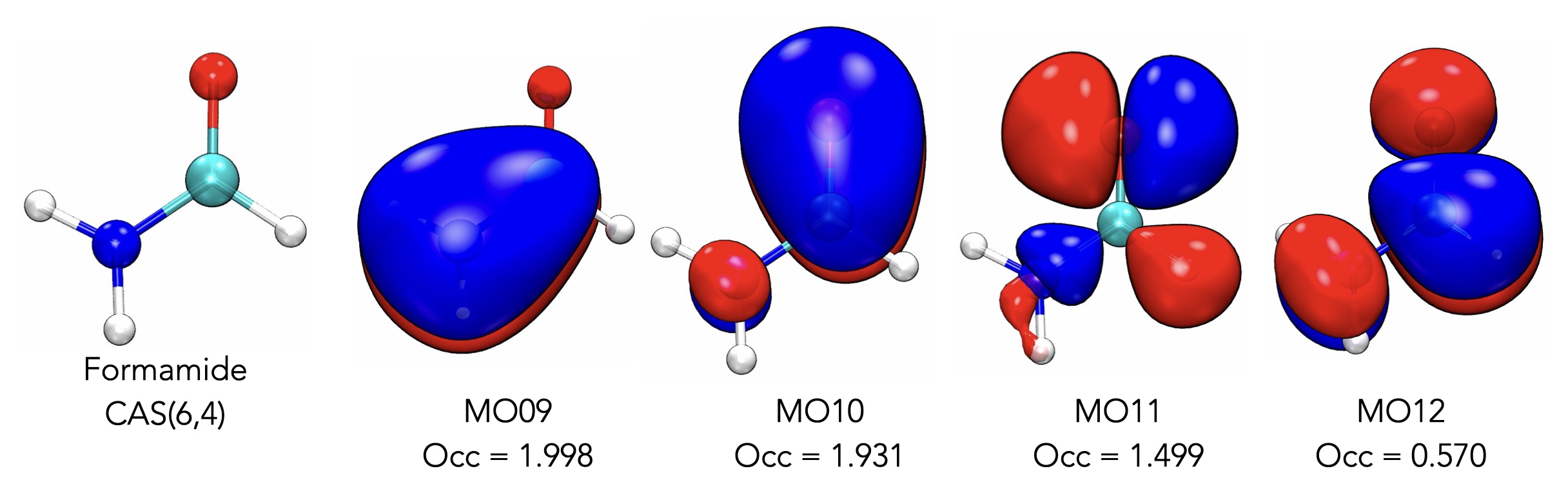}
\caption{The agent chose a valence CAS(6,4) containing the O lone pair and the three-orbital amide $\pi$ system before production. MP2 natural orbitals exposed a wrong provisional swap, which was corrected before the target-basis calculation. A truncation control and state diagnostics supported the selected $n_{\mathrm O}\pi^*$ root. The first SC-NEVPT2 output terminated normally but computed the excited root twice and omitted the ground-root correction; the agent refused that output and obtained a complete correlated pair in recovery jobs. This was a sound evidence gate even though its explanation of the root-selection masks was inaccurate. The final gap is reproducible for the chosen two-root model. Other valence or Rydberg singlets were not screened, so the global ordering rests partly on chemical expectation rather than a direct calculation in this run.}
\label{fig:si-active-007}
\end{figure}

\subsubsection{008: All-\textit{E}-hexatriene, Thiel geometry}

\begin{figure}[H]
\centering
\captionsetup{font=small}
\includegraphics[width=0.75\linewidth,height=0.3\textheight,keepaspectratio]{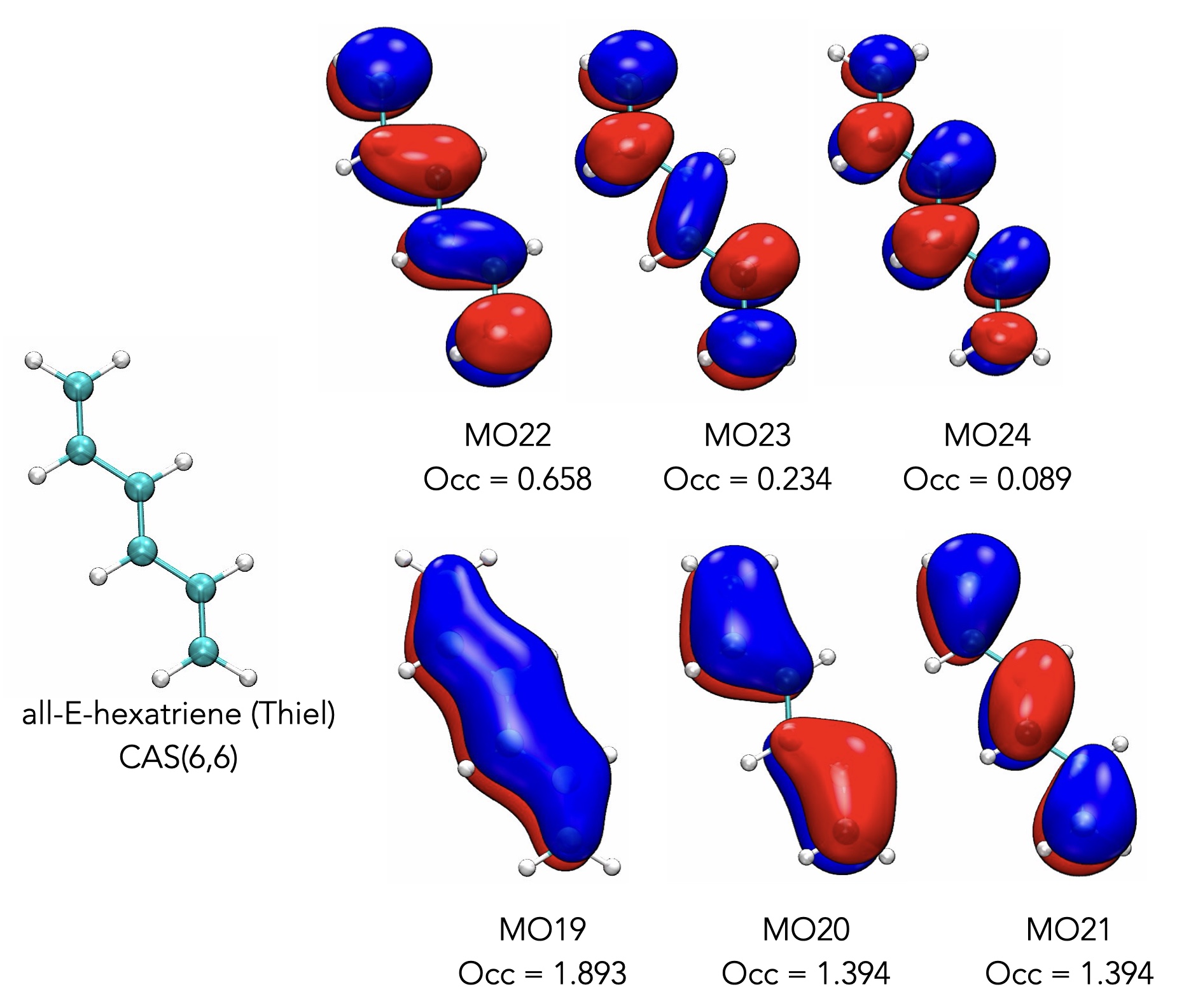}
\caption{Orbital scouting and explicit swaps assembled the full six-orbital valence $\pi$ manifold, avoiding interleaved non-$\pi$ orbitals. The agent also examined a nearby competing root and repaired numerical inconsistencies before accepting a converged two-root CASSCF/SC-NEVPT2 result. Occupations and the selected root's mixed configurations point to a dark, double-excitation-rich excited state. These observations support the internal identity of the computed root more strongly than the final energy alone. They do not exclude an unrepresented diffuse or $\sigma$-derived singlet, and a $\pi$-only multiroot screen cannot settle the ordering of states outside that space. This entry is distinct from the second hexatriene geometry in case 024.}
\label{fig:si-active-008}
\end{figure}

\subsubsection{009: Norbornadiene}

\begin{figure}[H]
\centering
\captionsetup{font=small}
\includegraphics[width=0.75\linewidth,height=0.3\textheight,keepaspectratio]{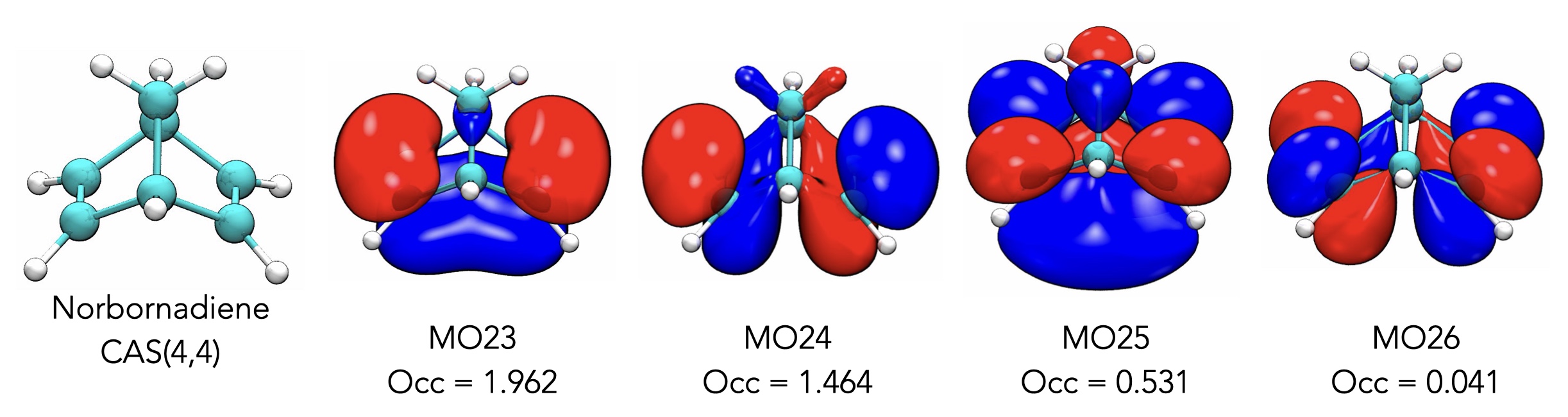}
\caption{Diffuse orbitals initially interrupted the valence active window. The agent used orbital symmetry and explicit swaps to restore the two bonding and two antibonding $\pi$ directions in CAS(4,4), corrected a root-selection issue, and obtained a converged two-root SC-NEVPT2 result. The first excited root is dipole forbidden and its configurations are consistent with the valence $^1A_2$ transition. Correlated valence--Rydberg ordering was not tested in one diffuse-inclusive calculation.}
\label{fig:si-active-009}
\end{figure}

\subsubsection{010: All-\textit{E}-octatetraene}

\begin{figure}[H]
\centering
\captionsetup{font=small}
\includegraphics[width=1\linewidth,height=0.4\textheight,keepaspectratio]{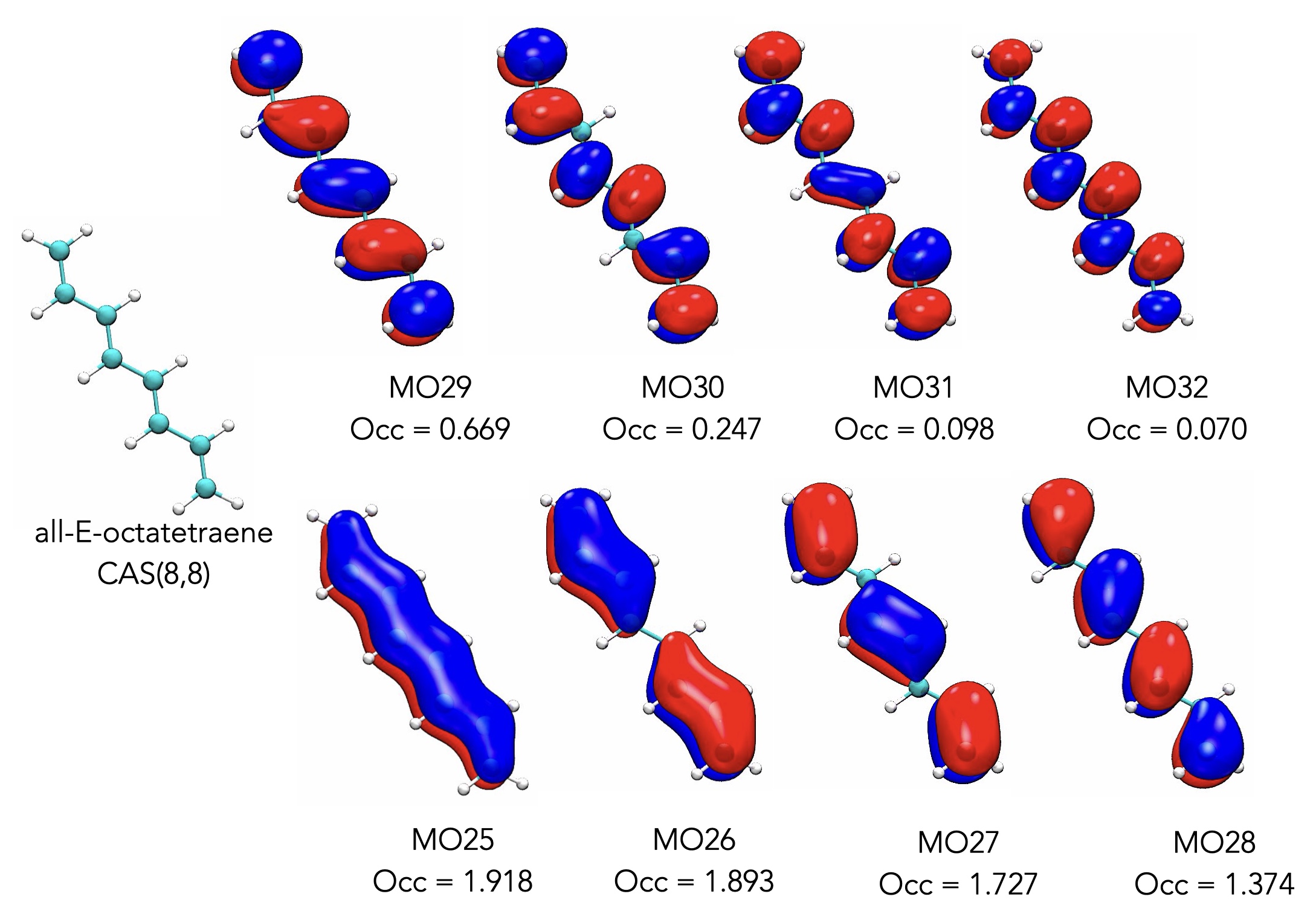}
\caption{A contiguous canonical frontier window would have included several $\sigma$ virtual orbitals. The agent instead selected four occupied carbon $\pi$ orbitals and four separated $\pi^*$ orbitals, then explicitly reordered them into a CAS(8,8). The final state-average occupations and a dominant double-excitation contribution support the intended conjugated manifold and a dark first excited root. The prescribed two-root CASSCF/SC-NEVPT2 calculation converged and produced a reproducible gap. However, no alternative active space or wider-root correlated treatment was run.}
\label{fig:si-active-010}
\end{figure}

\subsubsection{011: Propanamide}

\begin{figure}[H]
\centering
\captionsetup{font=small}
\includegraphics[width=1\linewidth,height=0.4\textheight,keepaspectratio]{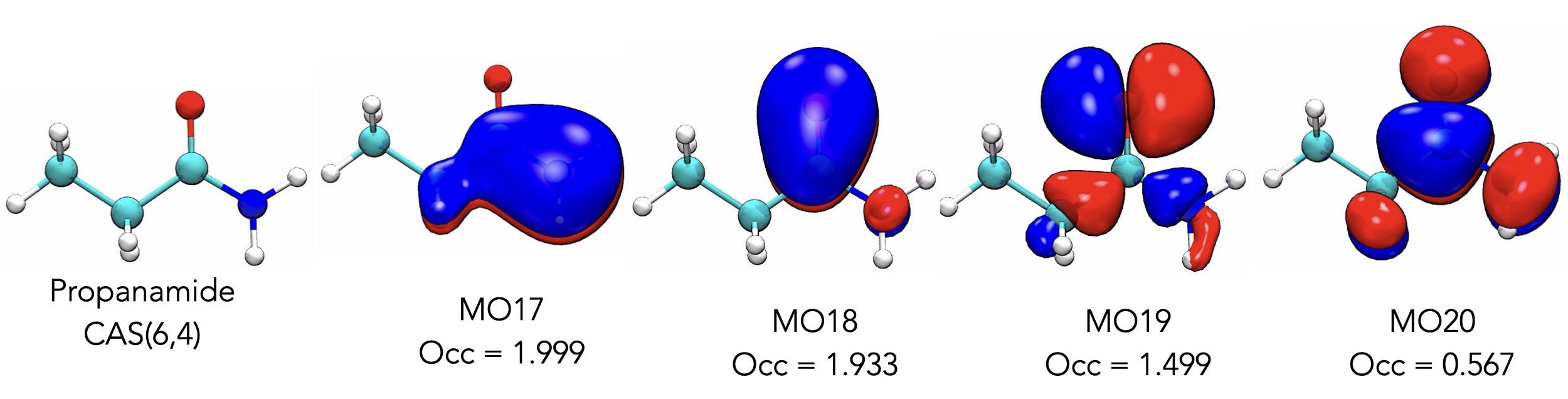}
\caption{TDA transition characters and natural transition orbitals led the agent to include the O lone pair in CAS(6,4) and to reject a low-lying diffuse virtual as the valence $\pi^*$ acceptor. Four-state then two-state CASSCF checks established the desired dark $n_{\mathrm O}\pi^*$ character before the final correlated calculation, and the orbital transfer into production was checked for continuity. No submitted job failed. This is a well-evidenced example of selecting an orbital by character instead of orbital energy alone.}
\label{fig:si-active-011}
\end{figure}

\subsubsection{012: Pyrrole}

\begin{figure}[H]
\centering
\captionsetup{font=small}
\includegraphics[width=0.75\linewidth,height=0.3\textheight,keepaspectratio]{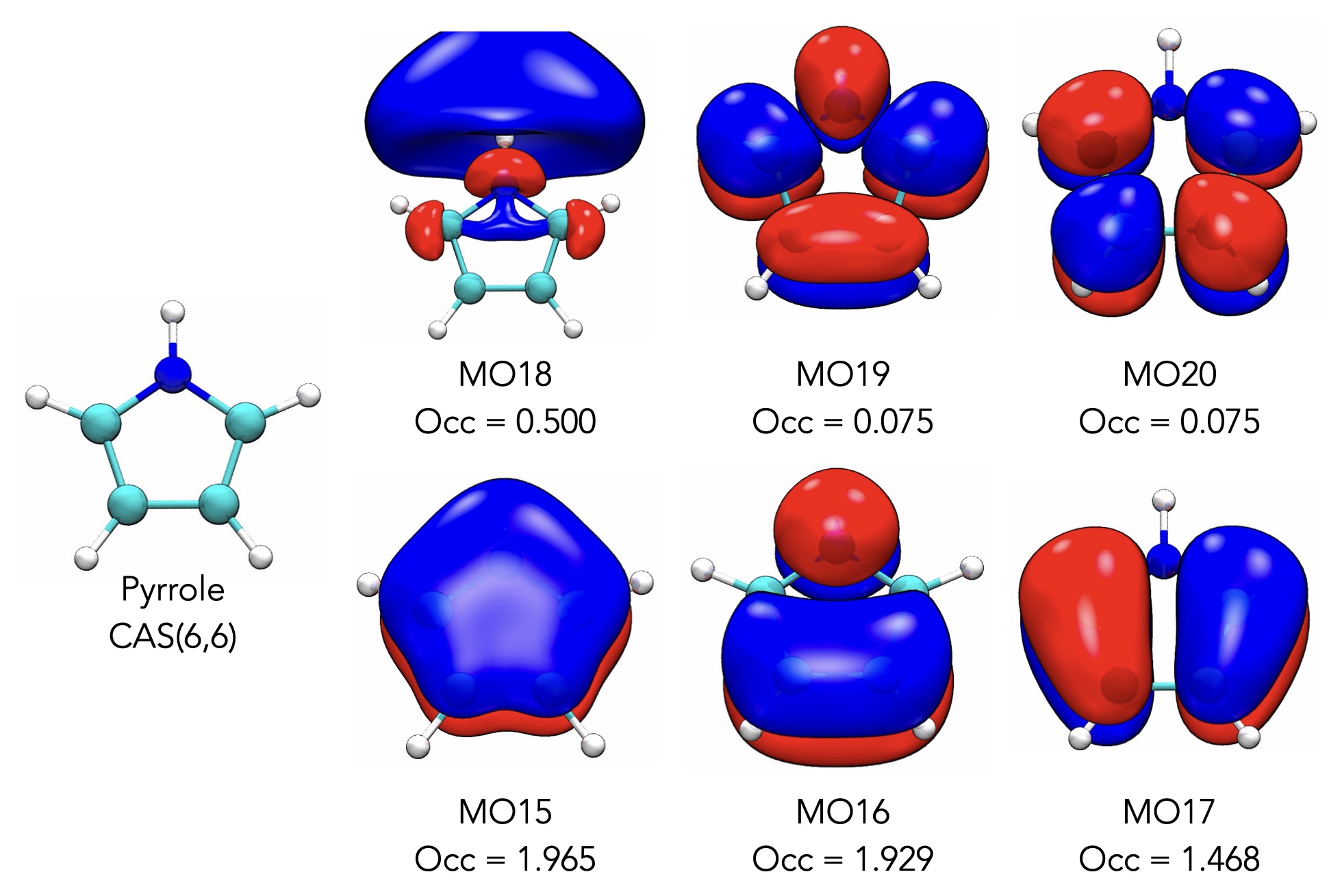}
\caption{The agent compared valence and diffuse-inclusive candidates and selected CAS(6,6), retaining a diffuse orbital after scouting. Deliberate initial rotations are documented, and the final two-root SC-NEVPT2 energy is reproducible from a converged CASSCF orbital set. The final report nevertheless described the excited root as valence $\pi\pi^*$, whereas optimized orbital character and configuration evidence indicate substantial Rydberg character. This is a case in which testing an extra orbital was scientifically useful but interpreting the resulting state was less reliable. The reported gap describes the selected two-root model; its relation to the lowest physical singlet and to the valence state behind the numerical comparator was not resolved by matched correlated calculations.}
\label{fig:si-active-012}
\end{figure}

\subsubsection{013: Thymine}

\begin{figure}[H]
\centering
\captionsetup{font=small}
\includegraphics[width=0.75\linewidth,height=0.3\textheight,keepaspectratio]{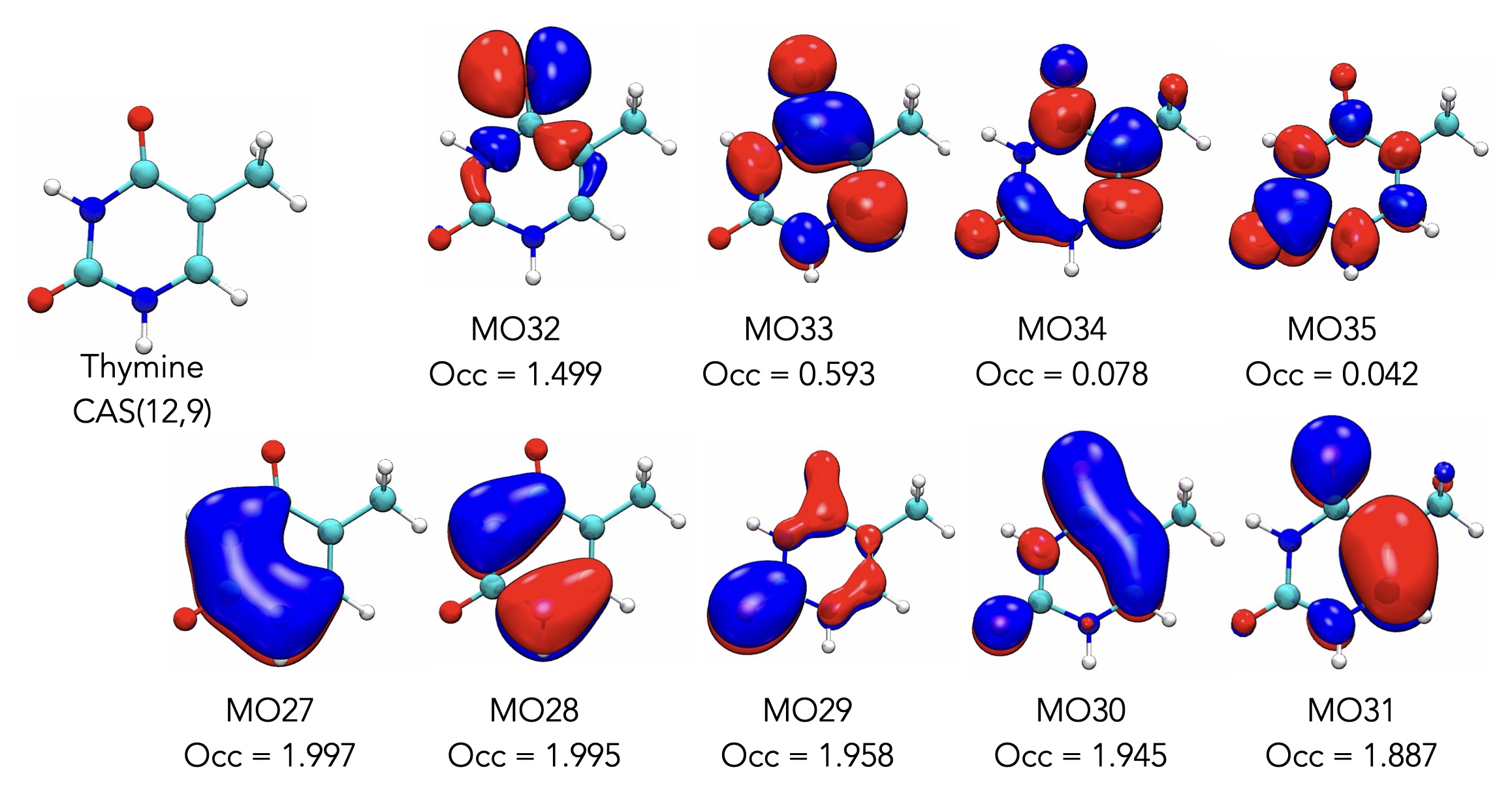}
\caption{The agent wrote active-space acceptance criteria before running calculations, targeted an O in-plane lone pair through AVAS, and repaired an AVAS setup error. A four-root CASCI diagnostic then checked orbital identity and state character before production. The final converged CAS(12,9) two-root calculation retained an $n_{\mathrm O}\!\to\!\pi^*$ first excited root, consistent with the reference state's character. The apparent reordering of active natural orbitals arose during optimization, not from an undocumented manual swap. This trajectory offers unusually clear evidence that a prospective orbital hypothesis was tested. It does not include a correlated comparison with the nearby bright $\pi\pi^*$ state, and the other O lone-pair direction was not independently tested; thus the result supports its selected transition more firmly than a comprehensive lowest-state ordering claim.}
\label{fig:si-active-013}
\end{figure}

\subsubsection{014: $s$-Triazine}

\begin{figure}[H]
\centering
\captionsetup{font=small}
\includegraphics[width=0.75\linewidth,height=0.3\textheight,keepaspectratio]{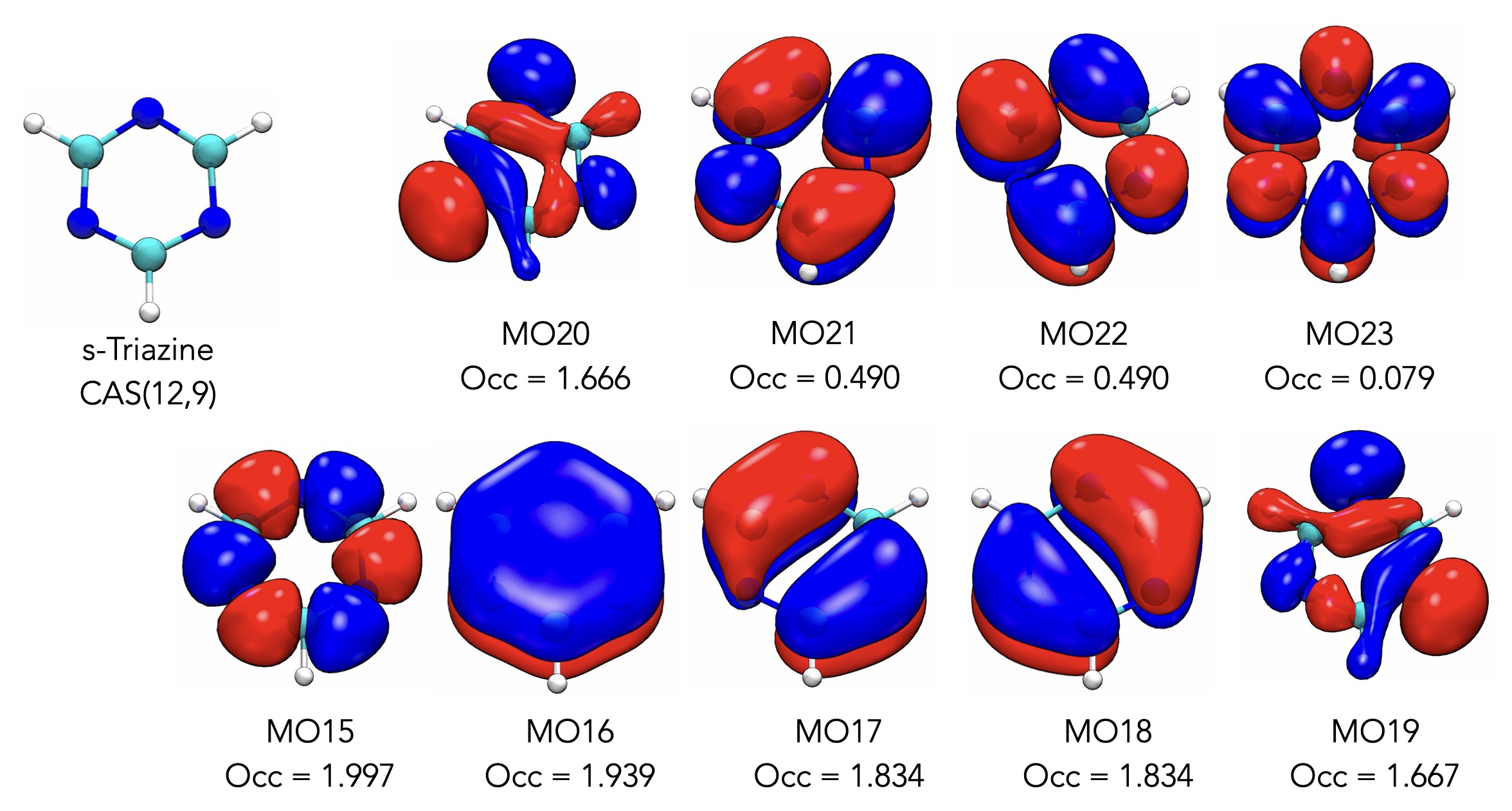}
\caption{The agent compared $\pi$-only and progressively lone-pair-enriched spaces, ultimately retaining three N lone-pair and six $\pi/\pi^*$ directions in CAS(12,9). A diffuse competitor was rejected by orbital-character evidence, while broader root checks exposed low valence character missing from a narrower average. After a calculation failure, the agent obtained a converged six-root CASSCF reference and SC-NEVPT2 corrections for all six states. Its reported low gap is a real result for that manifold, but the equal-weight two-singlet orbital optimization required by the prompt was never completed. A two-root average could relax into another orbital basin or change the ordering, so the six-root number cannot be treated as method-equivalent to the benchmark protocol. The trajectory shows useful adaptation to state complexity together with a clear limit on the reported comparison.}
\label{fig:si-active-014}
\end{figure}

\subsubsection{015: Uracil}
\label{app:case-015}

\begin{figure}[H]
\centering
\captionsetup{font=small}
\includegraphics[width=0.75\linewidth,height=0.3\textheight,keepaspectratio]{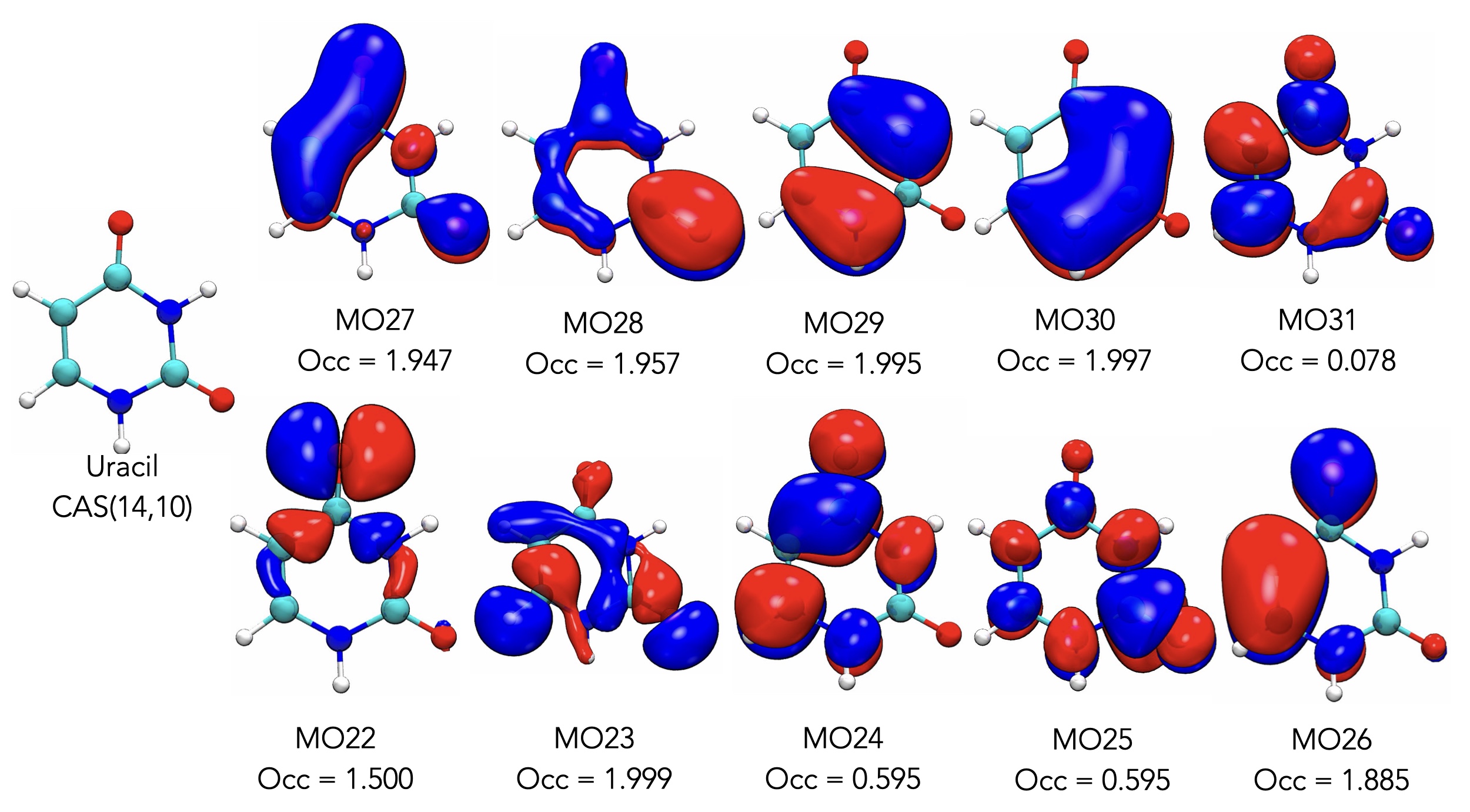}
\caption{The first planar calculation converged to a $\pi\pi^*$ excited root. The agent rejected that plausible numerical answer after recognizing that it had not reached the intended $n\pi^*$ state, then used symmetry-blocked averaging to obtain an O-lone-pair-to-$\pi^*$ candidate. State energies and configurations support the change of character. The reasoning was imperfect: early orbital audits mislabeled $\sigma$ orbitals as lone pairs, and the agent initially asserted that no nearby $n\pi^*$ root existed. The final assignment is better supported than the original one, but its comparison with the $\pi\pi^*$ calculation uses separately optimized orbital sets; diffuse or other low states were not fully surveyed. This case illustrates a consequential correction without making every step in the explanation reliable.}
\label{fig:si-active-015}
\end{figure}

\subsubsection{016: Acetaldehyde}

\begin{figure}[H]
\centering
\captionsetup{font=small}
\includegraphics[width=0.75\linewidth,height=0.3\textheight,keepaspectratio]{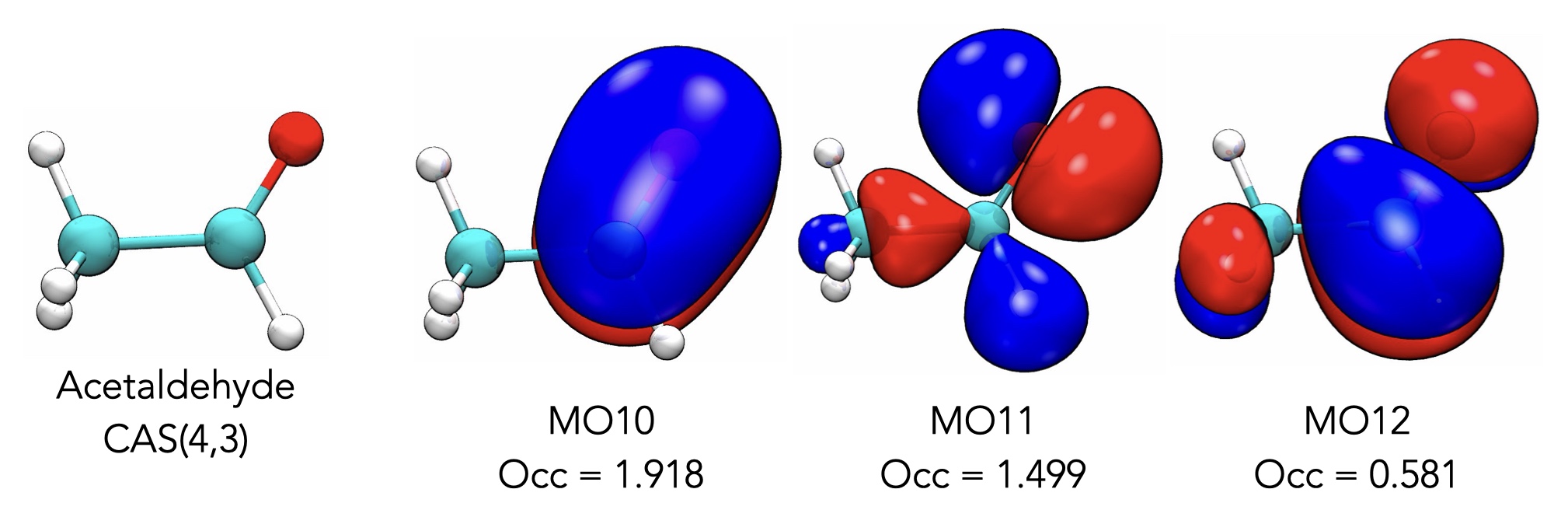}
\caption{The agent proposed a compact carbonyl CAS(4,3) containing $\pi$, O lone pair, and $\pi^*$, then checked their populations and dominant configurations in a small-basis pilot and in the production basis. Those diagnostics support an $n_{\mathrm O}\pi^*$ first excited root in the space actually used. A malformed root-selection request initially produced an unusable correlated output; two repair jobs then gave consistent state-specific totals and the reported gap. The successful recovery establishes numerical provenance but did not test the claim that this is the smallest balanced space, nor did it examine wider singlet or diffuse candidates. The strongest evidence is the agreement between the prospective orbital inventory and the observed state character; the root-selection misunderstanding added work without broadening the chemical test.}
\label{fig:si-active-016}
\end{figure}

\subsubsection{017: Acetone}

\begin{figure}[H]
\centering
\captionsetup{font=small}
\includegraphics[width=1\linewidth,height=0.4\textheight,keepaspectratio]{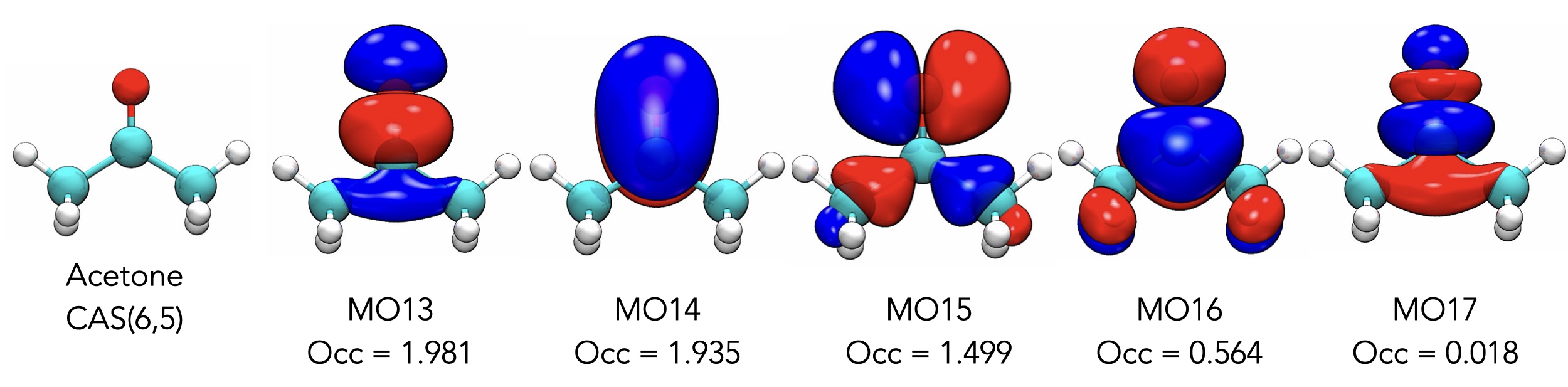}
\caption{The agent proposed a carbonyl $n,\pi,\pi^*$ starting space and, before seeing its result, specified a def2-SVP CASSCF test of adding the $\sigma/\sigma^*$ pair. The expanded CAS(6,5) shifted that pilot gap by 0.337~eV, so the agent retained the pair in the final def2-TZVPD space. Final populations and configurations support a dark $n\pi^*$ root. Two production optimizations failed; a damped rerun converged in energy, and a Newton--Raphson refinement then met the gradient criterion. The reported SC-NEVPT2 gap comes from that converged orbital set. This is a clear instance of a predeclared test changing a decision, but it establishes the sensitivity at CASSCF in the smaller basis only. No matched CAS(4,3) SC-NEVPT2 calculation in def2-TZVPD was performed, so the necessity or quantitative benefit of the extra pair at the final level remains unmeasured.}
\label{fig:si-active-017}
\end{figure}

\subsubsection{018: Benzene}
\label{app:case-018}

\begin{figure}[H]
\centering
\captionsetup{font=small}
\includegraphics[width=0.75\linewidth,height=0.3\textheight,keepaspectratio]{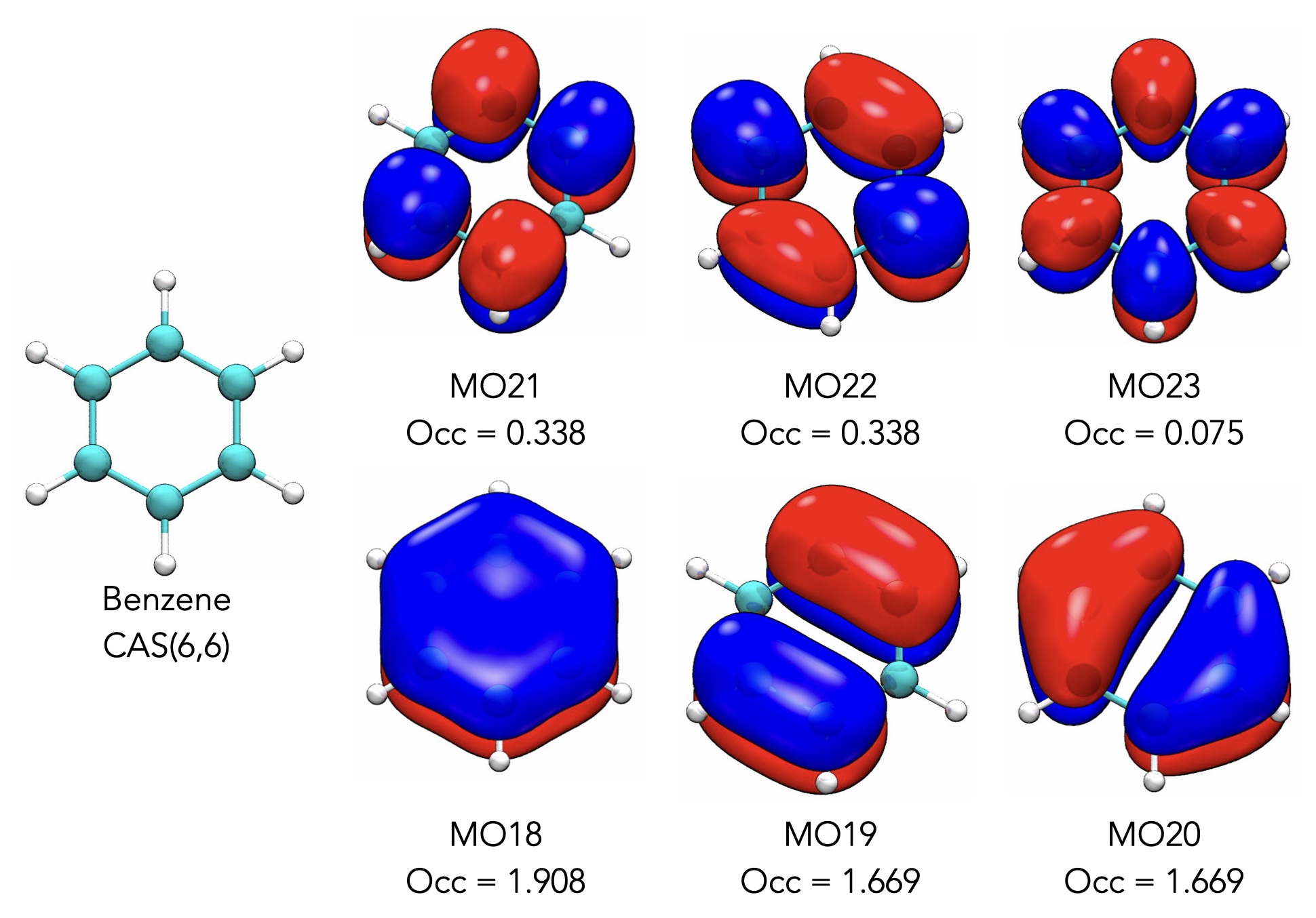}
\caption{The agent chose the complete six-orbital benzene $\pi/\pi^*$ space in advance and used orbital-population analysis to avoid non-$\pi$ orbitals interleaved in the canonical ordering. Its more consequential decisions came later: it refused a normally terminated CASSCF output whose orbital gradient was still too large, then rejected a perturbative-theory restart after the reference energy shifted. Recomputing CASSCF and SC-NEVPT2 together restored a consistent chain from orbitals to correlated root energies. The final gap is therefore supported by more than a normal-termination message. Only CAS(6,6) was seriously evaluated, and no independent extra-root or diffuse-state calculation tested whether the two-root $\pi$ model contains the globally lowest excited singlet.}
\label{fig:si-active-018}
\end{figure}

\subsubsection{019: $E$-Butadiene}

\begin{figure}[H]
\centering
\captionsetup{font=small}
\includegraphics[width=1\linewidth,height=0.4\textheight,keepaspectratio]{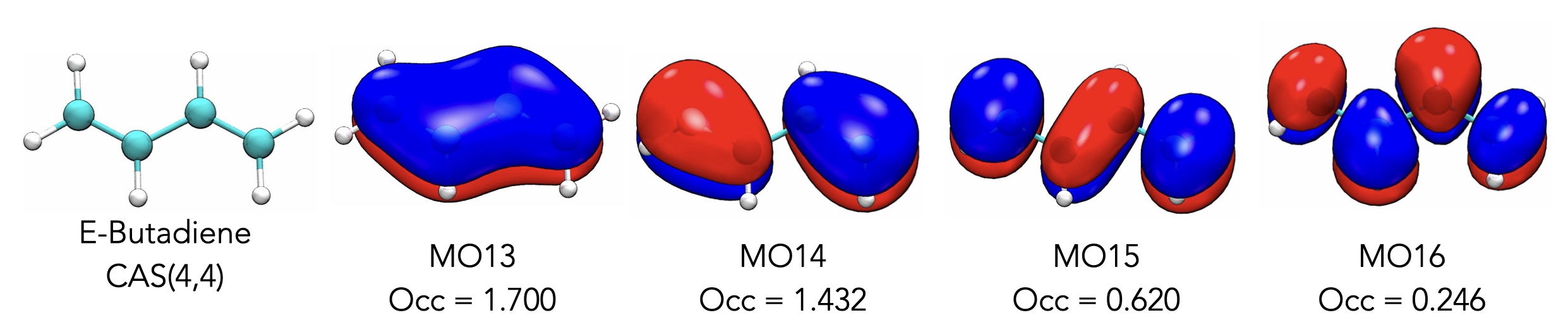}
\caption{AVAS selected a compact CAS(4,4) containing the full valence $\pi$ manifold, and the CASSCF/SC-NEVPT2 job completed. The computed first excited root was dark and $^1A_g$-like according to its configurations, whereas the tabulated comparison value represents a bright $^1B_u$ transition. This is an explicit state mismatch: numerical proximity between the two energies cannot establish agreement for the same excitation. The agent noted some state uncertainty but did not run the short multiroot or symmetry check that would have clarified the competing bright root. It also described its AVAS seed as RHF-derived, although the output indicates projection from a model-potential guess rather than a separate converged RHF calculation. The space selection itself is plausible; the weakness lies in treating a technically valid gap as the requested benchmark state.}
\label{fig:si-active-019}
\end{figure}

\subsubsection{020: 4-(Dimethylamino)benzonitrile}

\begin{figure}[H]
\centering
\captionsetup{font=small}
\includegraphics[width=1\linewidth,height=0.4\textheight,keepaspectratio]{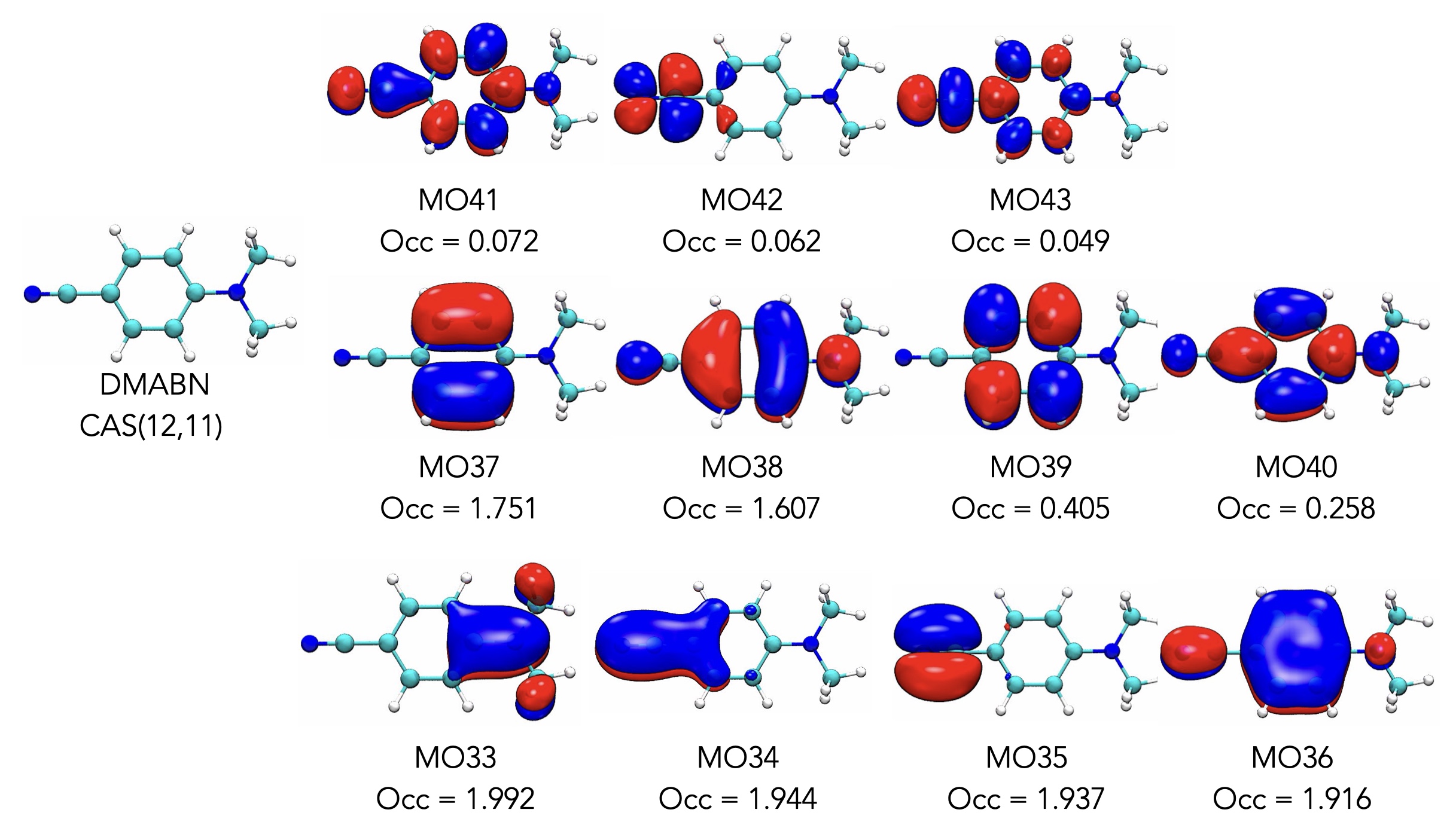}
\caption{For this donor--acceptor molecule, the agent built CAS(12,11) from amino and ring $\pi$ directions together with cyano acceptor directions. Orbital projection into the production basis initially imported the wrong coefficient set. A coefficient-identity check detected that mismatch, and the agent repaired the transfer before rerunning the two-root CASSCF/SC-NEVPT2 calculation. This is a strong provenance check: a normal numerical result from the wrong orbitals would not have tested the intended space. Final orbital populations support the donor--acceptor subspace as a whole, although individual optimized orbitals are mixed and should not be labeled as pure localized functions. The final gap is internally consistent, but the selected two-root calculation does not establish the global ordering in a molecule with several possible local and charge-transfer singlets. The numerical comparator alone cannot settle that state question.}
\label{fig:si-active-020}
\end{figure}

\subsubsection{021: Ethene}

\begin{figure}[H]
\centering
\captionsetup{font=small}
\includegraphics[width=0.75\linewidth,height=0.3\textheight,keepaspectratio]{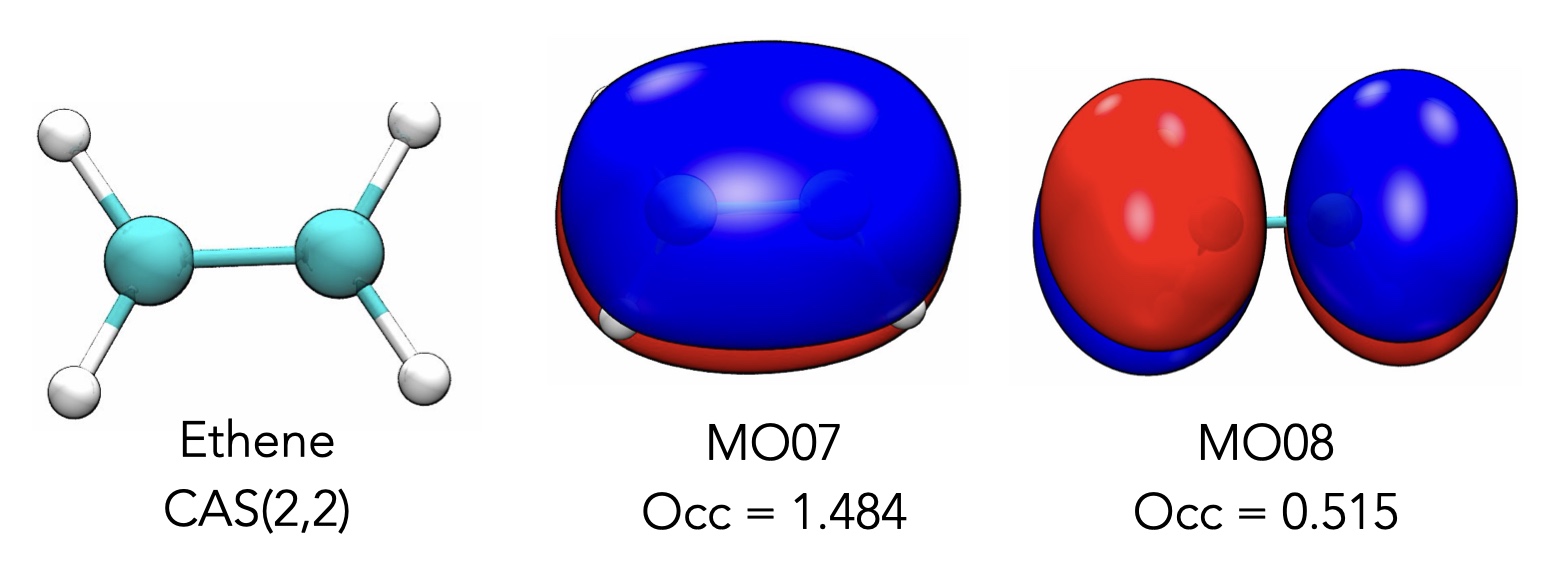}
\caption{The agent anticipated that diffuse orbitals in def2-TZVPD would displace the valence $\pi^*$ orbital from a default two-orbital active window. It explicitly rotated the $\pi/\pi^*$ pair into CAS(2,2), verified the resulting populations, and rejected a corrupted SC-NEVPT2 output before a successful retry. These actions support the calculation's orbital identity and final root-specific energy difference. The $\pi^*$ orbital nevertheless retained substantial diffuse and polarization character, while the small active space could not test additional $\sigma$--$\pi$ polarization or a separate diffuse state. A proposed CAS(2,3) fallback was never run. Consequently the work demonstrates successful correction of an orbital-index problem but offers limited evidence that the minimal CAS is accurate enough for this basis and transition.}
\label{fig:si-active-021}
\end{figure}

\subsubsection{022: Formaldehyde}

\begin{figure}[H]
\centering
\captionsetup{font=small}
\includegraphics[width=0.75\linewidth,height=0.3\textheight,keepaspectratio]{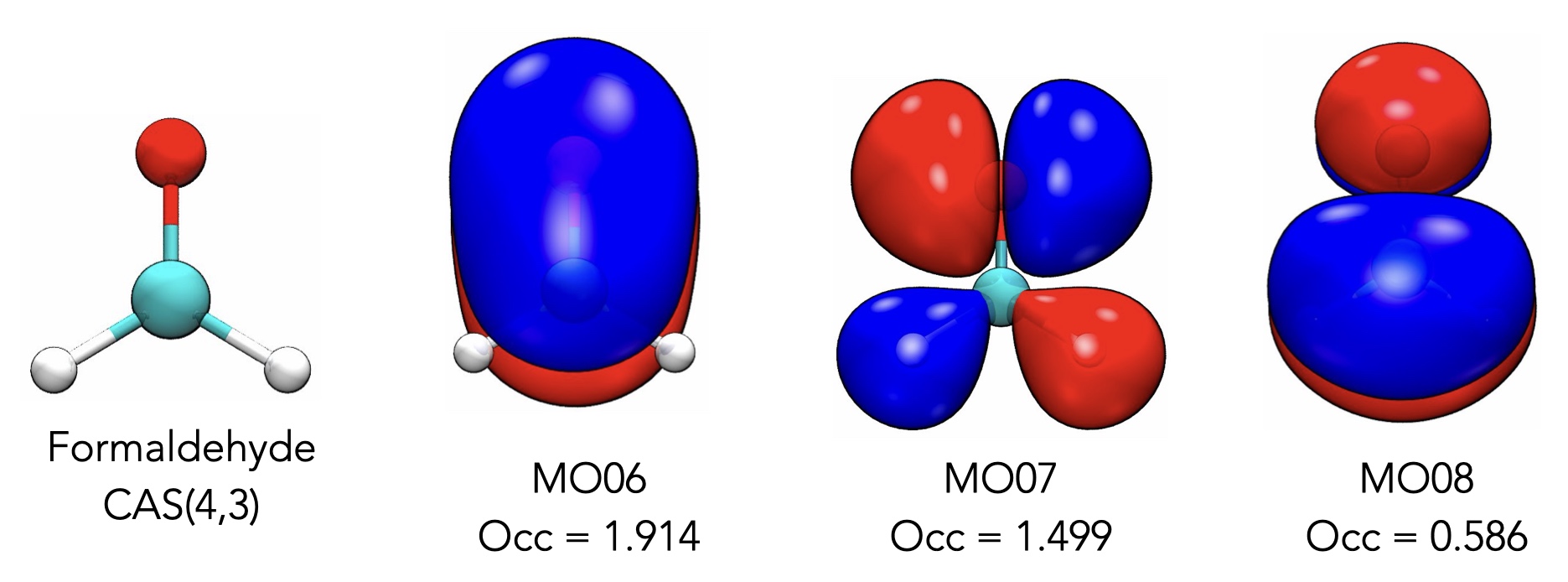}
\caption{The agent began with a chemically motivated $n_{\mathrm O},\pi,\pi^*$ CAS(4,3), inspected the RHF orbital characters, and used a small-basis state check before projecting the orbitals into def2-TZVPD. It then repeated a production-basis CASSCF gate rather than assuming the original ordering survived transfer. The final SC-NEVPT2 step used that previously converged orbital set, and both corrected roots support the reported gap. This ordered sequence gives direct evidence that the anticipated carbonyl transition was represented in the actual input. The larger CAS(6,5) alternative was only a contingency, not a computed comparison, and no broader root search was made. The result is therefore a credible prediction for the selected $n\pi^*$ model rather than a demonstration that its active space or global state ordering is unique.}
\label{fig:si-active-022}
\end{figure}

\subsubsection{023: Furan}

\begin{figure}[H]
\centering
\captionsetup{font=small}
\includegraphics[width=0.75\linewidth,height=0.24\textheight,keepaspectratio]{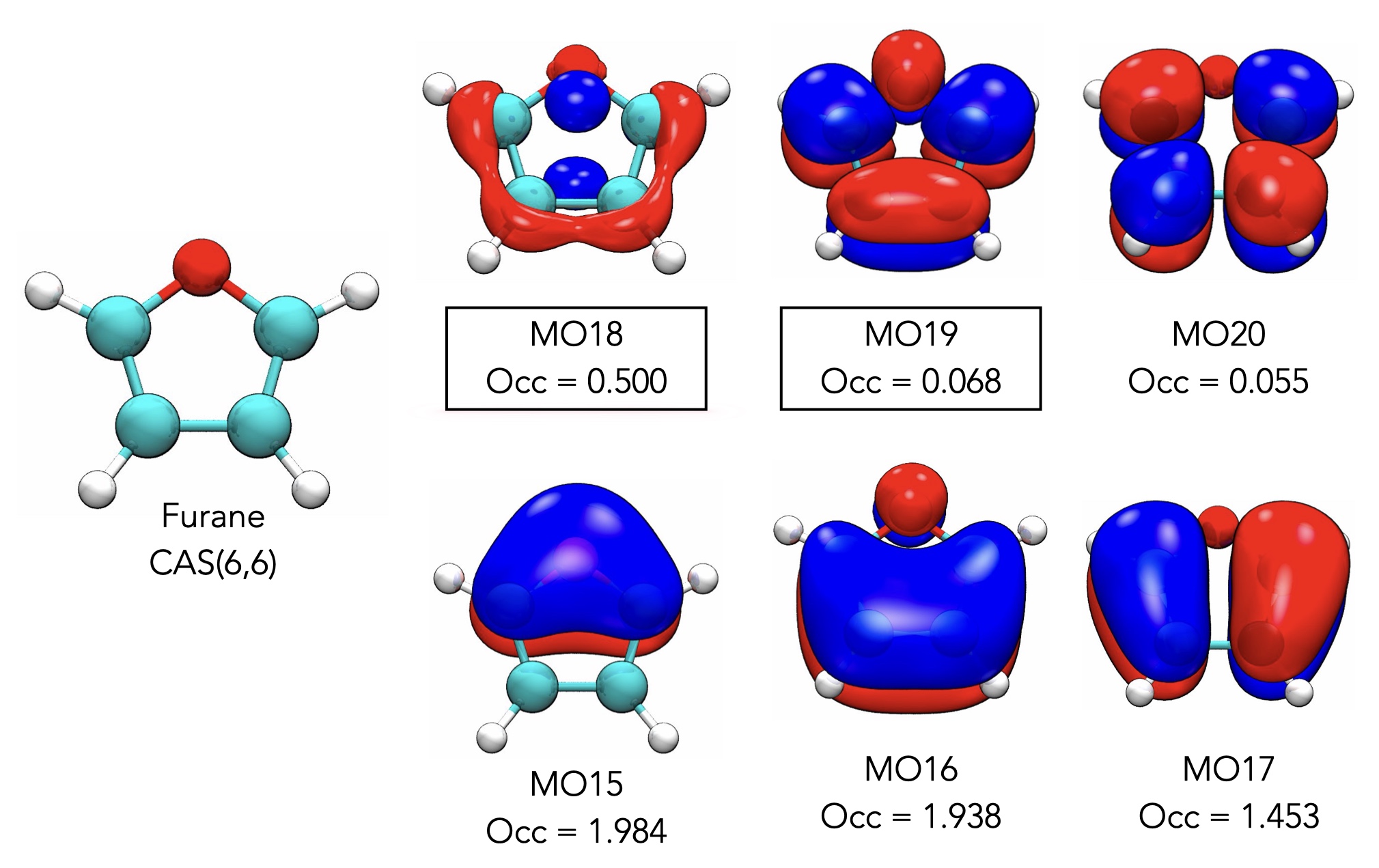}
\caption{The agent considered a diffuse 3s acceptor before calculation, then built CAS(6,6) from three occupied $\pi$ orbitals, two valence $\pi^*$ orbitals, and that diffuse direction. After a four-root validation, inspection of the final orbitals revealed that an in-plane orbital had entered the space; the agent repaired the map and repeated the check. The production first excited root is predominantly $\pi\!\to\!3s$ and dipole forbidden. The tabulated 6.32~eV reference is instead a CASPT2 value for a valence $^1B_2\,\pi\pi^*$ state, so the close numerical match compares different transitions. A four-root CASSCF screen showed the competing valence root, but both states were not corrected by matched SC-NEVPT2 calculations. The useful behavior is the explicit orbital and state diagnosis; the unresolved issue is their correlated ordering, not the arithmetic of the reported gap.}
\label{fig:si-active-023}
\end{figure}

\subsubsection{024: All-\textit{E}-hexatriene, Hoyer geometry}

\begin{figure}[H]
\centering
\captionsetup{font=small}
\includegraphics[width=0.75\linewidth,height=0.24\textheight,keepaspectratio]{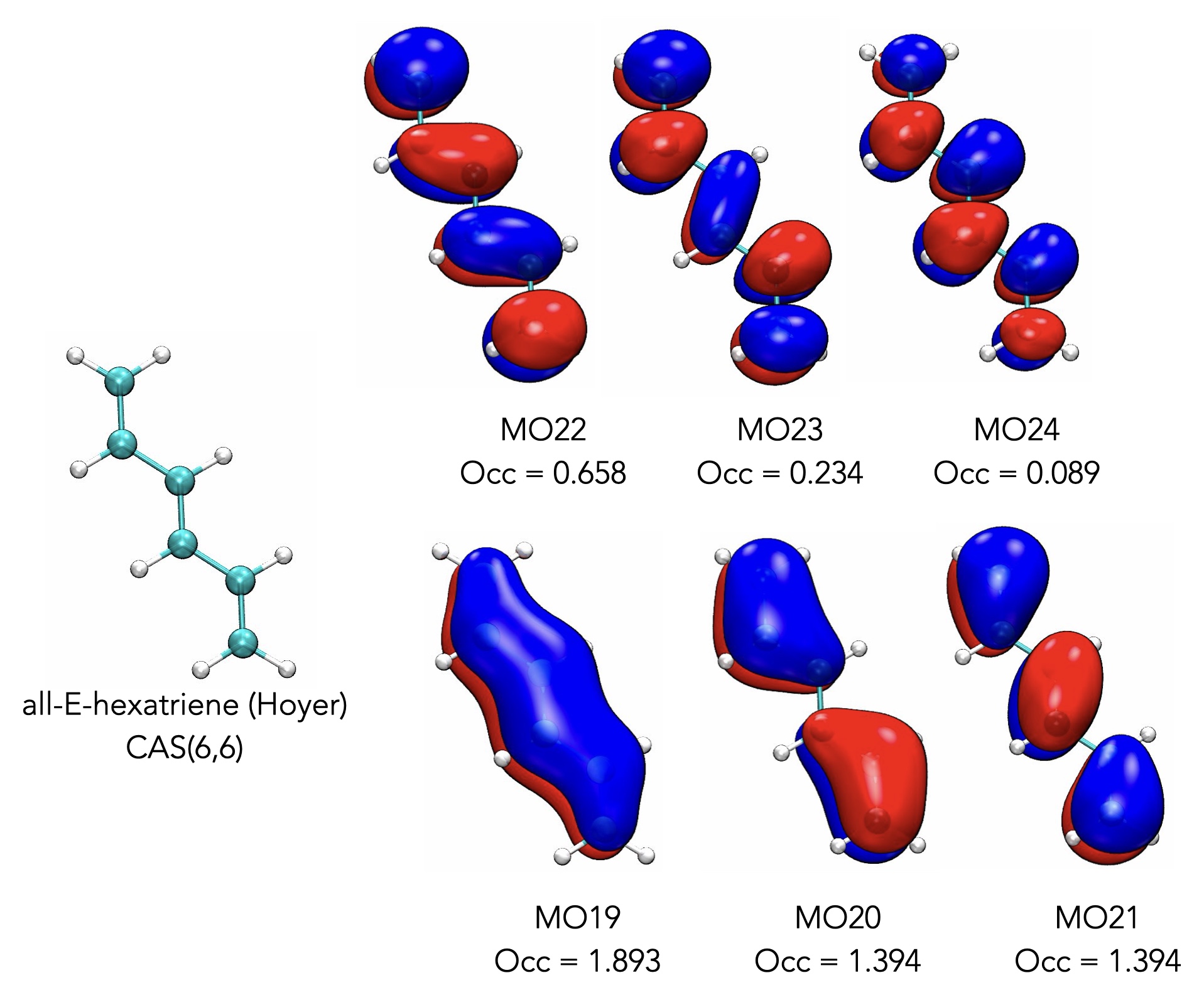}
\caption{At this second, distinct hexatriene geometry, the agent proposed the full valence $\pi$ CAS(6,6) before computing a gap. Canonical virtual $\pi^*$ orbitals were separated by $\sigma$-like orbitals, so it reordered the selected MOs, checked their carbon out-of-plane populations, and revalidated the space after projection into the production basis. The final equal-weight two-root CASSCF/SC-NEVPT2 calculation converged and reproduced the reported energy difference. This trajectory supports the chosen $\pi$ subspace more strongly than a raw orbital-index list would. It does not include an extra-root or diffuse-state comparison, and its description of the excited root as ``dark'' lacks an independent transition-strength analysis. One orbital-character threshold was relaxed after the result was visible, so that particular gate was not wholly prospective.}
\label{fig:si-active-024}
\end{figure}

\subsubsection{025: Naphthalene}
\label{app:case-025}

\begin{figure}[H]
\centering
\captionsetup{font=small}
\includegraphics[width=0.75\linewidth,height=0.3\textheight,keepaspectratio]{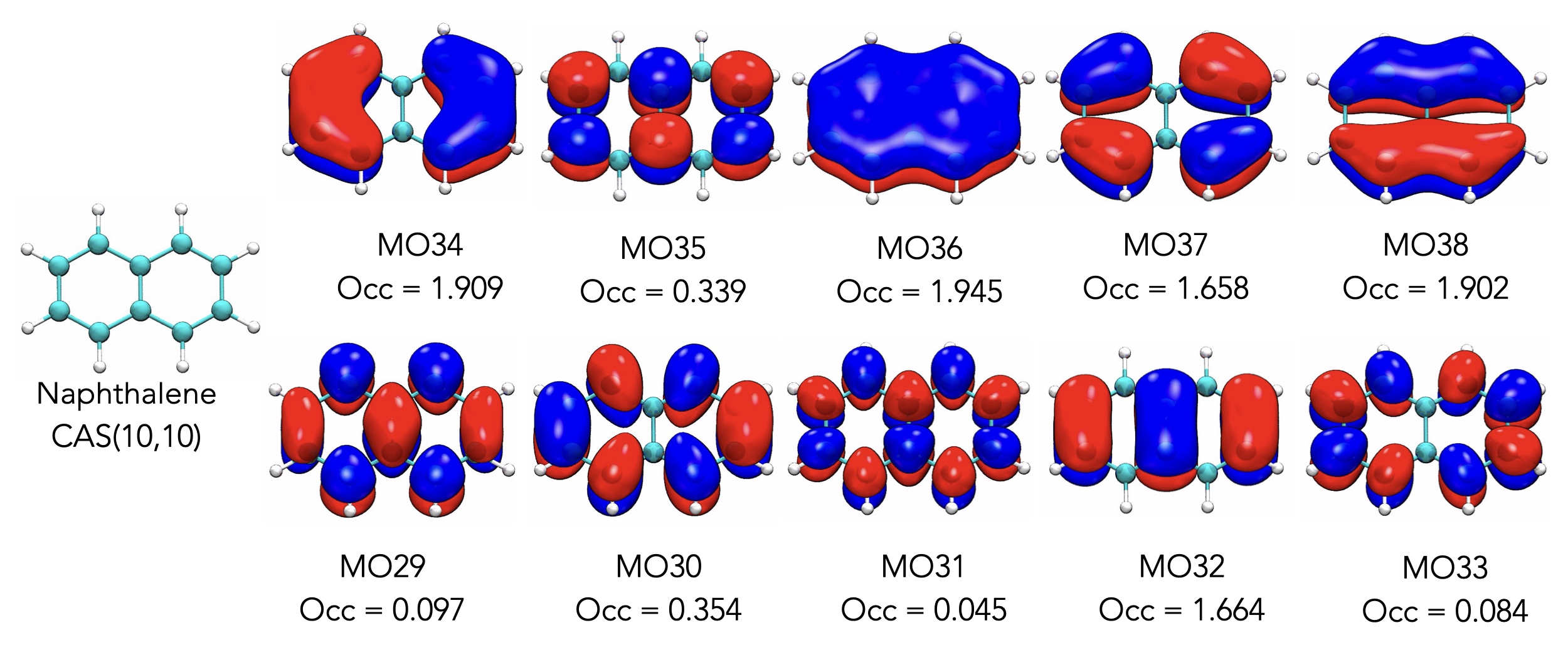}
\caption{The agent assembled the full ten-orbital valence $\pi$ system using symmetry labels and orbital plots. Its first state-targeted calculation yielded a $^1L_b$-consistent state. It then interpreted literature symmetry labels in a different molecular-axis convention and changed the assignment, ultimately reporting a $^1L_a$ state. The latter number lay closer to the tabulated reference, but the change did not make it the same transition: configurations and orbital occupations distinguish the two states. A parser error during initial orbital scouting was repaired, and the final numerical gap itself is traceable; the scientific problem is the late state-label reinterpretation. This case shows that a well-chosen active space can still lead to the wrong benchmark answer when symmetry conventions and root identity are mishandled. A single common-orbital lowest-two-root correlated comparison was not established.}
\label{fig:si-active-025}
\end{figure}

\subsubsection{026: $p$-Nitroaniline}
\label{app:case-026}

\begin{figure}[H]
\centering
\captionsetup{font=small}
\includegraphics[width=0.75\linewidth,height=0.3\textheight,keepaspectratio]{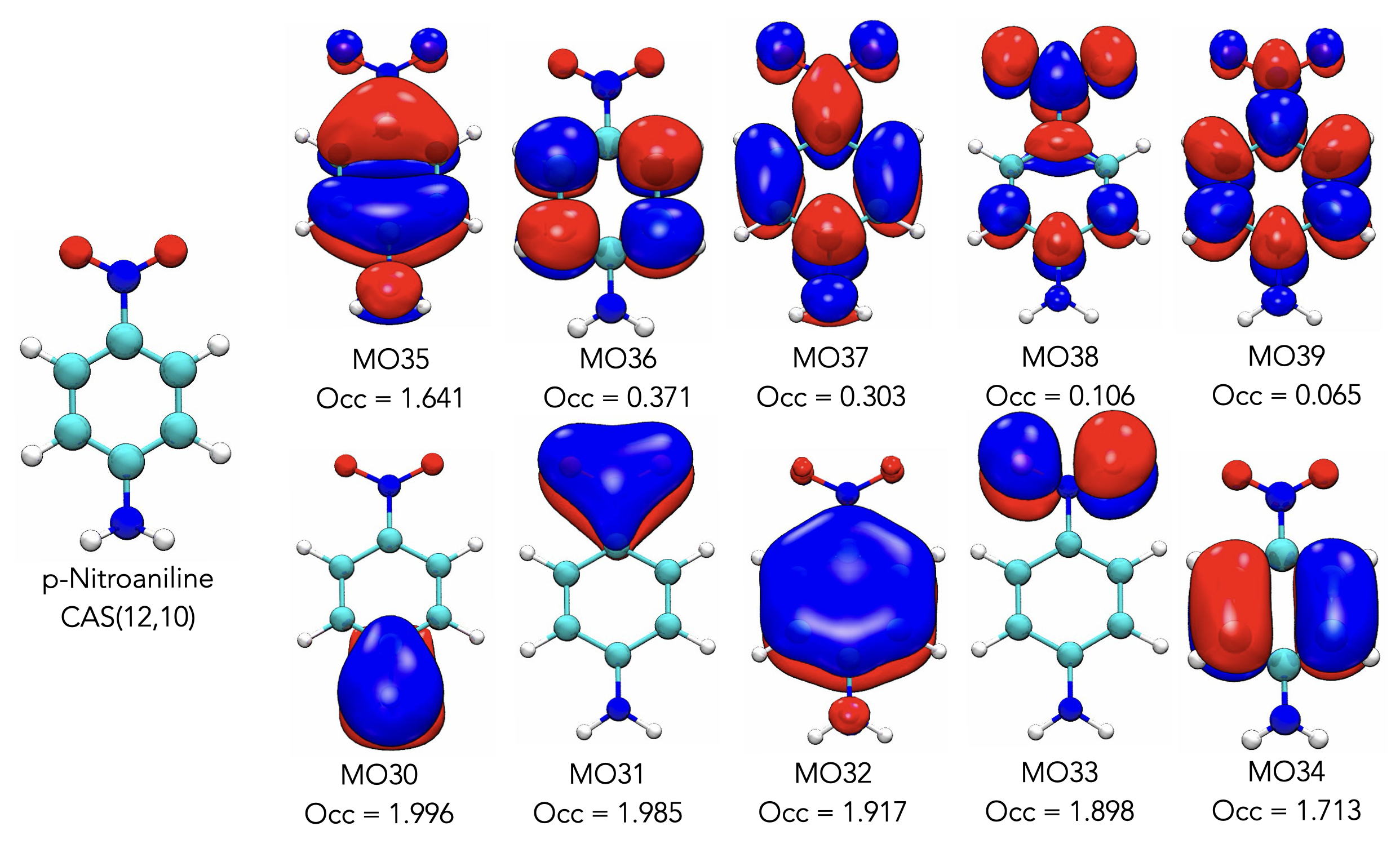}
\caption{The agent initially expected the donor-to-nitro charge-transfer band to be $S_1$. It rejected a canonical virtual selection, withdrew approval of a full-virtual-space projection, and rebuilt a bounded selection containing donor/ring and nitro-acceptor directions. A five-root CASSCF diagnostic then placed a weak first excited root below the brighter charge-transfer state, prompting a revised interpretation before the final two-root SC-NEVPT2 report. The five-root ordering is a CASSCF finding: the higher bright state did not receive matched correlated treatment, and diffuse mixing in one retained virtual leaves its balance uncertain.}
\label{fig:si-active-026}
\end{figure}

\subsubsection{027: Pyrazine}
\label{app:case-027}

\begin{figure}[H]
\centering
\captionsetup{font=small}
\includegraphics[width=0.75\linewidth,height=0.3\textheight,keepaspectratio]{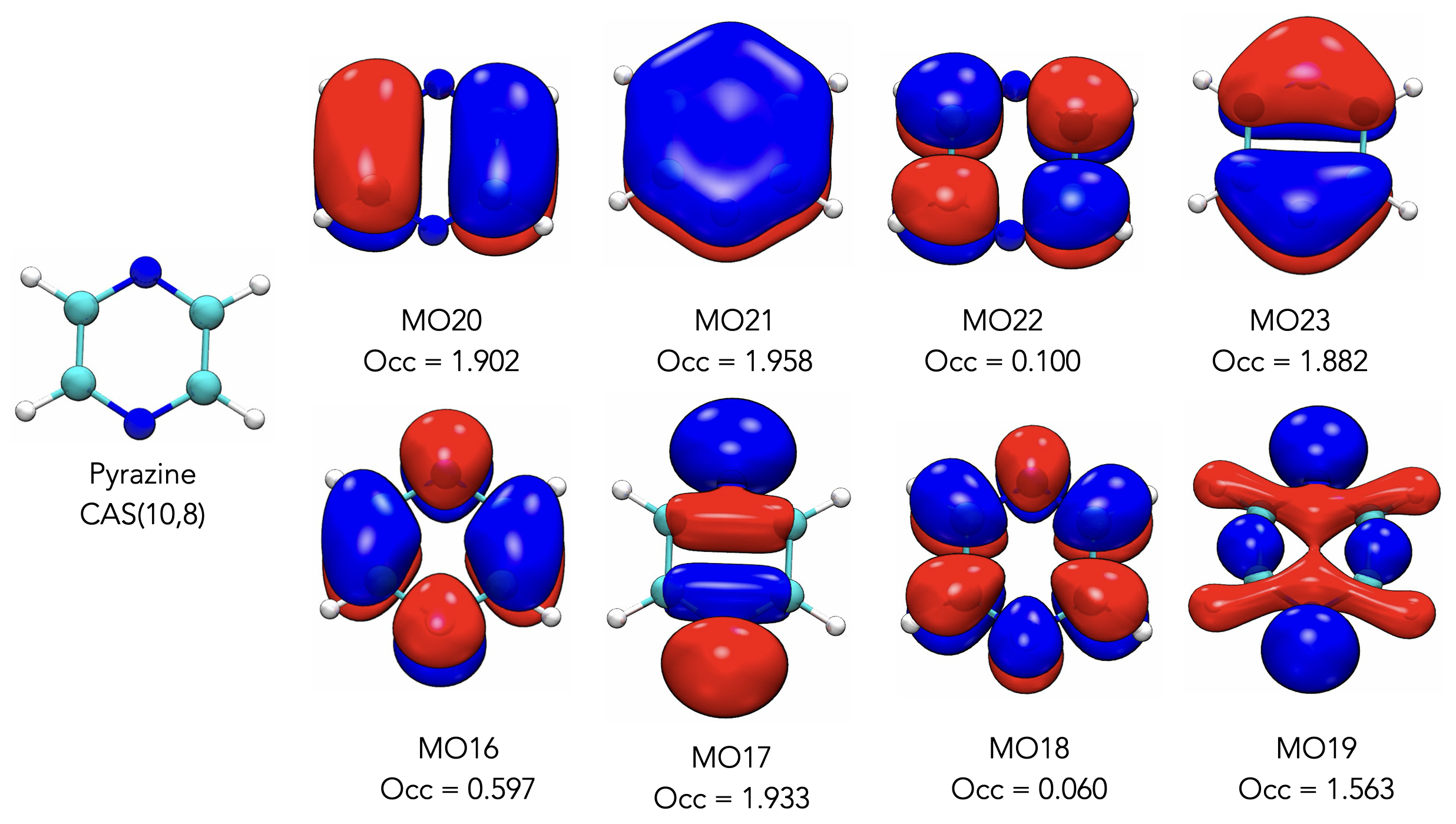}
\caption{The agent prospectively proposed a space containing two N lone pairs and the ring $\pi/\pi^*$ manifold, with a smaller $\pi$ space as a control. An initial two-root result converged to a $\pi\pi^*$ state, which the agent identified from configurations and occupations rather than reporting as the desired $n\pi^*$ excitation. A symmetry-targeted CAS(10,8) calculation recovered the lone-pair-to-$\pi^*$ character, and its correlated gap is reproducible. This is strong evidence for corrective state reasoning. It was accompanied by procedural inefficiencies: repeated RHF restarts responded to small printed convergence differences, and an unsupported production keyword required a parser-error retry. Other symmetry sectors were not tested under a single common-orbital correlated treatment, so the broader lowest-singlet ordering remains qualified.}
\label{fig:si-active-027}
\end{figure}

\subsubsection{028: Pyridazine}

\begin{figure}[H]
\centering
\captionsetup{font=small}
\includegraphics[width=0.75\linewidth,height=0.3\textheight,keepaspectratio]{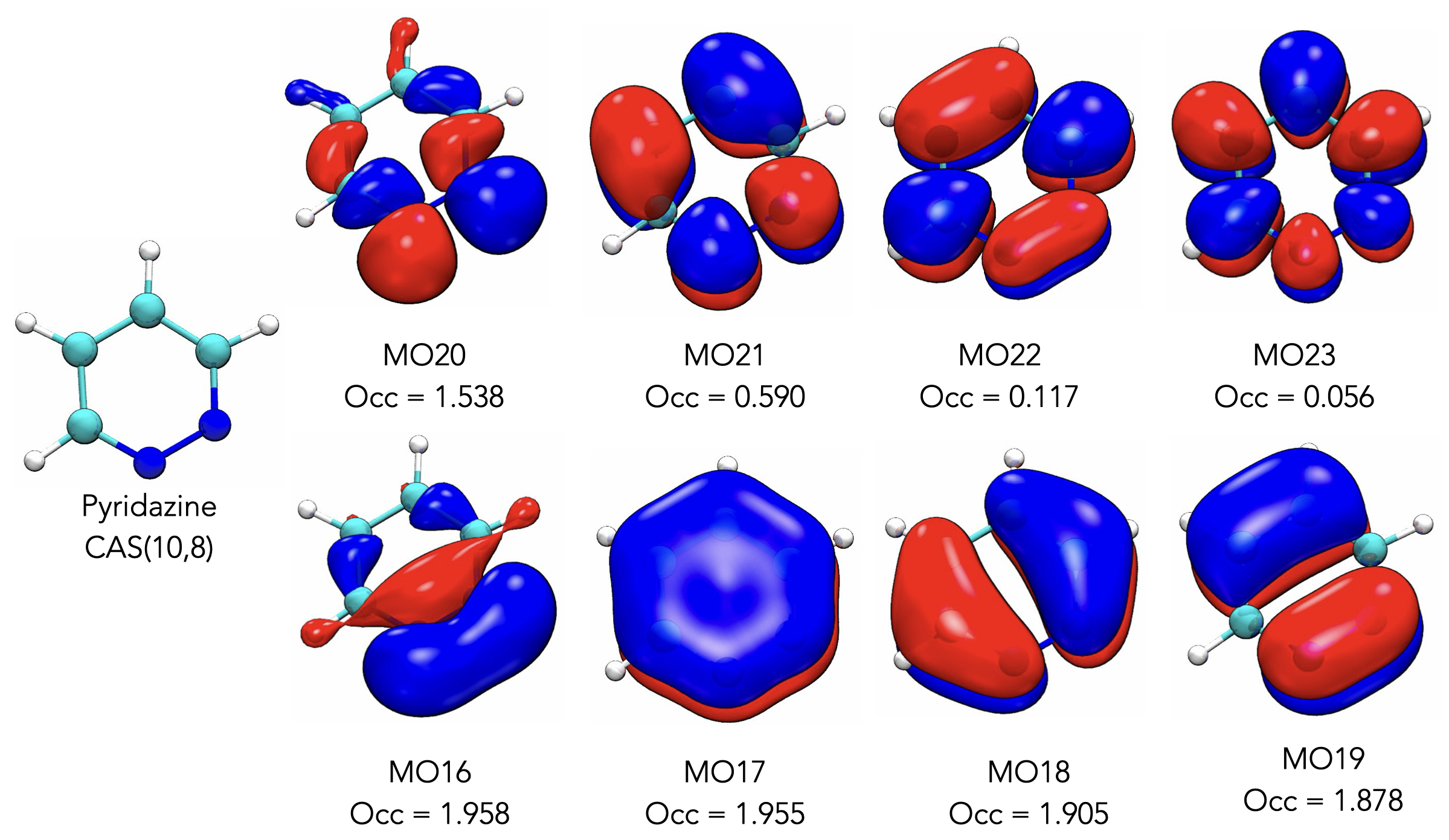}
\caption{The agent compared two candidate spaces prospectively and chose CAS(10,8) to retain both N lone-pair directions alongside the ring $\pi/\pi^*$ manifold. Final populations place the lone pairs and $\pi$ orbitals in the intended subspace, while dominant configurations support an $n\pi^*$ first excited root. A restart error caused a technical retry without altering the chemical selection; the converged production calculation and two root-specific corrections support its reported gap. This evidence favors the chosen model over the particular smaller candidate that was screened. It does not demonstrate that both lone pairs are necessary, because only one composition of the smaller space was tried, and no wider-root or diffuse-state study settled global ordering.}
\label{fig:si-active-028}
\end{figure}

\subsubsection{029: Pyridine}

\begin{figure}[H]
\centering
\captionsetup{font=small}
\includegraphics[width=0.75\linewidth,height=0.24\textheight,keepaspectratio]{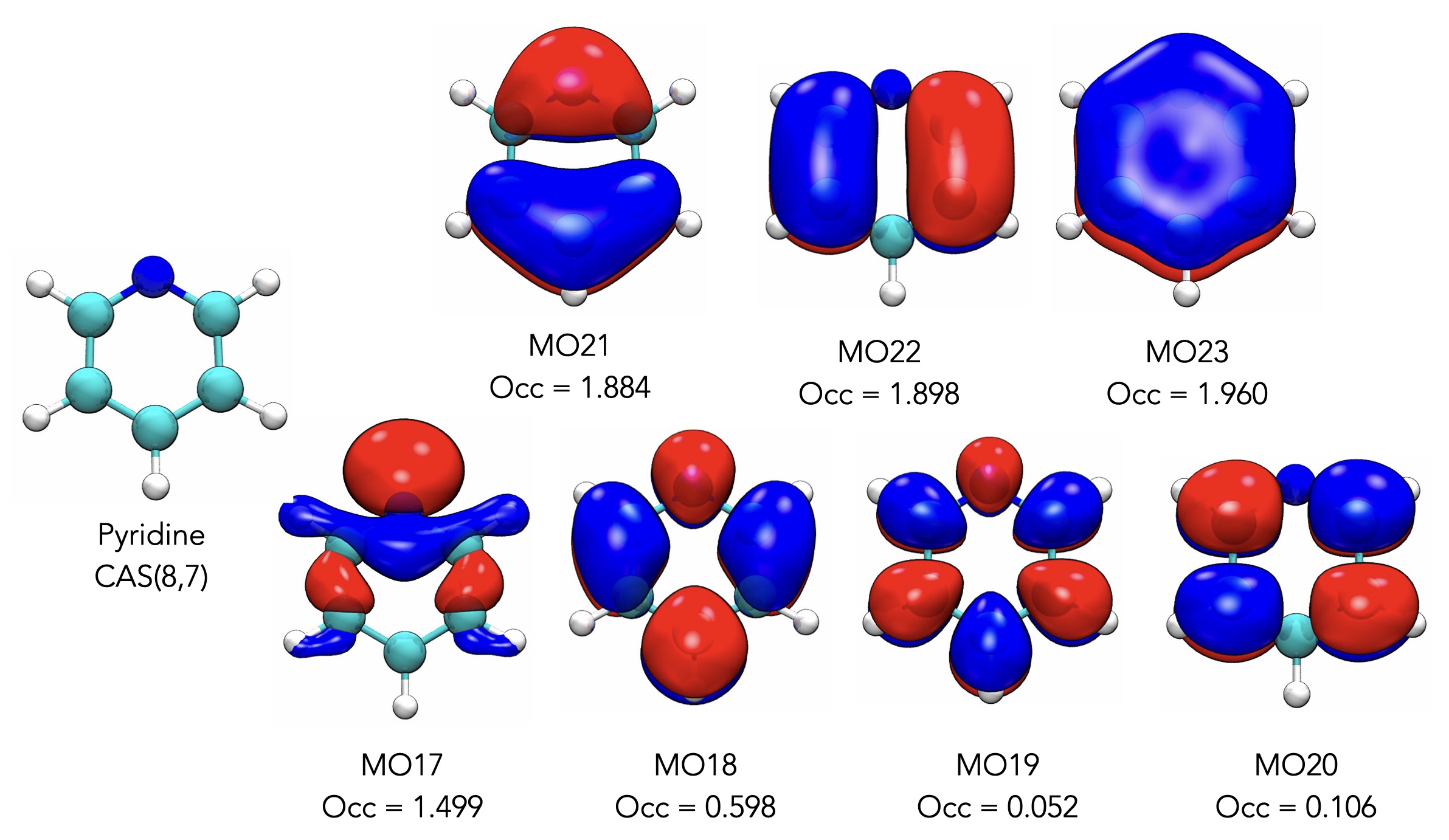}
\caption{From the outset, the agent considered both $n\pi^*$ and $\pi\pi^*$ states and included the N lone pair in CAS(8,7). An unconstrained three-root test lost the $n\pi^*$ character, so it used separately symmetry-blocked two-root averages. Their CASSCF gaps put $\pi\pi^*$ lower by 0.178~eV. The agent initially selected that order and prepared production for it, but before any SC-NEVPT2 job ran it added a matched test on the $n\pi^*$ pair. Correlation reversed the ordering: the $n\pi^*$ gap was 5.186~eV and the $\pi\pi^*$ gap 5.402~eV. The extra calculation answered a specific state question and changed the final assignment. Its limitation is methodological: separately optimized state pairs are not one equal-weight average of the lowest two singlets, so the exact requested common-orbital protocol remains untested.}
\label{fig:si-active-029}
\end{figure}

\subsubsection{030: Pyrimidine}

\begin{figure}[H]
\centering
\captionsetup{font=small}
\includegraphics[width=0.75\linewidth,height=0.24\textheight,keepaspectratio]{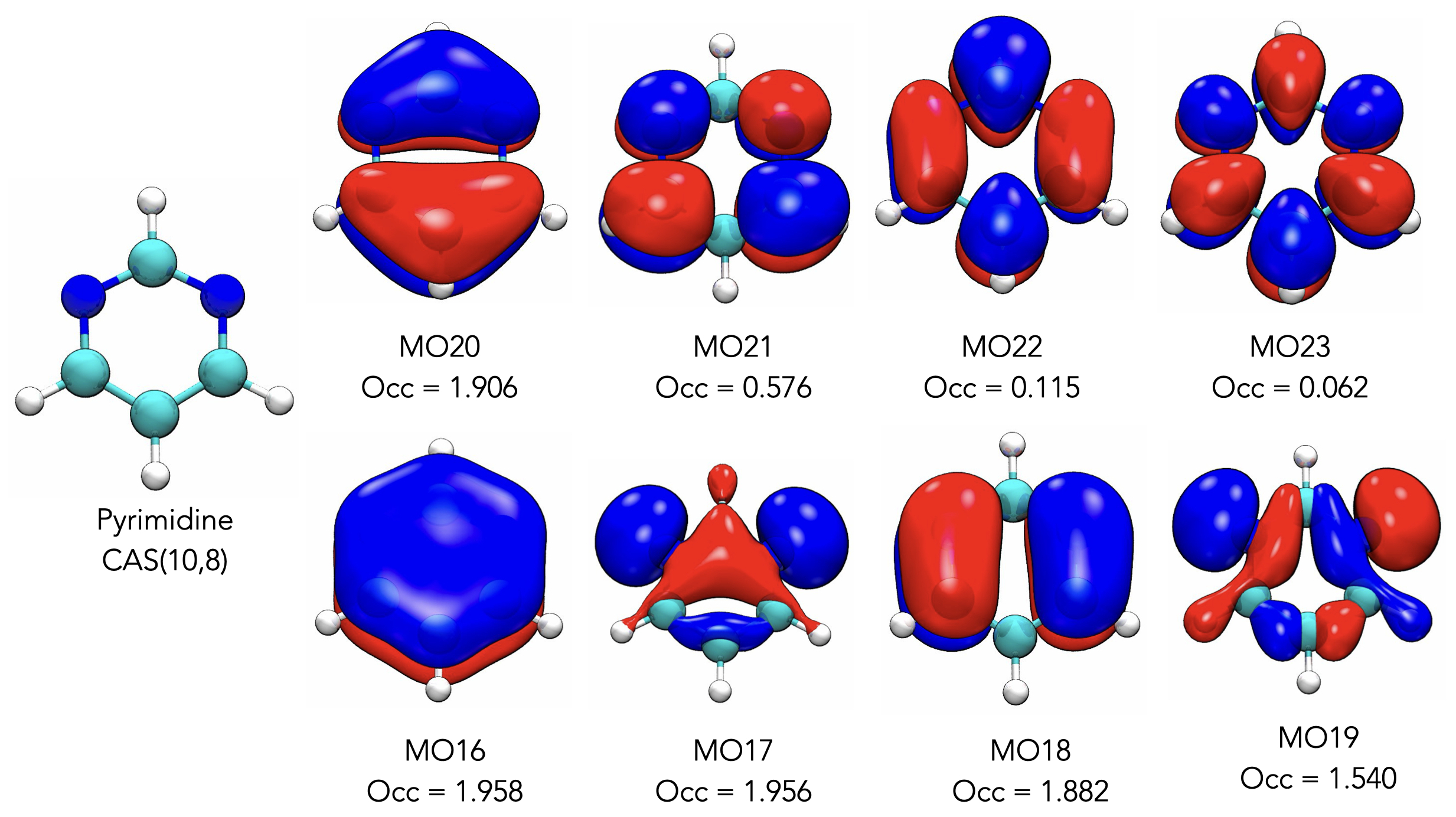}
\caption{The agent selected CAS(10,8) with two N lone-pair and six ring $\pi/\pi^*$ directions. A four-root fixed-orbital scout and subsequent small-basis relaxation put mixed $\pi\pi^*$ character below the clear $n\pi^*$ candidate; the order first reversed on projection into def2-TZVPD before target-basis orbital optimization. Rather than relying on orbital energies, the agent tracked configurations across these stages and tested the corrected root ordering. It rejected a GBW restart after finding an inconsistent reference energy, then reconverged CASSCF and performed SC-NEVPT2 in a single consistent production job. That job established the reported $n\pi^*$ gap as the lowest corrected excitation among its four represented roots. The record does not establish the cause of the restart mismatch or the response of a larger or diffuse-inclusive space; several extra replay and density-analysis steps added effort without resolving those questions.}
\label{fig:si-active-030}
\end{figure}

\subsubsection{031: $s$-Tetrazine}
\label{app:case-031}

\begin{figure}[H]
\centering
\captionsetup{font=small}
\includegraphics[width=0.75\linewidth,height=0.3\textheight,keepaspectratio]{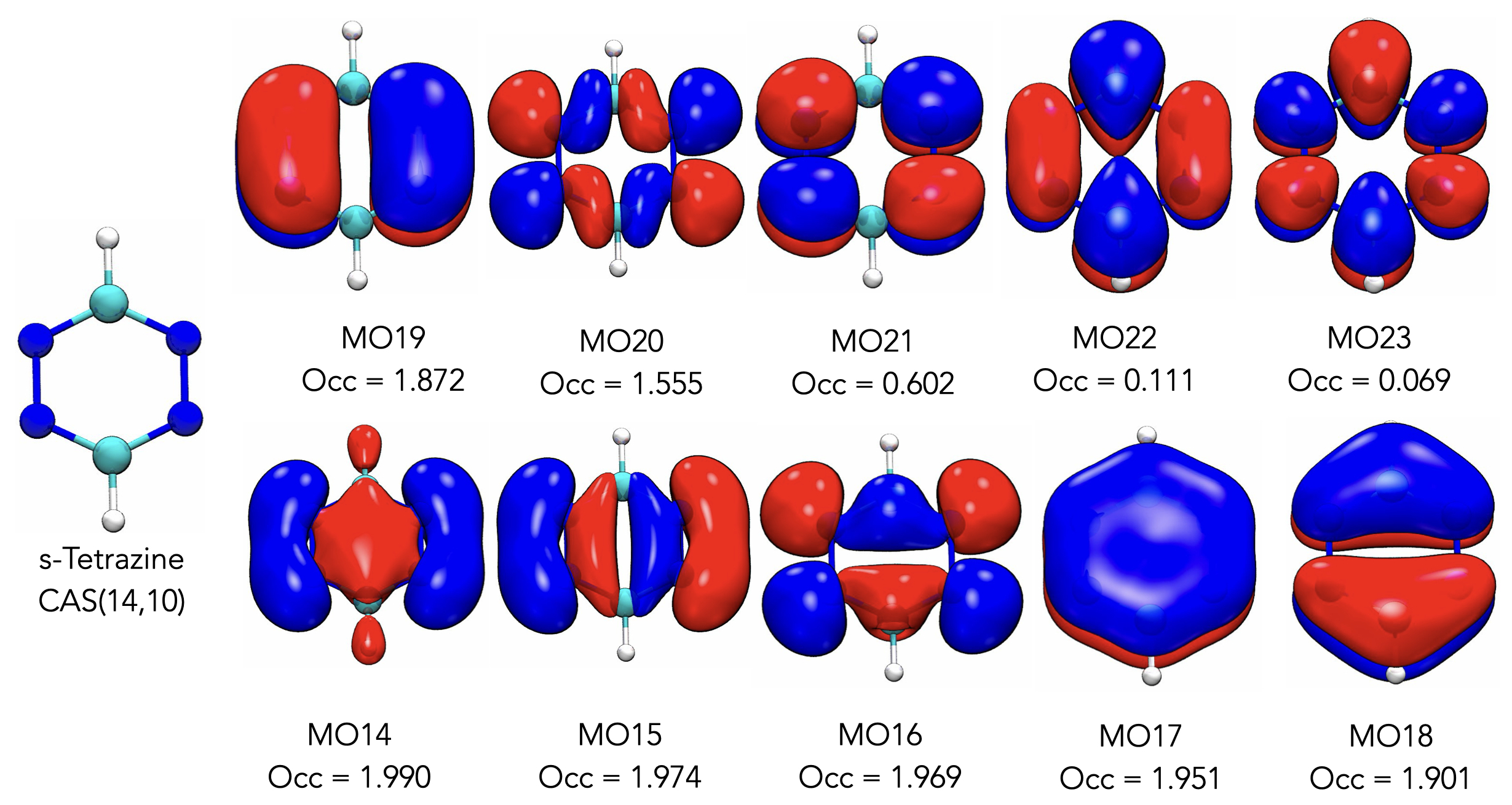}
\caption{The agent proposed CAS(14,10) before orbital scouting: all four N lone-pair combinations together with the full six-orbital ring $\pi$ system. Symmetry and population analysis caught a diffuse intruder, and a validation calculation checked that the selected orbitals and $n\pi^*$ character survived into production. One input-keyword error was repaired before the final converged two-root CASSCF/SC-NEVPT2 calculation. The continuity of configurations supports the assignment within this space. The trajectory validates a plausible literature-informed choice but does not show that this CAS is uniquely needed: no smaller space, larger diffuse-inclusive space, or broader state manifold was computed. The retrieved literature content supporting the initial choice is not preserved in full, and a negative subspace-gap diagnostic was explained rather than tested through a distinct orbital start.}
\label{fig:si-active-031}
\end{figure}

\subsubsection{032: Water}

\begin{figure}[H]
\centering
\captionsetup{font=small}
\includegraphics[width=0.75\linewidth,height=0.3\textheight,keepaspectratio]{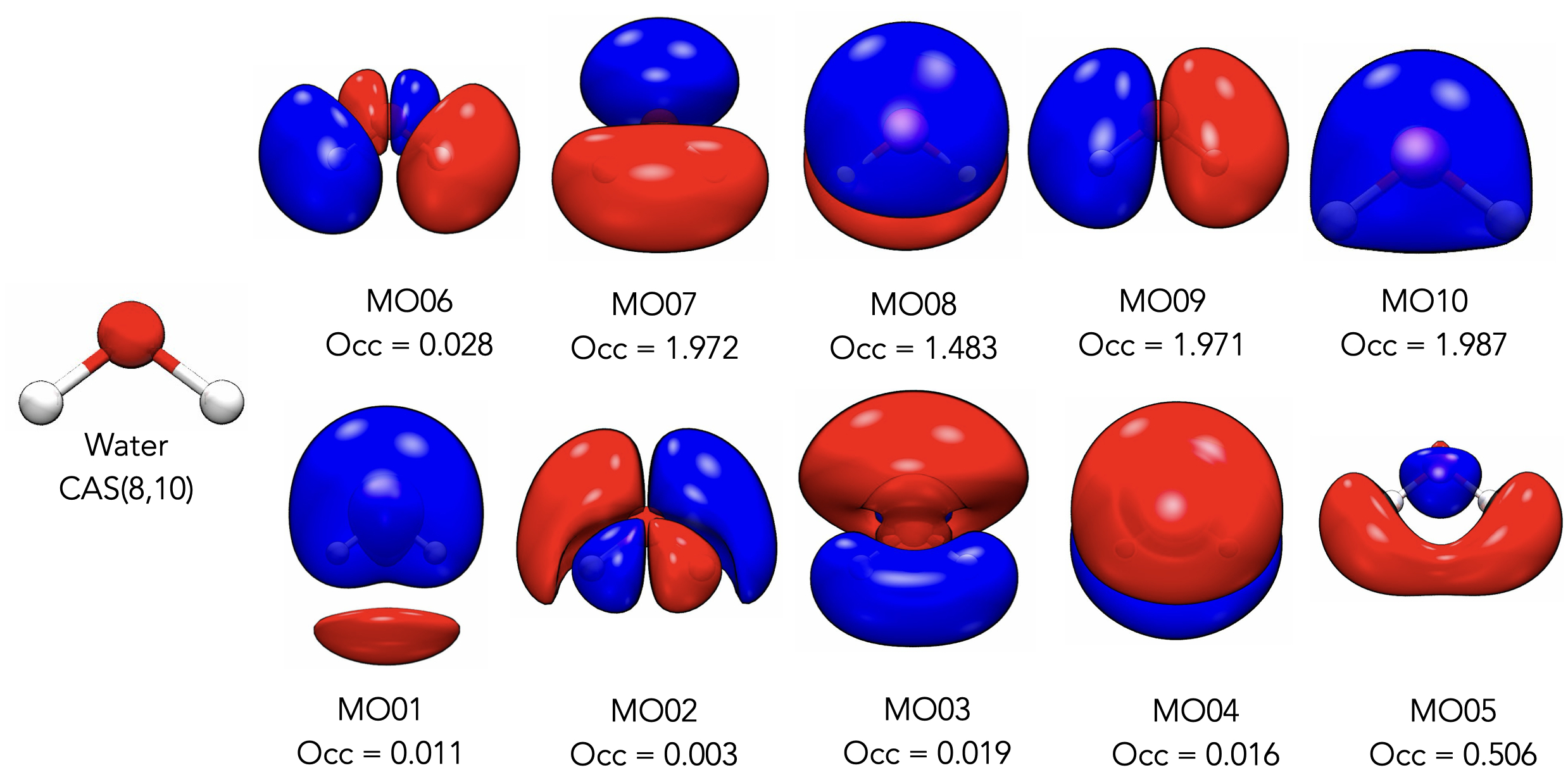}
\caption{The agent selected CAS(8,10) before detailed scouting to include four occupied valence orbitals and low virtual directions expected to contain the $4a_1$-like acceptor. After a technical retry, a symmetry-targeted equal-weight two-root CASSCF/SC-NEVPT2 calculation converged. Active-orbital occupations and leading configurations support a $1b_1\!\to\!4a_1$ candidate, and both correlated totals reproduce the reported gap. This establishes a coherent result for the represented state, but only one excited symmetry/root was treated. The calculation therefore cannot prove that it is the globally lowest singlet, nor does the available orbital analysis fully settle the acceptor's radial and diffuse character or the stability of the broad ten-orbital active space. The prospective proposal was chemically sensible; the run supplies less evidence for excluding competing states than for describing the one it computed.}
\label{fig:si-active-032}
\end{figure}


\newcommand{\captionPNARepOne}{\textbf{$p$-Nitroaniline (026), rep1: enlarging the initial valence space.} The agent began with CAS(12,10), then used CAM-B3LYP/TDA in two bases and MP2 natural orbitals to examine the low states and candidate active orbitals. It chose CAS(14,11), including a compact nitro lone-pair direction, and tested a smaller CAS(12,10) control. The smaller space lost its intended orbital character during CASSCF optimization. When the larger-space calculation produced a different $S_1$ character than the TDA screen had suggested, a four-root calculation led the agent to retain the lowest mixed valence state rather than its initial lone-pair assignment. It also rejected an SC-NEVPT2 output that duplicated an excited root. The corrected gap was 4.948~eV; the final orbital gradient was close to, but slightly above, the requested threshold.}

\newcommand{\captionPNARepTwo}{\textbf{$p$-Nitroaniline (026), rep2: retaining a directly selected space.} Starting from CAS(12,10), the agent used PBE0 orbitals and AVAS to select six occupied and four virtual orbitals spanning the amino donor, ring, and nitro acceptor. A four-root CASCI screen placed a weak, mixed $S_1$ below a brighter charge-transfer state, so the agent kept the selected space for two-root CASSCF. The first optimization stopped on energy before meeting its orbital-gradient criterion; a continuation converged. The first SC-NEVPT2 input selected the excited root twice, producing an invalid result despite normal program termination. Correcting the root indices gave a 4.945~eV gap. The run did not compare the final space against another optimized CAS.}

\newcommand{\captionPNARepThree}{\textbf{$p$-Nitroaniline (026), rep3: rebuilding the space after basis transfer failed.} The agent proposed CAS(12,10), surveyed low singlets with CAM-B3LYP/TDA, and used AVAS and multiroot calculations to test a conjugated valence space in a smaller basis. Attempts to transfer those orbitals to def2-TZVPD failed an orbital-projection check. The agent instead generated a new target-basis reference and rebuilt CAS(12,10) with AVAS; after unsuccessful exact-integral calculations, a density-fitted four-root screen ran successfully. A two-root CASSCF restart then met the orbital-gradient criterion, and SC-NEVPT2 gave 4.943~eV. The final state was weak and mixed within the selected space; the change from screened to optimized root character was not tracked by state overlaps.}

\newcommand{\captionPyrimidineRepOne}{\textbf{Pyrimidine (030), rep1: replacing a diffuse acceptor and targeting a state.} The agent proposed CAS(10,8) with both N lone pairs and the ring $\pi/\pi^*$ orbitals. PBE0/TDDFT exposed a diffuse-mixed candidate $\pi^*$ orbital; MP2 natural orbitals supplied a more compact alternative. Initial CASSCF trials with CAS(10,8) and reduced CAS(8,7) spaces did not retain the intended $n\pi^*$ state as the first excited root. A four-root calculation clarified the ordering, after which symmetry-targeted comparisons favored retaining both lone pairs in CAS(10,8). The final two-state SC-NEVPT2 gap was 4.571~eV for an $n\pi^*$ state. Because the production calculation selected one excited-state symmetry, it did not establish correlated ordering against other singlet symmetries.}

\newcommand{\captionPyrimidineRepTwo}{\textbf{Pyrimidine (030), rep2: testing low roots and a larger space.} The agent selected CAS(10,8) from a CAM-B3LYP/TDA state survey and orbital-character checks, explicitly placing two N lone pairs and six ring $\pi/\pi^*$ orbitals in the active window. Seven-root CASSCF found an $n\pi^*$ $S_1$ close to a mixed $\pi\pi^*$ state; a two-root calculation retained the $n\pi^*$ assignment. The agent also tried CAS(10,10) to admit diffuse challengers, but one intended orbital escaped during optimization, so this was not a complete larger-space test. Its first SC-NEVPT2 input selected the excited root twice; correcting the selector yielded a 4.572~eV gap.}

\newcommand{\captionPyrimidineRepThree}{\textbf{Pyrimidine (030), rep3: checking root order after orbital relaxation.} The agent proposed CAS(10,8) and used PBE0/TDA to identify both N lone pairs and the ring $\pi/\pi^*$ set. Because the desired orbitals were not consecutive in energy, it explicitly moved them into the active window. A six-root fixed-orbital CASCI screen placed mixed $\pi\pi^*$ below $n\pi^*$, whereas six-root CASSCF orbital relaxation reversed that order and made $n\pi^*$ the first excited state. The first two-root production job completed SC-NEVPT2 but missed the requested orbital-gradient threshold. A same-method restart converged and gave a 4.571~eV gap. No alternative CAS size was tested.}

\subsection{Active spaces from repeated runs}
\label{app:variability-figures}

The six figures below show the final active orbitals from the independent repetitions of $p$-nitroaniline and pyrimidine. The Original runs appear in Figs.~\ref{fig:trajectory-pna} and \ref{fig:trajectory-pyrimidine}; Table~\ref{tab:run-variability} compares all four runs for each molecule.

\begin{figure}[H]
\centering
\captionsetup{font=small}
\includegraphics[width=0.80\linewidth,height=0.28\textheight,keepaspectratio]{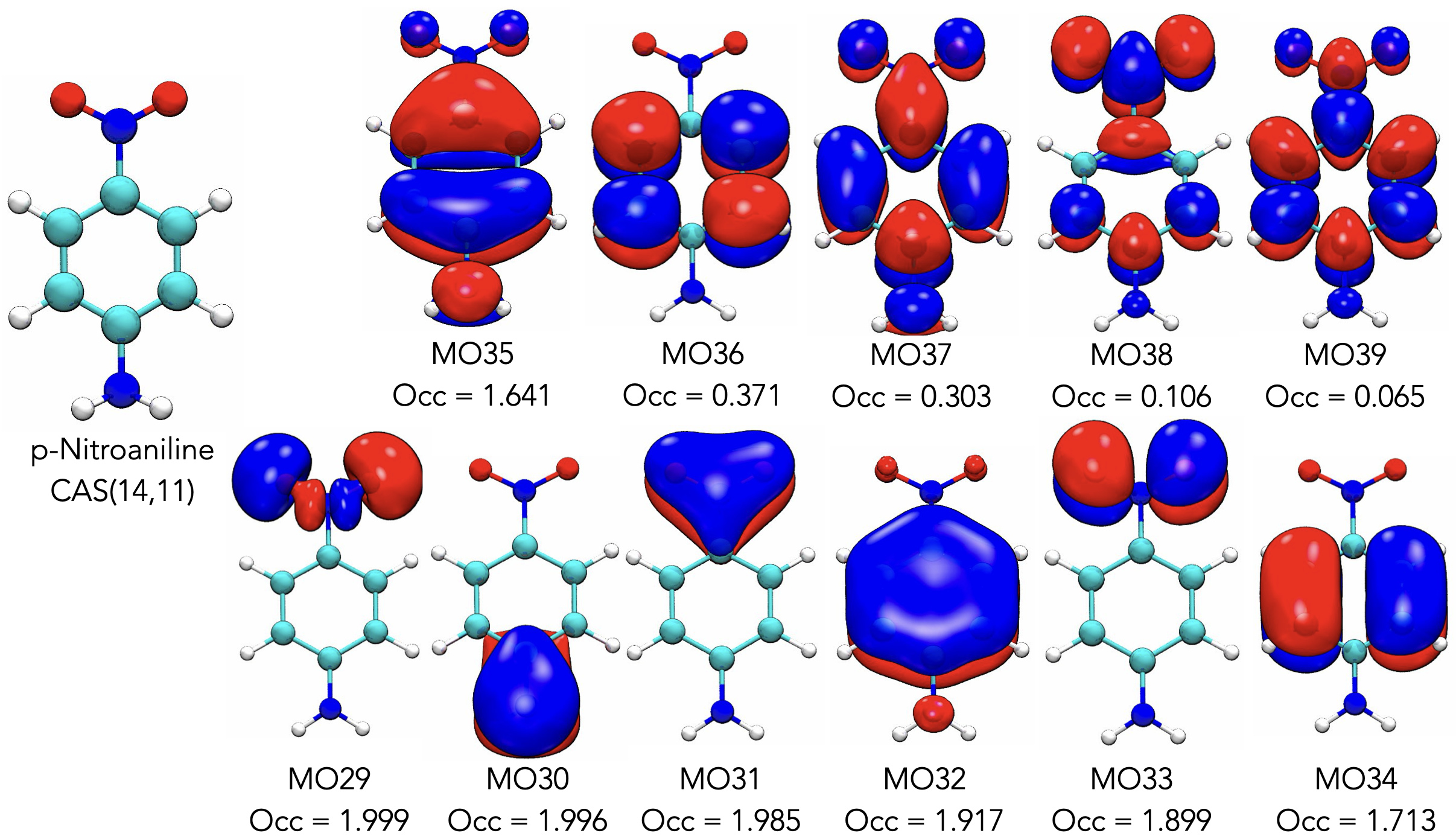}
\caption{\captionPNARepOne}
\label{fig:si-active-026-rep1}
\end{figure}

\begin{figure}[H]
\centering
\captionsetup{font=small}
\includegraphics[width=0.80\linewidth,height=0.28\textheight,keepaspectratio]{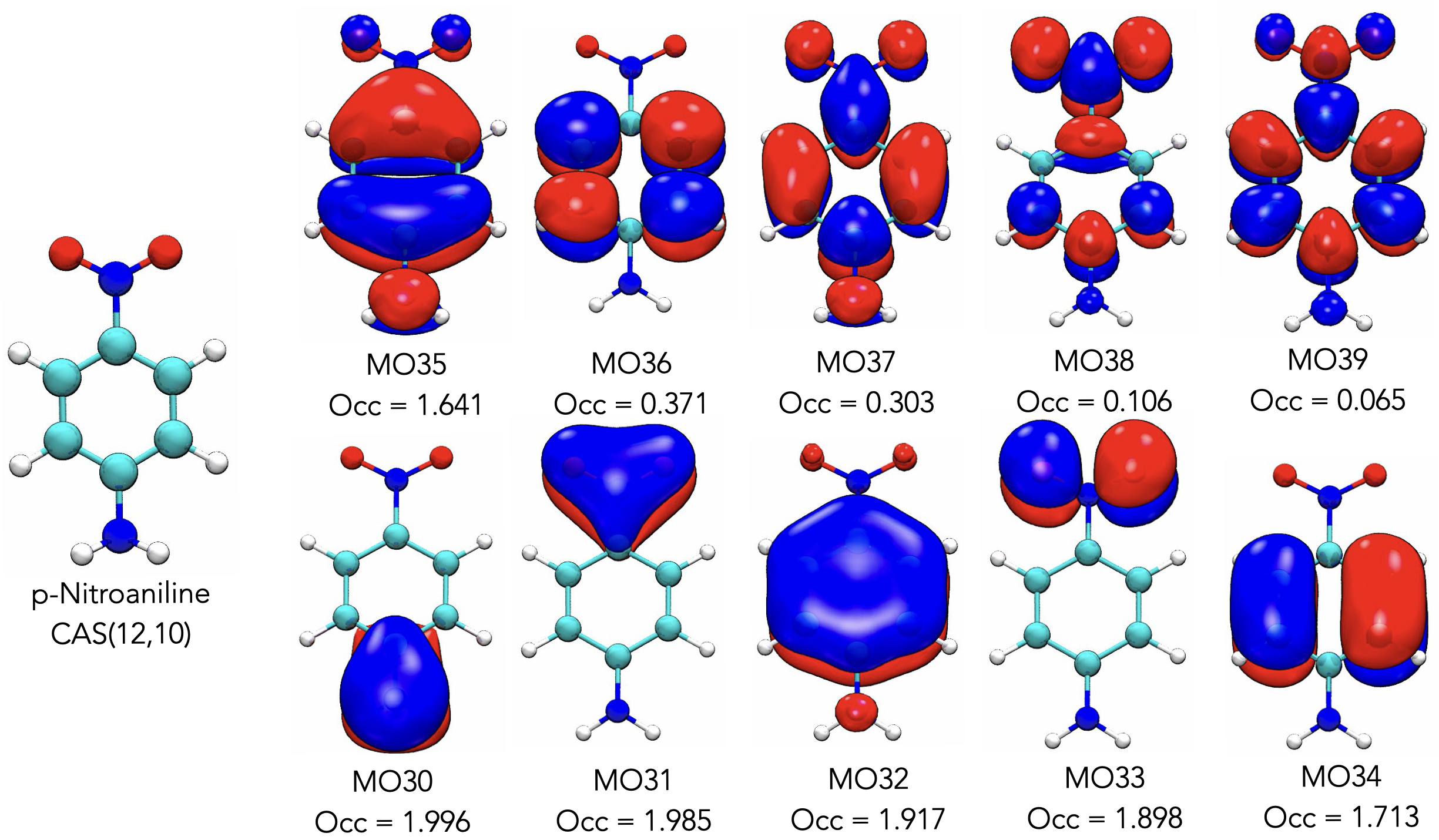}
\caption{\captionPNARepTwo}
\label{fig:si-active-026-rep2}
\end{figure}

\begin{figure}[H]
\centering
\captionsetup{font=small}
\includegraphics[width=0.80\linewidth,height=0.28\textheight,keepaspectratio]{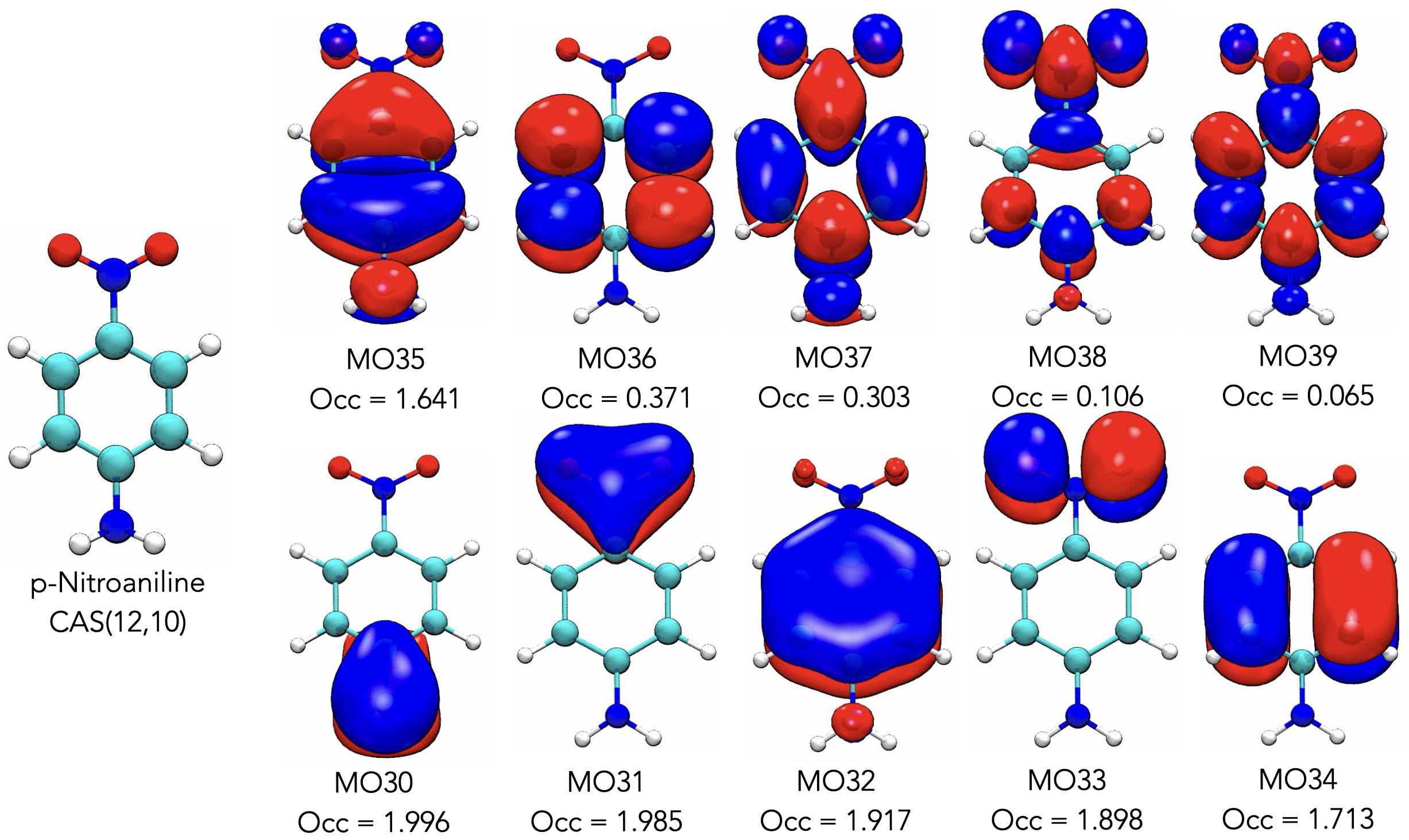}
\caption{\captionPNARepThree}
\label{fig:si-active-026-rep3}
\end{figure}

\begin{figure}[H]
\centering
\captionsetup{font=small}
\includegraphics[width=0.80\linewidth,height=0.28\textheight,keepaspectratio]{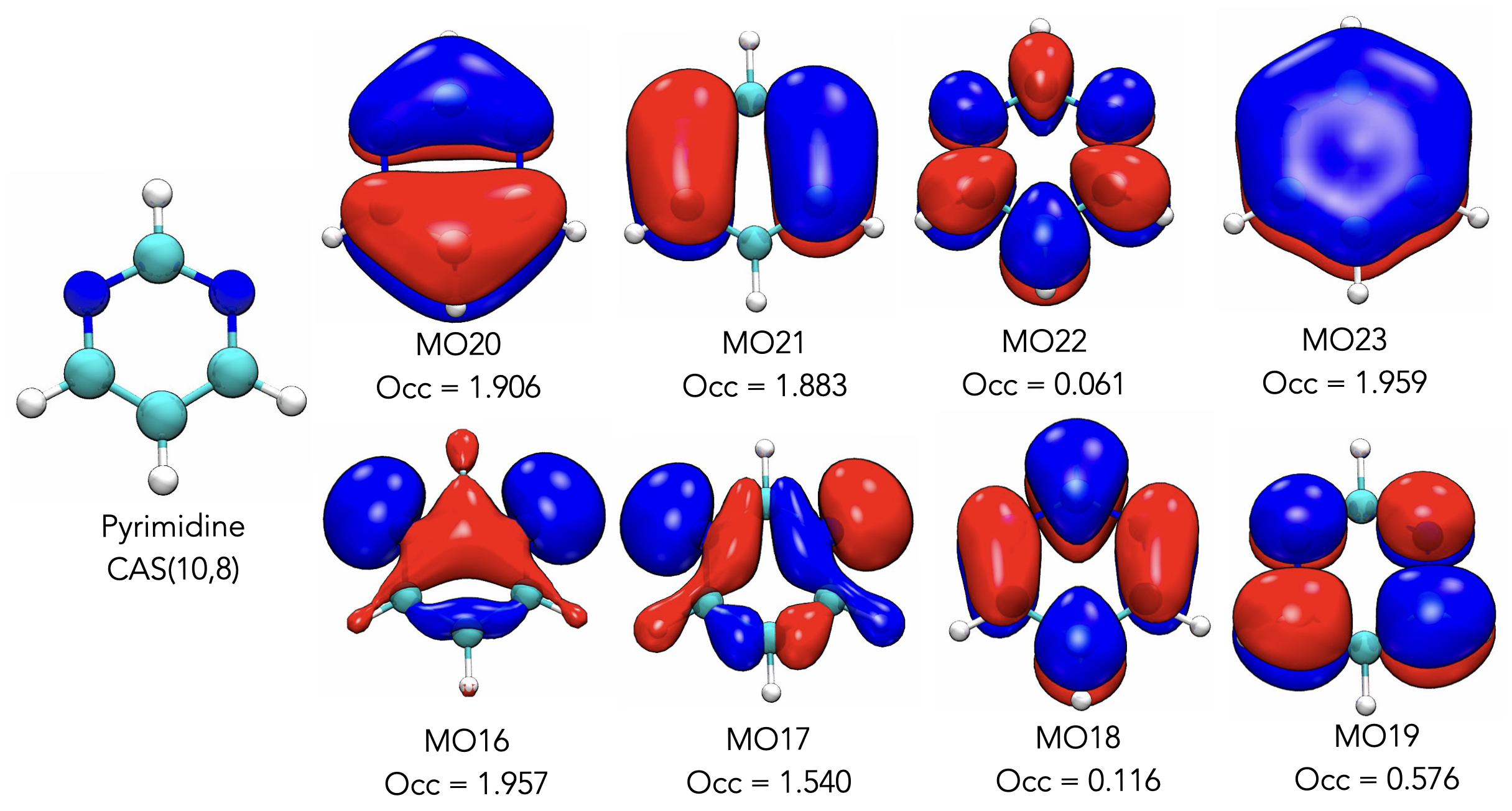}
\caption{\captionPyrimidineRepOne}
\label{fig:si-active-030-rep1}
\end{figure}

\begin{figure}[H]
\centering
\captionsetup{font=small}
\includegraphics[width=0.80\linewidth,height=0.28\textheight,keepaspectratio]{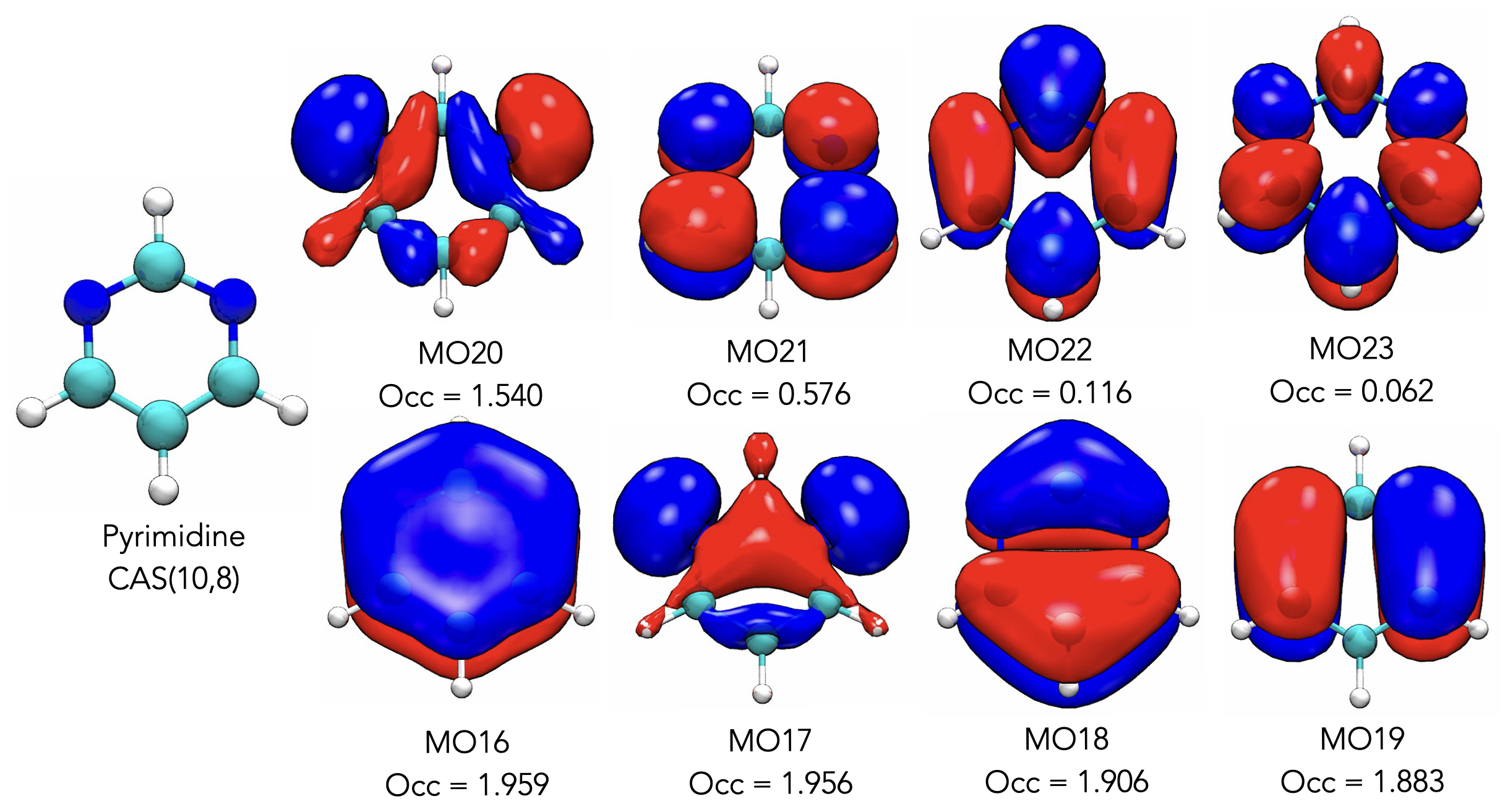}
\caption{\captionPyrimidineRepTwo}
\label{fig:si-active-030-rep2}
\end{figure}

\begin{figure}[H]
\centering
\captionsetup{font=small}
\includegraphics[width=0.80\linewidth,height=0.28\textheight,keepaspectratio]{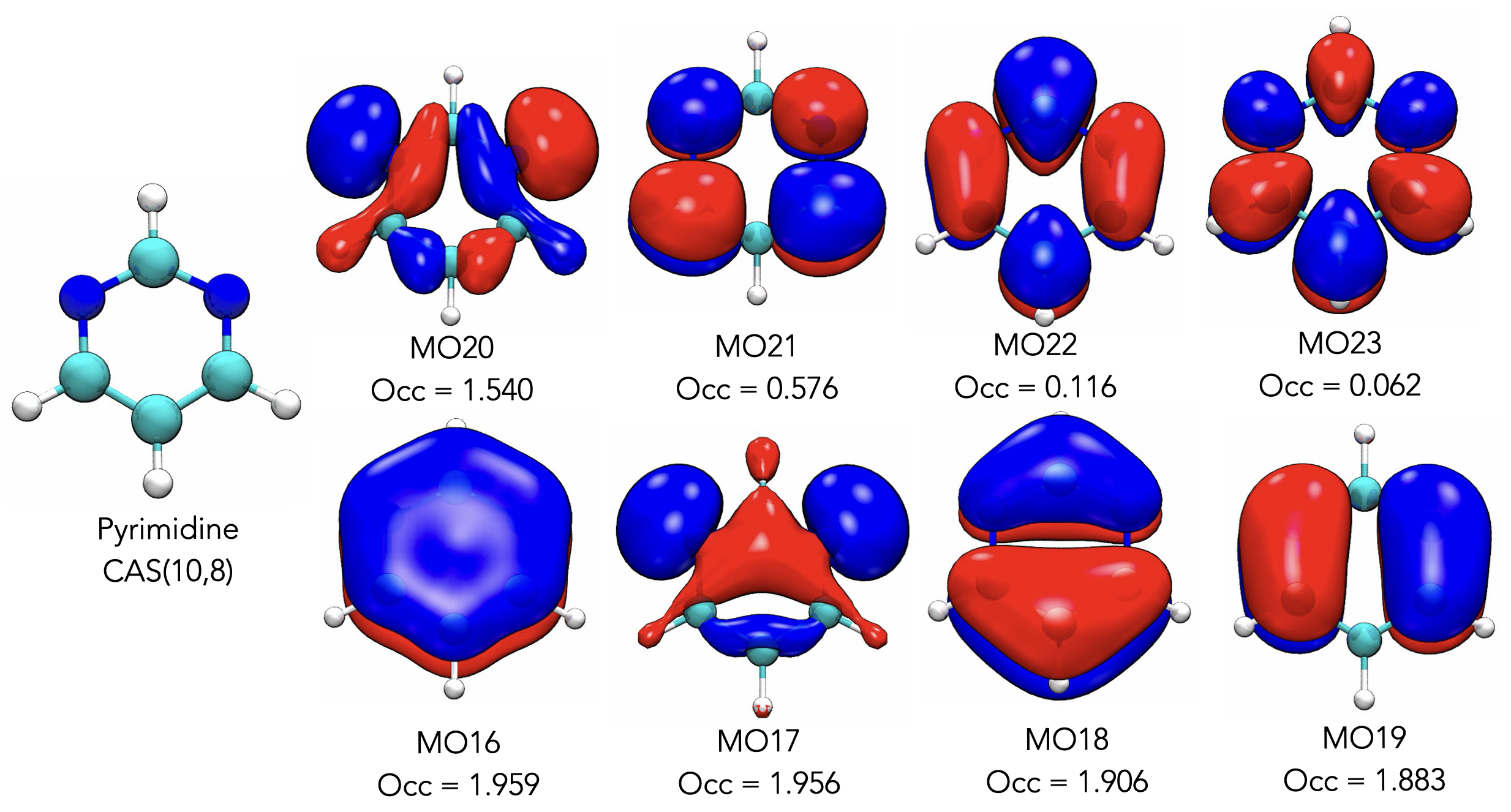}
\caption{\captionPyrimidineRepThree}
\label{fig:si-active-030-rep3}
\end{figure}

\subsection{Per-case {\cinzel Quntur} usage}
\begingroup
\scriptsize
\setlength{\tabcolsep}{2.5pt}
\renewcommand{\arraystretch}{1.08}
\begin{longtable}{@{}r >{\raggedright\arraybackslash}p{0.23\linewidth} rrrrr@{}}
\caption{LangSmith-reported token and call usage for the 32 selected {\cinzel Quntur} final-answer sessions. Search tokens are attributed to matched detached web-search calls; model and tool calls are counted within the agent trace.}\label{tab:usage-ledger-32}\\
\toprule
ID & Molecule & In-trace tokens & Search tokens & Combined tokens & Model calls & Tool calls \\
\midrule
\endfirsthead
\caption[]{Per-case LangSmith usage (continued).}\\
\toprule
ID & Molecule & In-trace tokens & Search tokens & Combined tokens & Model calls & Tool calls \\
\midrule
\endhead
\midrule\multicolumn{7}{r}{\emph{Continued on next page}}\\
\endfoot
\bottomrule
\endlastfoot
001 & acetamide & 10,047,201 & 5,214,295 & 15,261,496 & 185 & 224 \\
\rowcolor{black!7}
002 & adenine & 21,931,383 & 11,332,908 & 33,264,291 & 372 & 431 \\
003 & $p$-benzoquinone & 11,917,463 & 8,480,229 & 20,397,692 & 258 & 341 \\
\rowcolor{black!7}
004 & cyclopentadiene & 18,059,796 & 16,282,491 & 34,342,287 & 324 & 415 \\
005 & cyclopropene & 16,801,617 & 17,001,286 & 33,802,903 & 311 & 402 \\
\rowcolor{black!7}
006 & cytosine & 62,677,852 & 10,935,735 & 73,613,587 & 730 & 952 \\
007 & formamide & 7,654,810 & 6,305,110 & 13,959,920 & 193 & 218 \\
\rowcolor{black!7}
008 & all-$E$-hexatriene (Thiel) & 20,807,286 & 5,636,650 & 26,443,936 & 386 & 433 \\
009 & norbornadiene & 11,379,984 & 3,752,437 & 15,132,421 & 233 & 326 \\
\rowcolor{black!7}
010 & all-E-octatetraene & 6,687,417 & 3,602,820 & 10,290,237 & 131 & 159 \\
011 & propanamide & 14,069,123 & 6,621,017 & 20,690,140 & 263 & 478 \\
\rowcolor{black!7}
012 & pyrrole & 5,333,926 & 5,415,260 & 10,749,186 & 155 & 232 \\
013 & thymine & 4,566,212 & 7,169,427 & 11,735,639 & 149 & 227 \\
\rowcolor{black!7}
014 & $s$-triazine & 9,039,725 & 10,115,919 & 19,155,644 & 209 & 311 \\
015 & uracil & 24,015,098 & 14,108,170 & 38,123,268 & 376 & 635 \\
\rowcolor{black!7}
016 & acetaldehyde & 5,184,945 & 4,570,706 & 9,755,651 & 171 & 213 \\
017 & acetone & 18,315,917 & 13,627,233 & 31,943,150 & 324 & 494 \\
\rowcolor{black!7}
018 & benzene & 7,259,216 & 7,543,557 & 14,802,773 & 148 & 213 \\
019 & $E$-butadiene & 1,763,220 & 6,552,420 & 8,315,640 & 73 & 108 \\
\rowcolor{black!7}
020 & DMABN & 26,241,516 & 6,187,900 & 32,429,416 & 365 & 537 \\
021 & ethene & 3,735,592 & 5,228,562 & 8,964,154 & 117 & 190 \\
\rowcolor{black!7}
022 & formaldehyde & 3,344,284 & 4,678,185 & 8,022,469 & 135 & 197 \\
023 & furan & 10,042,117 & 7,245,168 & 17,287,285 & 222 & 308 \\
\rowcolor{black!7}
024 & all-$E$-hexatriene (Hoyer) & 3,842,417 & 6,387,266 & 10,229,683 & 132 & 204 \\
025 & naphthalene & 19,258,177 & 13,678,382 & 32,936,559 & 336 & 505 \\
\rowcolor{black!7}
026 & $p$-nitroaniline & 50,885,516 & 8,121,188 & 59,006,704 & 768 & 1,468 \\
027 & pyrazine & 26,635,919 & 14,618,735 & 41,254,654 & 370 & 641 \\
\rowcolor{black!7}
028 & pyridazine & 5,796,096 & 5,295,383 & 11,091,479 & 169 & 273 \\
029 & pyridine & 24,004,020 & 11,838,477 & 35,842,497 & 353 & 569 \\
\rowcolor{black!7}
030 & pyrimidine & 31,339,631 & 14,815,730 & 46,155,361 & 460 & 651 \\
031 & $s$-tetrazine & 5,836,108 & 5,731,691 & 11,567,799 & 157 & 241 \\
\rowcolor{black!7}
032 & water & 2,483,337 & 5,052,976 & 7,536,313 & 97 & 160 \\
\midrule
All 32 & Total & 490,956,921 & 273,147,313 & 764,104,234 & 8,672 & 12,756 \\
\end{longtable}
\noindent\footnotesize\emph{Notes.} Input and output tokens are included in the in-trace total; cached-read and reasoning tokens are subtypes, not extra usage. Search tokens cover matched detached web-search model calls and are added once to the in-trace total. Runner-monitor usage, abandoned sessions, and case 026's post-answer third turn are excluded. Token counts do not include ORCA compute resources. DMABN denotes 4-(dimethylamino)benzonitrile.
\endgroup

\section{Decision evidence across the 32 trajectories}
\label{app:decisions}

Table \ref{tab:decision-matrix} records one consequential scientific decision from each trajectory. ``Evidence available'' describes the observation on which the agent could act; ``agent decision and observed update'' describes the executed choice and its consequence. A statement that a choice preceded a result is made only where the audit identified a preceding trace, plan, or input. Each numbered row maps to the corresponding problem ID in Table~\ref{tab:asf-quntur-results-32}.

\begingroup
\scriptsize
\setlength{\tabcolsep}{3pt}
\renewcommand{\arraystretch}{1.14}
\begin{longtable}{@{}r >{\raggedright\arraybackslash}p{0.25\linewidth} >{\raggedright\arraybackslash}p{0.65\linewidth}@{}}
\caption{Audited decision evidence for all 32 original-prompt cases.}\label{tab:decision-matrix}\\
\toprule
ID & Evidence available & Agent decision and observed update \\
\midrule
\endfirsthead
\caption[]{Audited decision evidence (continued).}\\
\toprule
ID & Evidence available & Agent decision and observed update \\
\midrule
\endhead
\midrule\multicolumn{3}{r}{\emph{Continued on next page}}\\
\endfoot
\bottomrule
\endlastfoot
01 & Valence-orbital map and four-root diagnostic. & Explicit swaps assembled a carbonyl valence CAS(6,4); configurations and occupations then supported the selected root. The full action trace is absent, limiting the timing of the initial rationale. \\
\rowcolor{black!7}
02 & MP2 occupations and TDA transition characters. & The agent remapped orbitals into the production basis and used CAS(18,13), including N lone-pair and $\pi$ directions. Scouting ran before production; the exact prior motivation for the large size is not fully reconstructable. \\
03 & Orbital compositions and multiroot screens. & Reordered a quinone CAS(12,10), checked its target-basis transfer, and rejected an inconsistent post-CASSCF output before accepting the retry. The full decision trace is unavailable. \\
\rowcolor{black!7}
04 & Initial high-root result and a symmetry/root audit. & Replaced the first assignment with an $A_1/B_2$ state pair after checking orbital character and CASCI ordering. The recovery calculation produced the reported lower gap. \\
05 & A plausible diffuse 3s candidate and a nearby valence state. & Chose a Rydberg-capable CAS(2,3), then recovered from input, convergence, and restart problems. The computed $\pi\!\to\!3s$ state differs from the benchmark's valence transition. \\
\rowcolor{black!7}
06 & Different orbital basins in four-root screens; lone-pair loss in two-root optimization. & Recognized the active-space escape, withheld the escaped result, and reported a qualified four-root common-orbital CASCI/SC-NEVPT2 fallback. \\
07 & Carbonyl orbital and root diagnostics; one corrupted PT output. & Selected CAS(6,4) with an O lone pair and antibond, rejected the corrupted output, and recomputed a consistent root-specific energy pair. \\
\rowcolor{black!7}
08 & Full $\pi$ manifold and a nearby competing state. & Reordered a six-orbital $\pi$ space and repaired numerical inconsistencies before accepting the two-root result. \\
09 & Diffuse intrusion in the frontier orbitals and a root-selection error. & Explicitly restored valence orbitals, corrected root selection, and reported a dark valence state. A separate abandoned attempt offered only partial 3s-state evidence. \\
\rowcolor{black!7}
10 & Eight $\pi$ orbitals and a dark, double-excitation-rich root. & Assembled CAS(8,8) through scouting and deliberate swaps; the reported state followed from its configuration and zero oscillator strength. \\
11 & TDA/NTO character and a low diffuse orbital. & Rejected the diffuse orbital as a valence $\pi^*$, staged four-state then two-state averages, and kept a traceable orbital chain. The agent had consulted the ASF preprint before selection. \\
\rowcolor{black!7}
12 & Valence and diffuse active-space screens. & Tested CAS(6,5), CAS(6,6), and expanded candidates; selected the diffuse-inclusive space. Final orbital populations indicate a Rydberg-like root despite a later valence label. \\
13 & An advance acceptance rule and O-lone-pair AVAS target. & Repaired the AVAS setup, checked multiple roots, and retained a CAS(12,9) whose selected root has $n\!\to\!\pi^*$ character. \\
\rowcolor{black!7}
14 & Lone-pair/$\pi$ competition and close symmetry partners. & Compared several spaces and expanded the state manifold to retain the low-state character; the final six-root average is a qualified result relative to the prescribed two-root calculation. \\
15 & A converged $\pi\pi^*$ result in a planar calculation. & Declined to report that root as $S_1$; a symmetry-blocked average recovered an O-lone-pair-to-$\pi^*$ state. Some claimed lone-pair orbital labels were not supported by the orbital audit. \\
\rowcolor{black!7}
16 & A carbonyl CAS(4,3) pilot and a malformed root-selection request. & Corrected input validation and repeated-root PT settings; two repair jobs produced the same state-specific energy pair. Without a full trace, the repair rationale's timing is uncertain. \\
17 & A predeclared carbonyl-space test gave a $+0.337$~eV CASSCF gap shift in def2-SVP. & Retained $\sigma/\sigma^*$ with $n,\pi,\pi^*$, then revised convergence settings until the production orbitals and gradient passed its checks. \\
\rowcolor{black!7}
18 & $\pi$ orbital inspection, a large final gradient, and a shifted PT restart reference. & Rejected both superficially completed outputs and recomputed CASSCF and SC-NEVPT2 together, obtaining a consistent source for the gap. \\
19 & A full valence $\pi$ CAS and a dark first computed root. & Used an AVAS-prepared CAS(4,4) and identified the reported root as $^1A_g$-like. The benchmark value refers to the bright $^1B_u$ transition. \\
\rowcolor{black!7}
20 & A wrong orbital import and root-selection setting. & Used a coefficient-identity gate to reject an otherwise completed transfer, repaired it, and produced a consistent CAS(12,11) calculation. \\
21 & Diffuse orbitals intruding into the default active window. & Anticipated the issue, explicitly rotated in the ethene $\pi/\pi^*$ pair, and replaced a corrupt PT output. \\
\rowcolor{black!7}
22 & Formaldehyde orbital scouting and state diagnostics. & Checked the compact CAS(4,3) before PT and repeated the gate in the production basis; the final root-specific energies were reproducible. \\
23 & Near-lying 3s and valence states; an initially incorrect orbital map. & Built a Rydberg-capable CAS(6,6), inspected the final orbitals after a converged four-root validation, repaired the map, and reported the $\pi\!\to\!3s$ root. The benchmark value belongs to $\pi\pi^*$. \\
\rowcolor{black!7}
24 & $\pi$-orbital populations and basis-transfer checks. & Tested the six-orbital $\pi$ selection before production and revalidated it after basis projection. One relaxed character threshold was set after inspecting the result. \\
25 & Full $\pi$ CAS and competing symmetry-labelled states. & First obtained a $^1L_b$-consistent result, then remapped literature axes and reported a $^1L_a$ state. The orbital and CI records explain the change of assignment. \\
\rowcolor{black!7}
26 & Donor/ring and nitro orbitals, plus diffuse competitors. & Rejected a canonical selection, checked transferred orbitals, and screened five CASSCF roots. It revised an initial expectation that the donor-to-nitro charge-transfer state was $S_1$ after finding a weak energetic $S_1$. \\
27 & Two N lone pairs, the ring $\pi$ manifold, and a misleading first two-root result. & Diagnosed the $\pi\pi^*$ root from CI vectors and occupations, then targeted the $n\pi^*$ symmetry in a new calculation. \\
\rowcolor{black!7}
28 & Orbital populations and a prospective comparison of two spaces. & Retained both N lone pairs in CAS(10,8), checked the $n\pi^*$ configuration, and recovered from a restart error without changing the chemical choice. \\
29 & Close $n\pi^*$ and $\pi\pi^*$ candidates in CAS(8,7). & Recognized that the CASSCF root choice could reverse after correlation; matched SC-NEVPT2 tests favored $n\pi^*$ over $\pi\pi^*$. \\
\rowcolor{black!7}
30 & Four-root scout and a later GBW-reference mismatch. & Expanded the root check, rejected an unreliable restart, and reconverged CASSCF with all-root PT in one job. A tapered average tested sensitivity. \\
31 & Symmetry/population map and a diffuse intruder. & Chose CAS(14,10) in advance, validated the orbital characters and state continuity, then ran the prescribed two-root production calculation. \\
\rowcolor{black!7}
32 & Proposed water acceptor orbital and one symmetry-targeted excited root. & Kept the $4a_1$-like acceptor in CAS(8,10) and obtained a consistent $1b_1\!\to\!4a_1$ candidate after a technical retry. \\
\end{longtable}
\endgroup

\section{{\cinzel Quntur} agent roles}
\label{app:quntur-agents}

Table~\ref{tab:quntur-agents} lists the 59 roles in the supplied {\cinzel Quntur} agent graph and their configured model tiers: Sol denotes \texttt{gpt-5.6-sol} and Terra denotes \texttt{gpt-5.6-terra}. Across the 32 selected final-answer sessions, LangSmith records 5,471 in-trace model calls to Sol and 3,201 to Terra. The graph describes available delegation paths; participation in an individual benchmark session is established by its trace, not by this registry. Only 24 named roles in the table have attributable model calls in these sessions; one Terra call has no attributed role.

\begingroup
\small
\setlength{\tabcolsep}{3pt}
\renewcommand{\arraystretch}{1.14}
\begin{longtable}{@{}p{0.30\linewidth}p{0.56\linewidth}l@{}}
\caption{Registered {\cinzel Quntur} agents and their configured model tier.}\label{tab:quntur-agents}\\
\toprule
Name & Description & Model \\
\midrule
\endfirsthead
\caption[]{Registered {\cinzel Quntur} agents and model tiers (continued).}\\
\toprule
Name & Description & Model \\
\midrule
\endhead
\midrule\multicolumn{3}{r}{\emph{Continued on next page}}\\
\endfoot
\bottomrule
\endlastfoot
\texttt{autoci\_expert} & AUTOCI correlation and multireference settings. & Sol \\
\rowcolor{black!7}
\texttt{basis\_expert} & Orbital and auxiliary basis sets and effective core potentials. & Terra \\
\texttt{casresp\_expert} & CP-CASSCF response-solver settings. & Sol \\
\rowcolor{black!7}
\texttt{casscf\_expert} & Active-space, state-averaging, and multireference advice. & Sol \\
\texttt{cim\_expert} & Cluster-in-Molecule correlation settings. & Terra \\
\rowcolor{black!7}
\texttt{cis\_tddft\_expert} & CIS, TDDFT, and spin-flip TDA settings. & Sol \\
\texttt{compound\_expert} & ORCA Compound scripting and multistep logic. & Sol \\
\rowcolor{black!7}
\texttt{computational\_chemist} & Scientific planning, user approval, interpretation, and review. & Sol \\
\texttt{configuration\_recommender} & Recommends calculation configurations from the research objective. & Sol \\
\rowcolor{black!7}
\texttt{conical\_expert} & Conical-intersection optimization settings. & Terra \\
\texttt{cosmors\_expert} & OpenCOSMO-RS solvation settings. & Terra \\
\rowcolor{black!7}
\texttt{cpcm\_expert} & CPCM implicit-solvation settings. & Sol \\
\texttt{crest\_expert} & CREST conformer and rotamer searches. & Sol \\
\rowcolor{black!7}
\texttt{docker\_expert} & Host--guest docking settings. & Terra \\
\texttt{elprop\_expert} & Electric-property and response settings. & Terra \\
\rowcolor{black!7}
\texttt{eprnmr\_expert} & EPR, NMR, and magnetic-response settings. & Sol \\
\texttt{esd\_expert} & Vibronic spectra and excited-state dynamics settings. & Sol \\
\rowcolor{black!7}
\texttt{freq\_expert} & Frequency and thermochemistry settings. & Terra \\
\texttt{geom\_expert} & Geometry optimization, constraints, scans, and transition states. & Sol \\
\rowcolor{black!7}
\texttt{geometry\_generation} & Molecular geometry creation and manipulation. & Sol \\
\texttt{geometry\_section} & Molecular geometry specification in ORCA inputs. & Terra \\
\rowcolor{black!7}
\texttt{goat\_expert} & ORCA GOAT conformer-search settings. & Terra \\
\texttt{ice\_expert} & Iterative Configuration Expansion CI settings. & Terra \\
\rowcolor{black!7}
\texttt{imaginary\_frequency\_expert} & Imaginary-mode analysis and displaced geometries. & Terra \\
\texttt{input\_file\_service} & Specialist routing and backend-specific artifact creation. & Sol \\
\rowcolor{black!7}
\texttt{input\_keywords\_expert} & Standalone ORCA input directives. & Terra \\
\texttt{interact\_with\_os\_and\_files} & Post-calculation parsing and numerical analysis. & Terra \\
\rowcolor{black!7}
\texttt{irc\_expert} & Intrinsic reaction coordinate settings. & Terra \\
\texttt{loc\_expert} & Molecular-orbital localization settings. & Sol \\
\rowcolor{black!7}
\texttt{mcrpa\_expert} & CASSCF-based MC-RPA excitation settings. & Sol \\
\texttt{md\_expert} & Molecular-dynamics settings. & Terra \\
\rowcolor{black!7}
\texttt{mdci\_expert} & MDCI correlation and excited-state solver settings. & Sol \\
\texttt{mecp\_expert} & Minimum-energy crossing-point optimization settings. & Terra \\
\rowcolor{black!7}
\texttt{method\_expert} & Global method, DFT, and external-method settings. & Terra \\
\texttt{mp2\_expert} & MP2-family correlation settings. & Terra \\
\rowcolor{black!7}
\texttt{mrcc\_expert} & External MRCC interface settings. & Terra \\
\texttt{mrci\_expert} & Multireference CI and perturbation settings. & Sol \\
\rowcolor{black!7}
\texttt{mtr\_expert} & Vibrational mode-tracking scans. & Terra \\
\texttt{nbo\_expert} & External NBO interface settings. & Terra \\
\rowcolor{black!7}
\texttt{ndoparas\_expert} & Semiempirical NDO model parameters. & Terra \\
\texttt{neb\_expert} & Nudged elastic band reaction-path settings. & Terra \\
\rowcolor{black!7}
\texttt{numgrad\_expert} & Numerical-gradient controls. & Terra \\
\texttt{opi\_runner} & OPI/ORCA script submission and execution checks. & Terra \\
\rowcolor{black!7}
\texttt{orca\_utilities} & ORCA postprocessing utilities. & Terra \\
\texttt{output\_expert} & ORCA output and auxiliary-file settings. & Terra \\
\rowcolor{black!7}
\texttt{pal\_expert} & Parallel-resource and memory recommendations. & Terra \\
\texttt{paras\_expert} & Named input parameters and scan values. & Terra \\
\rowcolor{black!7}
\texttt{plots\_expert} & Orbital and density plot settings. & Terra \\
\texttt{pyscf\_runner} & PySCF script submission and execution checks. & Terra \\
\rowcolor{black!7}
\texttt{quantum\_chemistry\_calculation\_expert} & Backend choice, artifact coordination, and execution checks. & Sol \\
\texttt{rel\_expert} & Relativistic Hamiltonian and spin--orbit settings. & Sol \\
\rowcolor{black!7}
\texttt{rocis\_expert} & ROCIS spectroscopy settings. & Terra \\
\texttt{rr\_expert} & Resonance Raman settings. & Terra \\
\rowcolor{black!7}
\texttt{run\_orca} & Classic ORCA submission and execution checks. & Terra \\
\texttt{run\_xtb} & xTB and g-xTB submission and execution checks. & Terra \\
\rowcolor{black!7}
\texttt{runtypes\_keyword\_expert} & Basic calculation and run-type keywords. & Terra \\
\texttt{scf\_expert} & SCF convergence and wavefunction settings. & Terra \\
\rowcolor{black!7}
\texttt{solvator\_expert} & Automated explicit-solvent placement settings. & Terra \\
\texttt{symmetry\_expert} & Explicit molecular-symmetry controls. & Terra \\
\end{longtable}
\endgroup

\section{Registered {\cinzel Quntur} tools}
\label{app:quntur-tools}

Table~\ref{tab:quntur-tools} lists the 77 tools in the supplied registry. Registration does not imply that every tool was connected to every agent or used in a benchmark session; the case traces provide evidence of actual calls.

\begingroup
\small
\setlength{\tabcolsep}{3pt}
\renewcommand{\arraystretch}{1.14}
\begin{longtable}{@{}p{0.35\linewidth}p{0.60\linewidth}@{}}
\caption{Tools registered by {\cinzel Quntur}; registration alone does not imply an agent connection.}\label{tab:quntur-tools}\\
\toprule
Name & Description \\
\midrule
\endfirsthead
\caption[]{{\cinzel Quntur} registered tools (continued).}\\
\toprule
Name & Description \\
\midrule
\endhead
\midrule\multicolumn{2}{r}{\emph{Continued on next page}}\\
\endfoot
\bottomrule
\endlastfoot
\texttt{submit\_python\_job} & Submit a general Python compute job. \\
\rowcolor{black!7}
\texttt{submit\_crest\_job} & Run CREST and collect conformers. \\
\texttt{python\_repl} & Run bounded session-local Python code. \\
\rowcolor{black!7}
\texttt{pyscf\_python\_repl} & Use a PySCF-capable Python environment. \\
\texttt{opi\_python\_repl} & Use the OPI Python environment. \\
\rowcolor{black!7}
\texttt{validate\_opi\_workflow} & Validate an OPI workflow contract. \\
\texttt{run\_static\_workflow} & Dry-run or execute a reusable workflow. \\
\rowcolor{black!7}
\texttt{list\_session\_workflows} & List session workflow drafts. \\
\texttt{inspect\_session\_workflow} & Inspect one workflow draft. \\
\rowcolor{black!7}
\texttt{xtb\_geometry\_optimization} & Optimize geometries with xTB. \\
\texttt{constraint\_xtb} & Run constrained xTB optimization. \\
\rowcolor{black!7}
\texttt{codex\_workflow\_engineer} & Create or update a reusable workflow draft. \\
\texttt{submit\_slurm\_jobs} & Submit classic ORCA jobs. \\
\rowcolor{black!7}
\texttt{manage\_compute\_jobs} & Check, harvest, or cancel compute batches. \\
\texttt{submit\_opi\_slurm\_jobs} & Submit saved OPI/ORCA scripts. \\
\rowcolor{black!7}
\texttt{submit\_pyscf\_slurm\_jobs} & Submit saved PySCF scripts. \\
\texttt{get\_cluster\_resources} & Read available cluster resources. \\
\rowcolor{black!7}
\texttt{submit\_xtb\_jobs} & Submit batched xTB or g-xTB jobs. \\
\rowcolor{black!7}
\texttt{local\_system\_operations} & Alias for the session Python REPL. \\
\texttt{make\_metal\_complex} & Build metal-complex geometries with Architector. \\
\rowcolor{black!7}
\texttt{find\_ligand} & Look up a ligand by name or SMILES. \\
\texttt{check\_imaginary\_frequency} & Detect imaginary vibrational frequencies. \\
\rowcolor{black!7}
\texttt{parse\_distored\_xyz\_image} & Extract displaced XYZ geometry from mode data. \\
\texttt{create\_distorted\_geometry\_gaussian} & Generate Gaussian normal-mode displacements. \\
\rowcolor{black!7}
\texttt{generate\_trajectory\_files} & Generate vibrational-mode trajectories. \\
\texttt{update\_global\_memory} & Record shared session findings. \\
\rowcolor{black!7}
\texttt{update\_modular\_workflow\_artifacts} & Record reusable workflow artifact provenance. \\
\texttt{inorganic\_molecules\_xyz\_generation} & Generate XYZ geometry with Open Babel. \\
\rowcolor{black!7}
\texttt{organic\_molecules\_xyz\_generation} & Generate organic XYZ geometry with RDKit. \\
\texttt{inspect\_files} & Read, search, and query exact session files. \\
\rowcolor{black!7}
\texttt{manage\_session\_files} & Create, copy, move, or archive session files. \\
\texttt{read\_file\_content} & Read file content through the older interface. \\
\rowcolor{black!7}
\texttt{molecules\_system\_xyz\_generation} & Generate multicomponent XYZ geometry. \\
\texttt{get\_charge\_and\_multiplicity} & Estimate charge and multiplicity from SMILES. \\
\rowcolor{black!7}
\texttt{ask\_for\_clarifications} & Ask the user for a missing decision. \\
\texttt{verify\_orca\_input\_file} & Flag unrecognized ORCA input components. \\
\rowcolor{black!7}
\texttt{orca\_plot\_menu} & Inspect available orca\_plot options. \\
\texttt{orbital\_generation} & Export orbital or density volumetric data. \\
\rowcolor{black!7}
\texttt{orca\_vib} & Run Hessian-based vibrational postprocessing. \\
\texttt{orca\_2json} & Convert ORCA property output to JSON. \\
\rowcolor{black!7}
\texttt{orca\_mapspc} & Postprocess computed spectra. \\
\texttt{orca\_chelpg} & Compute CHELPG or RESP-style charges. \\
\rowcolor{black!7}
\texttt{orca\_2mkl} & Convert ORCA orbital containers. \\
\texttt{mokit\_fch2mkl} & Convert formatted checkpoint orbitals to MKL. \\
\rowcolor{black!7}
\texttt{Fukui\_cube} & Generate Fukui-function cube data. \\
\texttt{orbital\_render} & Render cube data to PNG. \\
\rowcolor{black!7}
\texttt{view\_png\_file} & Display a PNG image. \\
\texttt{view\_agent\_hierarchy} & Inspect agent delegation relationships. \\
\rowcolor{black!7}
\texttt{list\_available\_sections} & Locate sections in ORCA output. \\
\texttt{generate\_fchk\_files} & Generate Gaussian formatted checkpoint files. \\
\rowcolor{black!7}
\texttt{index\_fchk\_file} & Index formatted-checkpoint fields. \\
\texttt{list\_fchk\_fields} & List formatted-checkpoint fields. \\
\rowcolor{black!7}
\texttt{inspect\_fchk\_field\_metadata} & Describe a formatted-checkpoint field. \\
\texttt{retrieve\_fchk\_scalar} & Retrieve one formatted-checkpoint scalar. \\
\rowcolor{black!7}
\texttt{retrieve\_fchk\_array\_slice} & Retrieve a bounded checkpoint array slice. \\
\texttt{organic\_molecules\_conformer\_generation} & Generate and rank RDKit conformers. \\
\rowcolor{black!7}
\texttt{web\_search} & Search chemistry sources and documentation. \\
\texttt{view\_molecule} & Display XYZ molecular structures. \\
\rowcolor{black!7}
\texttt{zoom\_molecule} & Display a detailed molecular view. \\
\texttt{pdf\_reader} & Ask a question about a PDF. \\
\rowcolor{black!7}
\texttt{index\_pdf\_corpus} & Index PDFs for session-local retrieval. \\
\texttt{search\_pdf\_corpus} & Search indexed PDF passages. \\
\rowcolor{black!7}
\texttt{literature\_search} & Answer literature questions with citations. \\
\texttt{add\_reference} & Add DOI records to references.bibtex. \\
\rowcolor{black!7}
\texttt{view\_directory} & Inspect one session-directory level. \\
\end{longtable}
\endgroup

\fi


\end{document}